\documentclass[12pt, a4paper, twoside]{report}
\usepackage[bindingoffset=0.5cm]{geometry}	
\usepackage[utf8]{inputenc}
\usepackage{amsmath,amssymb,amsfonts}
\usepackage{hyperref}
\usepackage{textcomp}
\usepackage[dvips,pdftex]{graphicx}
\usepackage{flafter}
\usepackage{booktabs}
\usepackage{array}
\usepackage{paralist}
\usepackage{dsfont}
\usepackage{caption}
\usepackage{subcaption}
\usepackage{float}
\usepackage{fancyhdr}
\usepackage{pict2e}
\usepackage{lastpage}
\usepackage{mathtools}
\usepackage{amsthm}
\usepackage{IEEEtrantools}
\usepackage{bbold}
\usepackage[all]{nowidow}
\usepackage{color, soul}					
\definecolor{lightblue}{rgb}{.90,.95,1}	
\sethlcolor{lightblue}					
\usepackage{courier}
\usepackage{subcaption}					
\usepackage[german,english]{babel}		
\usepackage{pdfpages}					
\usepackage{xcolor}						
\usepackage{soul}						
\usepackage[makeroom]{cancel}			
\usepackage[nottoc,numbib]{tocbibind}	
\usepackage{framed}						
\usepackage{lscape}						

\newcommand{\specialcell}[2][c]{%
  \begin{tabular}[#1]{@{}c@{}}#2\end{tabular}}

\newcommand{\ud}{\mathrm{d}}

\theoremstyle{remark}	

\begin{document}

\pagenumbering{roman}

\title{Two-mode Phonon Squeezing in Bose-Einstein Condensates for Gravitational Wave Detection}
\author{Paul Juschitz}
\date{12.01.2021}

\maketitle

\begin{center}
	\chapter*{Abstract}
\end{center}

Squeezed, nonclassical states are an integral tool of quantum metrology due to their ability to push the sensitivity of a measurement apparatus beyond the limits of classical states. While their creation in light has become a standard technique, the production of squeezed states of the collective excitations in gases of ultracold atoms, the phonons of a Bose-Einstein condensate (BEC), is a comparably recent problem. This task is continuously gaining relevance with a growing number of proposals for BEC-based quantum metrological devices and the possibility to apply them in the detection of gravitational waves. The objective of this thesis is to find whether the recently described effect of an oscillating external potential on a uniform BEC can be exploited to generate two-mode squeezed phonon states, given present day technology. This question brings together elements of a range of fields beyond cold atoms, such as general relativity and Efimov physics. To answer it, the full transformation caused by the oscillating potential on an initially thermal phononic state is considered, allowing to find an upper bound for the magnitude of this perturbation as well as to quantify the quality of the final state with respect to its use in metrology. These findings are then applied to existing experiments to judge the feasibility of the squeezing scheme and while the results indicate that they are not well suited for it, a setup is proposed that allows for its efficient implementation and seems within experimental reach. In view of the vast parameter space leaving room for optimization, the considered mechanism could find applications not only in the gravitational wave detector that originally motivated this work, but more generally in the field of quantum metrology based on ultracold atoms.

\begin{center}
	\chapter*{Acknowledgements}
\end{center}

It may be a cliché, but there is no way around it: there are far more people that I have to thank than I can name here.

My deep gratitude goes to my supervisor, Ivette Fuentes. Thank you, Ivette, for sparking my interest in your field when you came to Vienna, for giving me the opportunity to come to Nottingham to work with you and finally for your ongoing support. I genuinely appreciate the way you had confidence in me and made me a part of your group from day one. Without your ideas, your passion and also your patience, this thesis could not have become what it is. Richard Howl has provided crucial advice and guidance far beyond what should be expected, bridging multiple time zones in doing so. I did not take this for granted and it played an essential part in me finishing this work.

The immediate helpfulness I encountered when writing to other groups on my quest for experimental data amazed me. I would particularly like to thank Lena Hannah Dogra, Randall Hulet, Peter van der Straten and Rudolf Grimm for their input and the fruitful discussions.

A big thank you goes to my friends and future or former colleagues who encouraged me to take the steps that led me here and kept me company along the way: Ana Lucía Báez Camargo, Jan Kohlrus, Daniel Hartley, Luis Cortés Barbado, Josh Morris and Maxime Debiossac. It made a tremendous difference knowing that you are there to help.

To Stephan Huimann: your friendship, the constant source of calm you are and finally the trail you have blazed to us finishing our theses lifted a big weight off my shoulders.

Lastly, I cannot express the gratitude I feel for my family. The sheer amount of support all of you gave me in the last few months, be it practical or emotional, was invaluable and incredible. It may have been me who wrote what follows, but I could not have done it without you.

\newpage

\tableofcontents

\newpage

\pagenumbering{arabic}

\chapter*{Introduction}
\addcontentsline{toc}{chapter}{Introduction}	

Perhaps the greatest endeavour in experimental physics over the past century was the measurement of gravitational waves (GWs). Even though they had been predicted by Einstein in 1916, their exceedingly small amplitudes allowed them to evade detection for almost exactly a hundred years, until a first signal of a black hole merger was finally captured in September 2015, ringing in a new phase of astronomy beyond electromagnetic radiation. The remarkable device to achieve this feat was LIGO, in its essence a Michelson interferometer with an arm length of $4$km. To this day, interferometers have remained the only apparatus to successfully observe GW events and in their ongoing development, quantum metrology\footnote{Quantum metrology is the application and exploitation of elements unique to quantum theory to devise and improve the sensitivity of measurements.} and in particular the injection of squeezed states (often also referred to as twin-beam states) of light plays a prominent role \cite{the_ligo_scientific_collaboration_enhanced_2013}. This class of states has the remarkable feature that the uncertainty in one of their observables can be pushed (squeezed) to, in principle, arbitrarily low values.

Only a short time before the first GW measurement, a proposal for a detector based on the collective excitations of an ultracold state of matter, the phonons on a Bose-Einstein condensate (BEC), was published \cite{sabin_phonon_2014}. This device, if realized, would not only allow for more compact observatories, but could also extend their frequency range beyond that of LIGO's peak sensitivity and thereby allow to capture signals originating from different sources. Because the imprint a gravitational wave leaves on the phonons populating a BEC is strongly tied to their degree of squeezing, the creation of an initial squeezed state plays a crucial part. This was the origin that sparked the idea for this thesis: finding \textit{how} and \textit{how well} a pair of frequency modes can be squeezed and thereby providing a piece of the puzzle that makes up the realization of the proposed detector.

While the generation of twin-beam states of light dates back to 1985 \cite{slusher_observation_1985} and has meanwhile become a standard technique in quantum optics experiments, the experimental realization of BECs \textit{themselves} only happened ten years later \cite{davis_bose-einstein_1995, anderson_observation_1995, bradley_evidence_1995}. The task of finding a feasible scheme to produce squeezed phonon modes is thus a young one that has, to the knowledge of the author, not been answered conclusively to this day.

The starting point for the present work was a recent article by R\"atzel et al. \cite{ratzel_dynamical_2018}, who found that one of the effects an oscillating mass exerts onto a nearby BEC is phonon squeezing. Their proposal was to detect this influence of the rather small gravitational potential and thereby measure the imprint left by the nearby mass. Here, however, the suggestion is made to \textit{exploit} this ability of a periodic potential to create phononic twin-beam states, using a source that is not gravitational and can be controlled more easily while producing higher amplitudes.

From this point on, the work that ensued for this thesis encompassed considering the full time evolution found in \cite{ratzel_dynamical_2018} and its application to an initial thermal state, resulting in a scheme that promises the efficient creation of two-mode phonon squeezing and the determination of a set of experimental parameters it would be well suited for. This scheme, if realized in a laboratory, could then not only be put into use for the gravitational wave detector that motivated it, but also in a much broader context of quantum metrology applications.

Because this thesis uses elements from a vast array of different fields, part \ref{part:1} is dedicated to providing the reader with the concepts most essential to the considerations that follow.

Chapter \ref{chap:BEC_basics} contains the basics of the description of Bose-Einstein condensates, with particular focus on the uniform, weakly interacting Bose gas, the model which will be used later in the proposal for the squeezing scheme. It is also where the first encounter with \textit{Bogoliubov transformations} and the phonon picture takes place, leading to limitations to phonon life times through damping and finally to a quick rundown of condensates in harmonic traps, which will be used later to approximate existing experiments as estimates for the squeezing scheme.

The subsequent chapters \ref{sec:feshbach_resonances} and \ref{sec:three-body collisions} are centered around the possibility of tuning the interaction strength between the condensate particles and the consequences of doing so. Feshbach resonances have become an important tool in BEC experiments in that they provide a mechanism to control the scattering length of the atoms. This ability, however, can result in enhanced three-body losses, where the collision of three particles causes the formation of a molecule and the redistribution of its binding energy as kinetic energy. As a result, the particles participating in the collision are lost from the trap. This raises the question of how the interaction strength and the three-body recombination rate are related, which leads to the field of Efimov physics. In his work, Efimov predicted the emergence of an attractive three-body potential in the regime of large scattering lengths, with the striking consequence of resonances, where (for positive scattering lengths) local minima in the three-body loss rate appear. These Efimov resonances can then be used to extend the life time of a condensate that would otherwise decay quickly at similar interaction strengths.

Since squeezed states are an integral concept to this thesis, a proper definition of what they are was indispensable, although it has proven more subtle than initially thought. An overview of the multitude of historical definitions can be found in chapter \ref{sec:squeezed_states}, as well as the most important features that distinguish single- and two-mode squeezed states from classical ones.

Part \ref{part:1} then concludes with chapter \ref{sec:gravitational_waves} and a very brief introduction to gravitational waves in the transversal traceless gauge as they are considered in the proposed GW detector.

The core of the thesis is made up by part \ref{part:2}, starting with the why and how in chapters \ref{sec:detector_proposal} and \ref{sec:squeezing mechanism}: in the former, the reader can find a description of the original idea of the phononic GW detector and the ongoing work involving \textit{frequency interferometry}. It is here where we will see the asymptotically exponential scaling of detector's sensitivity with the initial squeezing, necessitating a way to efficiently produce such states. Chapter \ref{sec:squeezing mechanism} then presents the how by revisiting the derivation of the effect a periodic external potential exerts onto the phonon states populating a BEC. In order to do so, an ansatz from quantum field theory is followed here.

Everything so far leads up to the findings of the author, which are presented in chapter \ref{sec:implementation}. It begins with a first approach to simply maximize the squeeze factor found in the previous chapter. This ansatz, however, would lead to the immediate destruction of the condensate, depending on the experimental parameters. Consequently, a phonon budget has to be set as a small fraction of the number of particles in the condensate. The full transformation of a state that is initially thermal under the time evolution given in the previous chapter is then considered. This leads to the derivation of the number of quasi-particles produced, given a desired squeeze factor. From the same transformation, the overlap of the final state with the ideal of the squeezed vacuum is quantified in terms of the temperature of the Bose gas and its frequency spectrum. The rather complicated expressions found for the final states did not allow for analytical or numerical optimization over the parameter space of conceivable experiments within the scope of this thesis. Nonetheless, some qualitative arguments are made on possible paths towards optimization and the bounds of said parameter space set by previously made approximations. Existing (present-day and discontinued) experimental setups are then investigated for their potential for the application of the found squeezing mechanism and finally, parameters for an experiment that would be well suited for its implementation are presented.

Chapter \ref{chap:conclusion} concludes with a few final remarks and a summary of questions that remain to be addressed in future work.


The interested reader can look up tools, techniques and derivations that were used throughout the thesis in the appendix, including references to literature providing more details. Finally, the Mathematica code that was written to find the state transformation can be found in the appendix chapter \ref{chap:app_mathematica_transformation}.

\part{Essential Concepts \label{part:1}}


\chapter{Bose-Einstein Condensates \label{chap:BEC_basics}}

\section{History}

Following the work of Bose \cite{bose_plancks_1924} on photon statistics from 1924, Einstein made the prediction of a phase transition in a noninteracting atomic gas in the next year \cite{einstein_quantentheorie_1925}: below a critical temperature, a major part of the particles \textit{condensate} in their energetic ground state and form a \textit{Bose-Einstein condensate} (BEC).

While it would take a long time until this phenomenon should be observed in an experiment, the theoretical steps characterizing the subsequent decades included the connection to superfluidity made by London in 1938 \cite{london_lambda-phenomenon_1938} and Bogoliubov's 1947 theory for interacting bosonic gases leading to a description of elementary excitations \cite{bogoliubov_theory_1947}. The latter is of utmost importance for the purposes of this thesis since, in the low energy limit, these excitations were found to behave as phonons. Derived in 1961, the Gross-Pitaevskii equation \cite{gross_structure_1961, pitaevskii_vortex_1961} describes the weakly interacting, non-uniform Bose gas in the zero temperature limit. Despite its early discovery, it gained most attention only after BECs had been experimentally observed.

The first atomic species that was considered a promising candidate and in fact brought close to Bose-Einstein condensation was spin-polarized hydrogen in the 1970s. With the arrival of advanced trapping and cooling techniques, however, alkali atoms took over, leading to the first experimental realization of a BEC in 1995. The first two groups to succeed were the ones of Ketterle at MIT and Cornell and Wieman at Boulder, using gases of $^{23}$Na and $^{87}$Rb, respectively \cite{davis_bose-einstein_1995, anderson_observation_1995}. In the same year, Hulet's group at rice university found evidence for a BEC in $^7$Li \cite{bradley_evidence_1995}. Over the next years, $^1$H, $^4$He$^*$\footnote{Atomic $^4$He in an excited state}, $^{39}$K, $^{41}$K, $^{52}$Cr $^{85}$Rb, $^{133}$Cs, $^{170}$Yb, $^{174}$Yb have followed suit.

\section{The Noninteracting Bose Gas: Critical Temperature and Condensation \label{sec:ideal_bose_gas}}

The following paragraphs provide a walk-through of the signature of Bose-Einstein condensation, starting from Bose statistics applied to a noninteracting gas: the non-negligible occupation number of the energetic ground state. For a more detailed derivation, the reader is referred to \cite{pitaevskij_bose-einstein_2010, pethick_bose-einstein_2008}.

As a first step, consider the grand canonical partition function

\begin{equation} \label{eq:ideal_bose_gas_grand_canonical_partition}
	Z ( \beta, \mu ) = \sum_{N=0}^\infty \sum_k \mathrm{e}^{\beta \left( \mu N  - E_k \right) } \; ,
\end{equation}

where $\mu$ is the chemical potential and $\beta=1/k_B T$ the inverse temperature. The first sum goes over all particle numbers $N$ and the second one over all eigenstates $k$ of the governing Hamiltonian with $N$ particles and energy $E_k$. For non-interacting particles, the Hamiltonian can be written as a sum of identical single-particle Hamiltonians and the single-particle eigenstates satisfy

\begin{equation}
	\hat{H}^{(1)} | i \rangle = \epsilon_i | i \rangle
\end{equation}

with single-particle energies $\epsilon_i$. Thus, for a global state characterized by the configuration $\{\tilde{n}_i\}$\footnote{The tilde in $tilde{n}_i$ is just to distinguish the particle mode occupation numbers from the particle density $n$ which will be used more often later on.}, \eqref{eq:ideal_bose_gas_grand_canonical_partition} can be rewritten as

\begin{equation}
	Z ( \beta, \mu ) = \sum_{\{\tilde{n}_i\}} \mathrm{e}^{
		\beta \left(\mu \sum_i \tilde{n}_i - \sum_i \epsilon_i \tilde{n}_i \right)
		}
	= \prod_i \sum_{\{\tilde{n}_i\}} \mathrm{e}^{
		\tilde{n}_i \beta \left( \mu - \epsilon_i \right)
		}
\end{equation}

and, for bosons,

\begin{equation}
	Z ( \beta, \mu ) = \prod_i \frac{1}{1- \mathrm{e}^{
		\beta \left( \mu - \epsilon_i \right)
		}} \; .
\end{equation}

The average occupation numbers and the total number of particles are then

\begin{equation} \label{eq:ideal_bose_gas_occupation_numbers}
	\langle \tilde{n}_i \rangle = \frac{1}{\mathrm{e}^{
		\beta \left(\epsilon_i - \mu \right)
		} - 1}
	\qquad
	\mathrm{and}
	\qquad
	N = \sum_i \langle \tilde{n}_i \rangle \; .
\end{equation}

This result is a good point to take a step back and make a few observations: firstly, \eqref{eq:ideal_bose_gas_occupation_numbers} shows that the chemical potential $\mu$ is determined by the temperature and the requirement that the total particle number $N$ be the sum of the average occupation numbers $\langle \tilde{n}_i \rangle$. Secondly, due to the nonnegativity of $\langle \tilde{n}_i \rangle $, the chemical potential is bounded from above as $\mu < \epsilon_0 $ by the lowest eigenvalue of $\hat{H}^{(1)}$. Thirdly and most importantly, the behaviour of the occupation numbers for different temperature regimes can be inferred from \eqref{eq:ideal_bose_gas_occupation_numbers}. For high temperatures $k_B T \gg \epsilon_0 - \mu$, the chemical potential remains far below $\epsilon_0$ and the Boltzmann distribution $\langle \tilde{n}_i \rangle = \mathrm{e}^{- \beta \left( \epsilon_i - \mu \right)}$ is recovered. Hence, more states can be thermally occupied than there are particles and all average occupation numbers remain small. If, on the other hand, the temperature is lowered towards a critical temperature $T_c$, $\mu$ increases and eventually converges towards its upper bound $\epsilon_0$, where the occupation number of the ground state $N_0$ diverges, whereas the the number particles in an excited state $N_T$ remains finite. This is the signature of Bose-Einstein condensation. Defining $N_0$ and $N_T$ as

\begin{equation} \label{eq:ideal_bose_gas_condensate_thermal}
	N = \underbrace{N_0}_{\equiv \langle \tilde{n}_0 \rangle} + \underbrace{N_T}_{\equiv \sum_{i \neq 0} \langle \tilde{n}_i \rangle} \; ,
\end{equation}

we can recognize the critical temperature as the point where

\begin{equation} \label{eq:ideal_bose_gas_def_critical_temperature}
	N_T \left( T_c, \mu = \epsilon_0 \right) = N \; .
\end{equation}

Below $T_c$, a significant part of the gas is in the ground state and $N_0$ becomes non-negligible.

One of the most simple settings to derive the critical temperature as well as the behaviour of $N_0$ and $N_T$ is a three-dimensional box, where $\epsilon_\mathbf{p} = p^2/2m$ and $\mathbf{p} = 2 \pi \hbar \mathbf{n} / L$ (with the length of the box $L$) to satisfy the boundary conditions. For the thermal part (the particles out of the condensate), the sum in \eqref{eq:ideal_bose_gas_condensate_thermal} can be replaced by an integral as a good approximation in the limit of a large box, resulting in

\begin{equation} \label{eq:ideal_bose_gas_thermal_number_box}
	N_T = \frac{V}{\lambda_T^3} g_{3/2} (z) \; ,
\end{equation}

where $V$ is the volume of the box, $z \equiv \mathrm{e}^{\beta \mu}$ is called the \textit{fugacity},

\begin{equation}
	\lambda_T \equiv \sqrt{\frac{2 \pi \hbar^2}{m k_B T}}
\end{equation}

the thermal de Broglie wavelength and $g_{3/2} (z)$ is a member of the Bose functions

\begin{equation}
	g_a (z) = \sum_{j=1}^{\infty} \frac{z^j}{j^a} \; .
\end{equation}

Plugging the result \eqref{eq:ideal_bose_gas_thermal_number_box} into the definition of the critical temperature \eqref{eq:ideal_bose_gas_def_critical_temperature}, $T_c$ can be derived as

\begin{IEEEeqnarray}{rCl}
	N_c \equiv N_T \big( T_c, \overbrace{ \mu = \epsilon_0}^{=0} \big)
	&=&
	V \left( \frac{m k_B T_c}{2 \pi \hbar^2} \right)^{3/2} \overbrace{g_{3/2} \left( 1 \right)}^{=2.612} = N
	\\*[3pt]
	\Rightarrow \quad T_c
	&=&
	\frac{2 \pi \hbar^2}{m k_B} \left( \frac{n}{g_{3/2} (1)} \right)^{2/3}
	= \frac{2 \pi \hbar^2}{m k_B} \, \lambda_{T_c}^2
\end{IEEEeqnarray}

with the particle density $n=N/V$. The last equality reveals that at the critical temperature, the thermal de Broglie wavelength assumes the same magnitude as the average interparticle separation $ \lambda_{T_c} \approx n^{-1/3}$. This illustrates the frequently quoted overlap of the individual particle wave functions leading to the formation of a coherent superposition.

For the thermal occupation number of the noncondensate part, we have

\begin{equation}
	N_T \left( T < T_c, \mu = 0 \right) = N \left( \frac{T}{T_c} \right)^{3/2}
\end{equation}

and for the condensate fraction

\begin{equation}
	N_0 \left( T < T_c \right) ) = N \left( 1 - \left( \frac{T}{T_c} \right)^{3/2} \right) \; .
\end{equation}

Through similar considerations, the same quantities can be found for a three-dimensional harmonic potential $V \left( \mathbf{x} \right) = m/2 \left( \omega_x^2 x^2 + \omega_y^2 y^2 + \omega_z^2 z^2 \right)$:\footnote{For a detailed derivation, see \cite{pethick_bose-einstein_2008}.}

\begin{IEEEeqnarray}{rCl}
	T_c
	&=&
	\frac{\hbar \bar{\omega}}{k_B} \left( \frac{N}{g_3 (1)} \right)^{1/3}
	\\
	N_T
	&=&
	N \left( \frac{T}{T_c} \right)^3
	\\
	N_0
	&=&
	N \left( 1 - \left( \frac{T}{T_c} \right)^3 \right) \; ,
\end{IEEEeqnarray}

where $\bar{\omega} = \left( \omega_x \omega_y \omega_z \right)^{1/3}$ is the cubic mean frequency of the potential.

\section{The Uniform, Weakly Interacting Bose Gas \label{sec:uniform_weakly_interacting}}

The treatment of the non-interacting Bose-Einstein condensate in the previous section contained some subtleties, such as its infinite compressibility \cite{pitaevskij_bose-einstein_2010} and the fact that the above considerations were made assuming thermal equilibrium which cannot be reached in the absence of interactions.

Here, a microscopic theory of the interacting Bose gas is presented. Following Bogoliubov's Ideas \cite{bogoliubov_theory_1947}, this theory is based on two central steps: the \textit{Bogoliubov approximation} \eqref{eq:bogo_approx_field_operator} and a \textit{Bogoliubov transformation}. In the former, the macroscopically occupied ground state of the Bose gas is treated classically while the quantum nature of the fluctuations about it are allowed to keep their quantum nature. A Bogoliubov transformation, which can be seen as a change of basis, will then be used to reveal a picture of quasi-particles which are non-interacting to first approximation.

In addition to \eqref{eq:bogo_approx_field_operator}, two important assumptions are made in this context. The Bose gas is considered to be in the \textit{dilute regime}

\begin{equation} \label{eq:diluteness_condition}
	r_0 \ll n^{-1/3} \; ,
\end{equation}

where $r_0$ is the range of the inter-particle interaction.\footnote{In fact, the typical particle density of Bose-Einstein condensates is one of their distinguishing features. While air at room temperature has $n \approx 10^{19} \mathrm{cm}^{-3}$ and liquids and solids are of the order $n \approx 10^{22} \mathrm{cm}^{-3}$, usual atomic densities for BECs are $n \approx 10^{13} - 10^{15} \mathrm{cm}^{-3}$.} In combination with the temperatures $T<T_c$ necessary for Bose-Einstein condensation, this means that any scattering processes other than pairwise interactions in the low-energy limit can be neglected. Hence, the scattering behaviour does not depend on the exact shape of the interaction potential and is completely determined by one single parameter, the s-wave scattering length\footnote{See subsection \ref{subsec:single-chanel_scattering}.} $a_s$. This is also called the domain of \textit{universality}. It will play a central role in the subsequent chapters \ref{sec:feshbach_resonances} and \ref{sec:three-body collisions} on Feshbach resonances and three-body losses.

The second assumption that is made is that of weak interactions, such that

\begin{equation} \label{eq:weak_interaction_approximation}
	|a_s| \ll n^{-1/3} \; .
\end{equation}

A remark has to be made that even in the dilute regime \eqref{eq:diluteness_condition}, the condition \eqref{eq:weak_interaction_approximation} might not be fulfilled. In particular, this can be the case near a Feshbach resonance, where $a_s$ can be tuned using an external magnetic field.

Consider the Hamiltonian of the system

\begin{IEEEeqnarray}{rCl} \label{eq:full_interacting_hamiltonian}
	\hat{H} &=& \int \ud \mathbf{r} \,
		\hat{\psi}^\dagger \left( \mathbf{r} \right)
		\left(
			-\frac{\hbar^2}{2m} \nabla^2 + \mathcal{V} \left( \mathbf{r} \right)
		\right)
		\hat{\psi} \left( \mathbf{r} \right) \nonumber
	\\
	& & +
	\int \ud \mathbf{r} \, \ud \mathbf{r}' \,
		\hat{\psi}^\dagger \left( \mathbf{r'} \right) \hat{\psi}^\dagger \left( \mathbf{r} \right) U \left( \mathbf{r}' - \mathbf{r} \right) \hat{\psi} \left( \mathbf{r'} \right) \hat{\psi} \left( \mathbf{r} \right) \; ,
\end{IEEEeqnarray}

where $\hat{\psi} \left( \mathbf{r} \right)$ and $\hat{\psi}^\dagger \left( \mathbf{r} \right)$ are the quantum field operators satisfying the canonical bosonic commutation relations, $\mathcal{V} \left( \mathbf{r} \right)$ is an external potential and $U \left( \mathbf{r} \right)$ is the interaction potential.

The Bogoliubov approximation can now be expressed in two equivalent ways, both of which are implying that the bulk of the Bose gas is in the classical ground state, whereas excitations are kept as quantum operators. The most intuitive way to formulate this idea is to rewrite $\hat{\psi}$ as

\begin{equation} \label{eq:bogo_approx_field_operator}
	\hat{\psi} \left( \mathbf{r} \right) = \psi \left( \mathbf{r} \right) + \delta \hat{\psi} \left( \mathbf{r} \right)
\end{equation}

with the ground state wave function $\psi \left( \mathbf{r} \right)$ and a small quantum perturbation $\delta \hat{\psi} \left( \mathbf{r} \right)$.

For a uniform Bose gas in a box with volume $V$ and cyclic boundary conditions, on the other hand, it is more convenient to consider the creation and annihilation operators $\hat{a}^\dagger_\mathbf{p}, \, \hat{a}_\mathbf{p}$ for a state with momentum $ \mathbf{p} $ satisfying the bosonic commutation relations. In this particular case, $\hat{\psi} \left( \mathbf{r} \right)$ is a sum of plane wave solutions \cite{pitaevskij_bose-einstein_2010}

\begin{equation} \label{eq:uniform_bose_gas_field operator}
	\hat{\psi} \left( \mathbf{r} \right) = \frac{1}{\sqrt{V}} \sum_\mathbf{p} \hat{a}_\mathbf{p} \mathrm{e}^{\frac{i}{\hbar} \mathbf{p} \mathbf{r}} \; , 
\end{equation}

leading to the Hamiltonian

\begin{equation} \label{eq:uniform_interacting_hamiltonian}
	\hat{H} = \sum_\mathbf{p} \frac{p^2}{2m} \hat{a}^\dagger_\mathbf{p} \hat{a}_\mathbf{p}
	 + \frac{1}{2V} 
	 \sum_{
	 	\mathbf{p}_1, \, \mathbf{p}_2, \, \mathbf{q}
		 } 
	 U_\mathbf{q} \hat{a}^\dagger_{\mathbf{p}_1 + \mathbf{q}} \hat{a}^\dagger_{\mathbf{p_2} - \mathbf{q}} \hat{a}_{\mathbf{p}_1} \hat{a}_{\mathbf{p}_2} \; ,
\end{equation}

with the Fourier transform of the interaction potential

\begin{equation}
	U_\mathbf{q} \equiv \int \ud \mathbf{r} \, U \left( \mathbf{r} \right) \mathrm{e}^{-\frac{i}{\hbar} \mathbf{qr}} \; .
\end{equation}

The shape of an actual interatomic potential $U \left( \mathbf{r} \right)$ would lead to problems when trying to apply perturbation theory. However, if only s-wave scattering needs to be considered, $U \left( \mathbf{r} \right)$ can be replaced by any effective potential $U_\mathrm{eff} \left( \mathbf{r} \right)$ that reproduces the same s-wave scattering length $a_s$ and is better suited for a perturbative treatment. We can therefore take

\begin{equation}
	U_\mathbf{q} = \int \ud \mathbf{r} \, U_\mathrm{eff} \left( \mathbf{r} \right) \mathrm{e}^{-\frac{i}{\hbar} \mathbf{qr}} \; .
\end{equation}

The combination of \eqref{eq:bogo_approx_field_operator} and \eqref{eq:uniform_bose_gas_field operator} clarifies the second way to express the Bogoliubov approximation:

\begin{equation}
	\hat{\psi} \left( \mathbf{r} \right) = \frac{1}{\sqrt{V}} \left( \hat{a}_0 + \sum_{\mathbf{p} \neq 0} a_\mathbf{p} \mathrm{e}^{\frac{i}{\hbar} \mathbf{p} \mathbf{r}} \right) \approx \frac{1}{\sqrt{V}} \left( \sqrt{N_0} + \sum_{\mathbf{p} \neq 0} \hat{a}_\mathbf{p} \mathrm{e}^{\frac{i}{\hbar} \mathbf{p} \mathbf{r}} \right) \; ,
\end{equation}

where the last equality is due to normalization. This means that the creation and annihilation operators for the ground state $\hat{a}_0$ and $\hat{a}^\dagger_0$ are replaced by their classical expectation value and their commutator is assumed to vanish under the approximation $N_0 \gg 0 \, \Rightarrow \,\sqrt{N_0 + 1} \approx \sqrt{N_0}$.

\begin{equation} \label{eq:bogo_approx_ladder_operators}
	\hat{a}_0 \approx \hat{a}^\dagger_0 \approx \sqrt{N_0}
\end{equation}

\subsection{Zeroth Order: Ground State Energy and Chemical Potential}

To first approximation, all terms in \eqref{eq:uniform_interacting_hamiltonian} that contain creation or annihilation operators with $\mathbf{p} \neq 0$ are discarded. Up to the same order, we can take $N_0=N$ and relate the $\mathbf{p}=0$ Fourier transform of the effective potential

\begin{equation} \label{eq:low_energy_limit_effective_potential}
	U_0 \equiv \int \ud \mathbf{r} \, U_\mathrm{eff} \left( \mathbf{r} \right)
\end{equation}

to the interaction strength $g$ and thus the s-wave scattering length as

\begin{equation} \label{eq:zeroth_order_interaction_strength}
	U_0 = g \equiv \frac{4 \pi \hbar^2 a_s}{m} \; .
\end{equation}

The only term left in \eqref{eq:uniform_interacting_hamiltonian} is then

\begin{equation}
	E_0 = \frac{N^2 U_0}{2 V} = \frac{1}{2} N g n \; .
\end{equation}

This, in turn, yields the speed of sound \cite{pitaevskij_bose-einstein_2010}

\begin{equation} \label{eq:zeroth_order_speed_of_sound}
	c_s = \sqrt{\frac{gn}{m}}
\end{equation}

and the chemical potential

\begin{equation} \label{eq:zeroth_order_chemical_potential}
	\mu = \frac{\partial E_0}{\partial N} = gn = mc_s^2 \; .
\end{equation}

\subsection{Second Order: Bogoliubov Transformation and Quasiparticles}

\begin{figure}[h!pbt]
	\centering
	\begin{subfigure}[b]{0.35\textwidth}
		\centering
		\includegraphics[width=\textwidth]{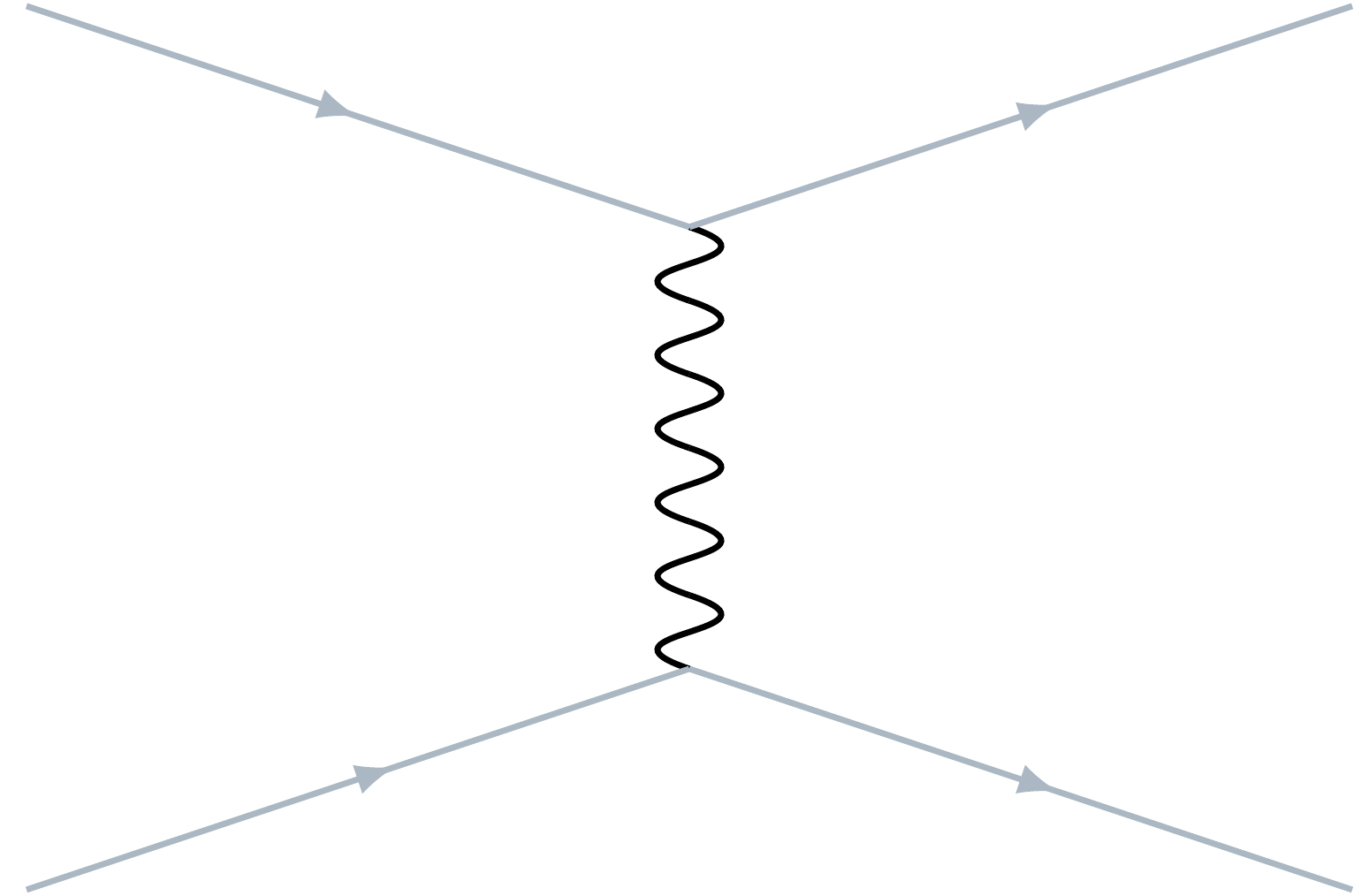}
		\caption{}
		\label{fig:ground_state_interaction}
	\end{subfigure}
	\\
	\begin{subfigure}[b]{0.35\textwidth}
		\centering
		\includegraphics[width=\textwidth]{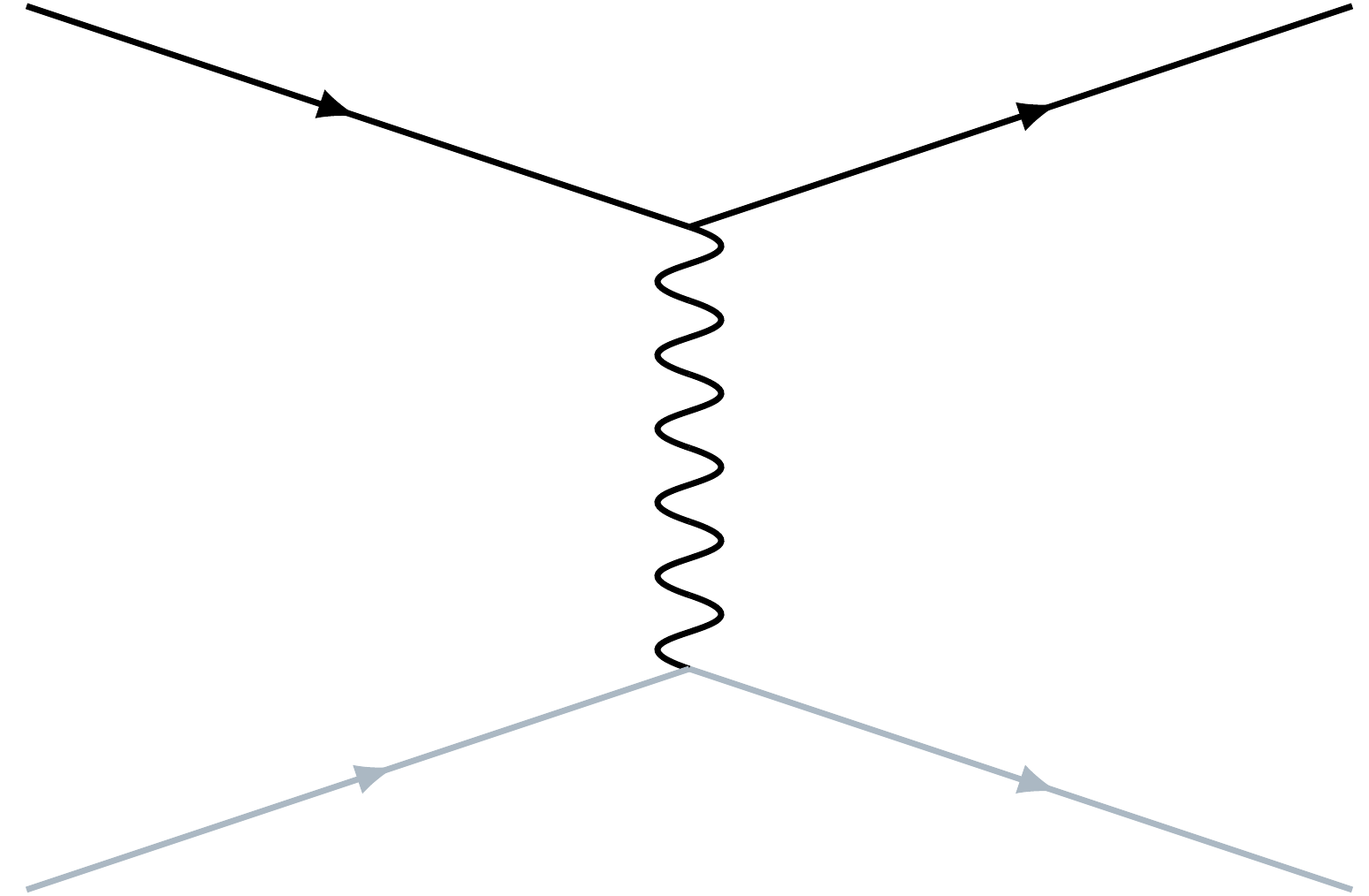}
		\caption{}
		\label{fig:hartree_interaction}
	\end{subfigure}
	\hfill
	\begin{subfigure}[b]{0.35\textwidth}
		\centering
		\includegraphics[width=\textwidth]{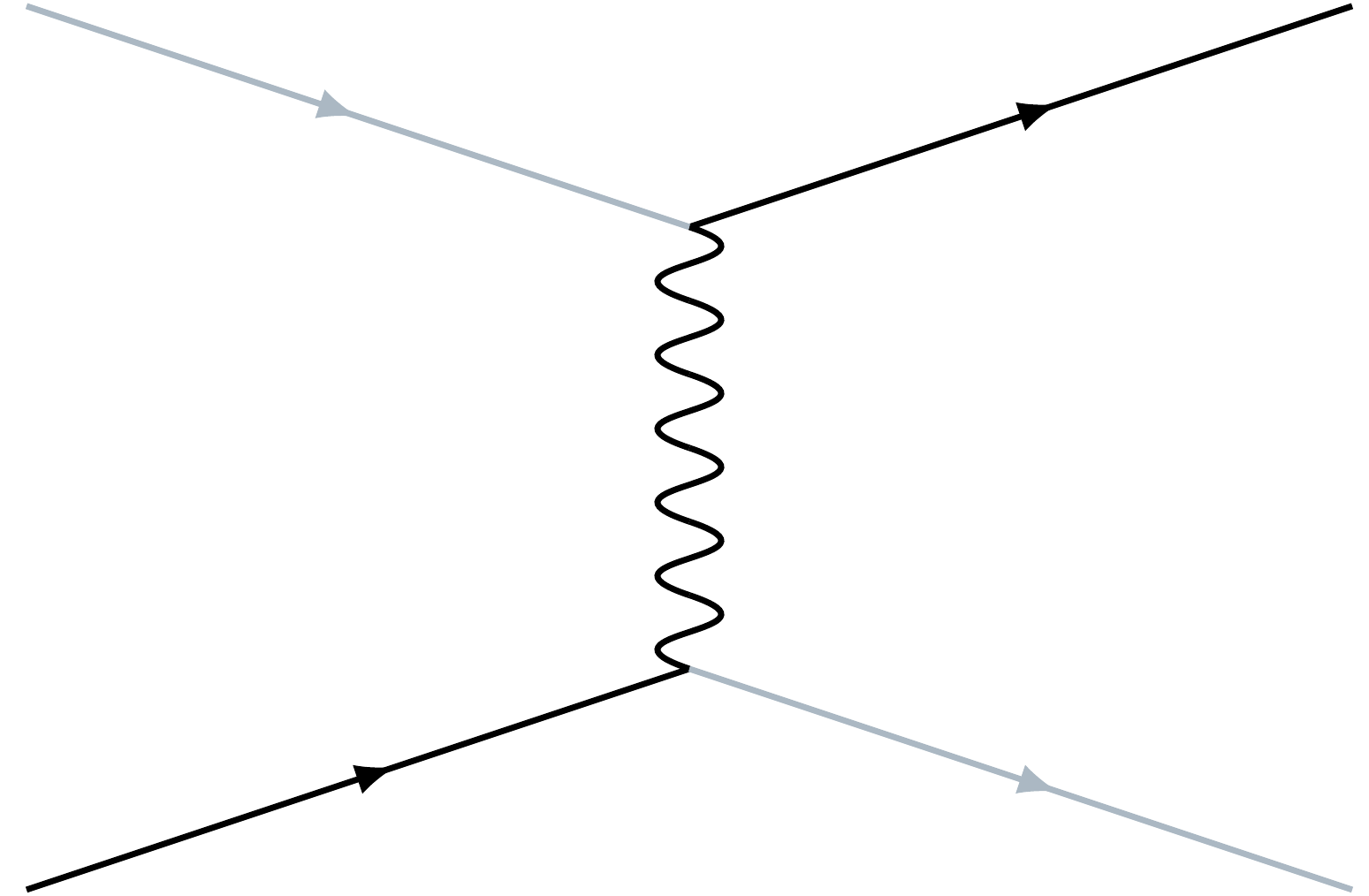}
		\caption{}
		\label{fig:fock_interaction}
	\end{subfigure}
	\hfill
	\begin{subfigure}[b]{0.35\textwidth}
		\centering
		\includegraphics[width=\textwidth]{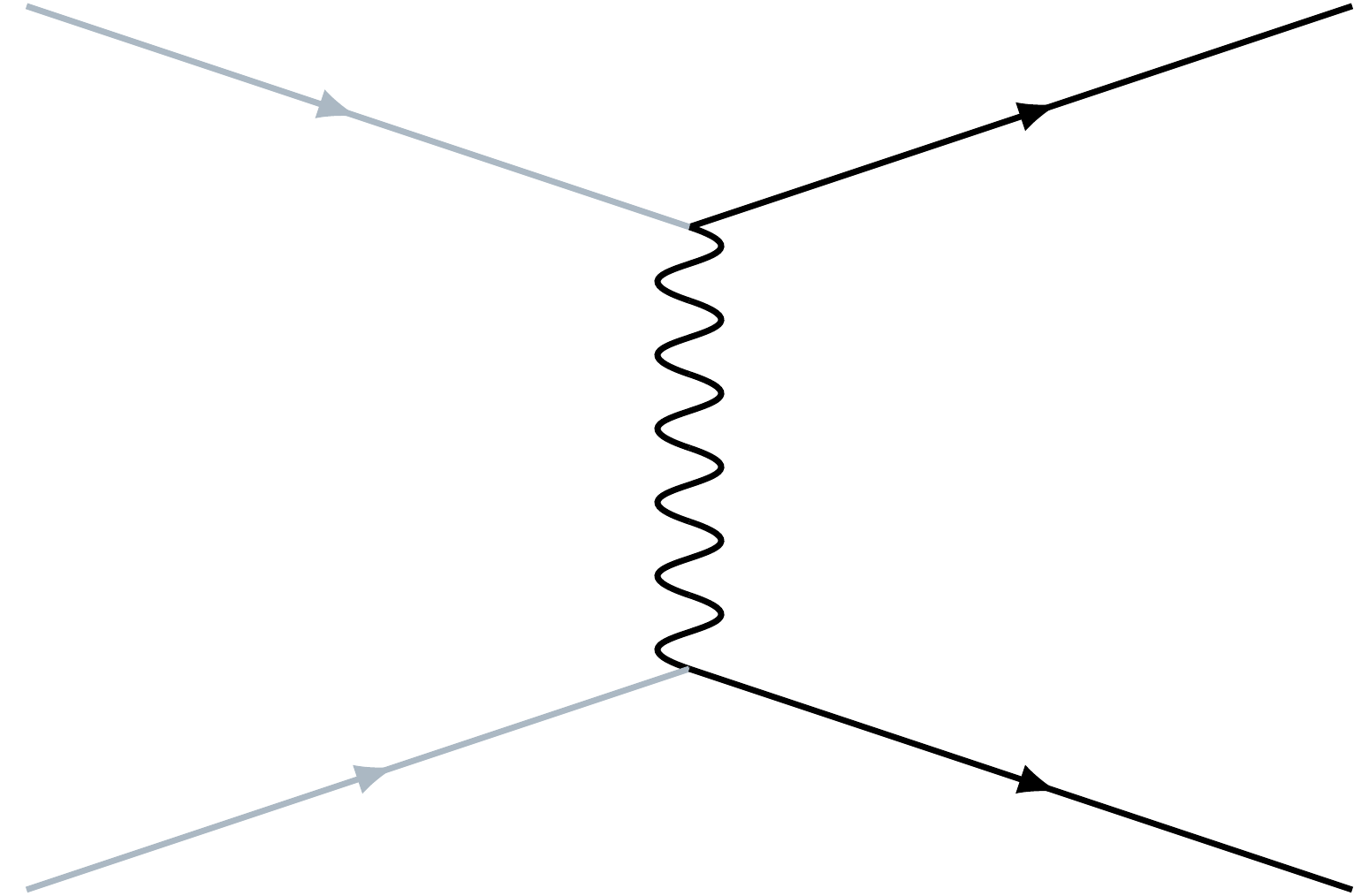}
		\caption{}
		\label{fig:symmetric_scattering_interaction_1}
	\end{subfigure}
	\hfill
	\begin{subfigure}[b]{0.35\textwidth}
		\centering
		\includegraphics[width=\textwidth]{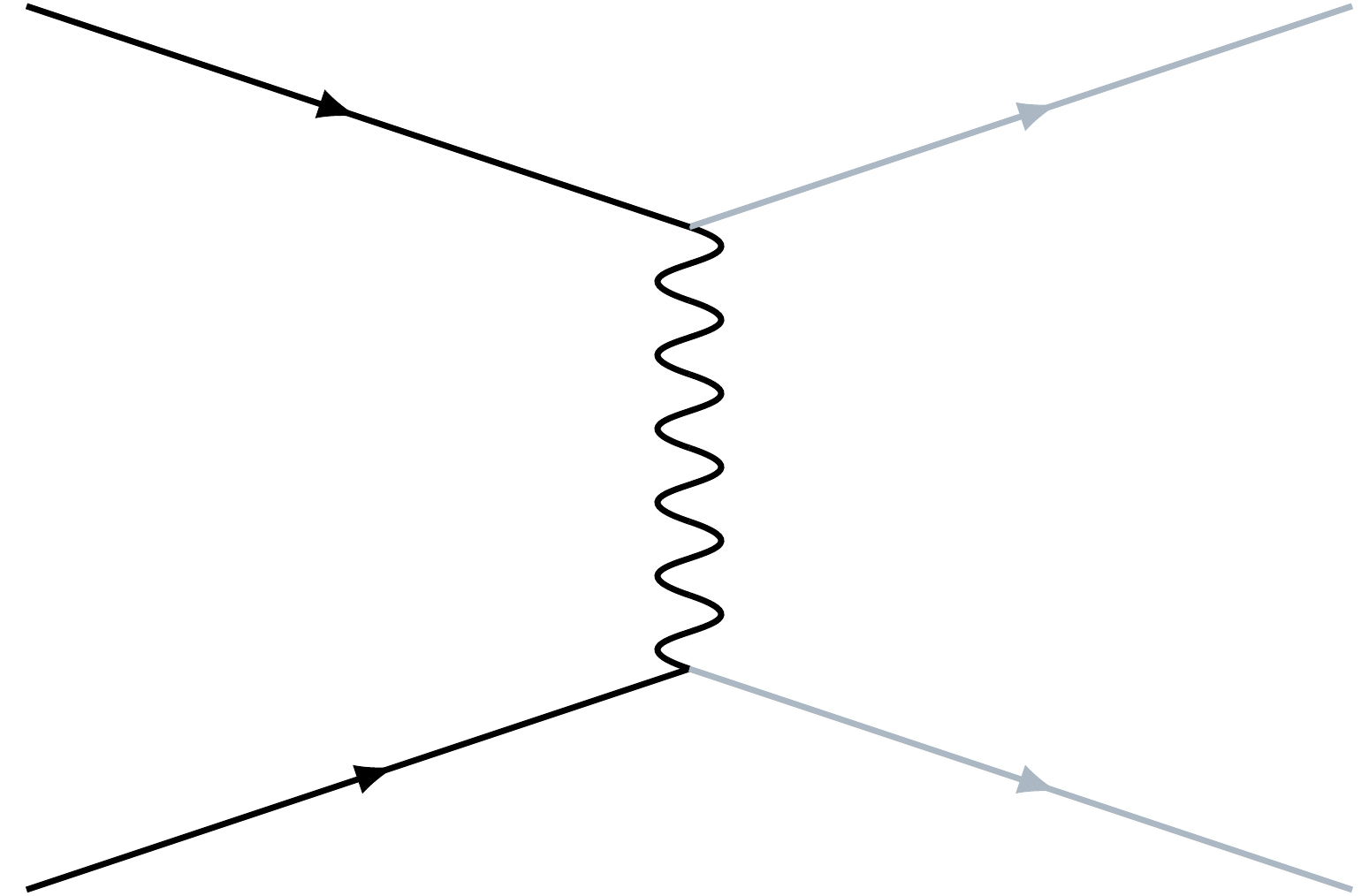}
		\caption{}
		\label{fig:symmetric_scattering_interaction_2}
	\end{subfigure}
	\caption{
		{\footnotesize
			Interactions in the Hamiltonian up to second order in $\hat{a}_\mathbf{p}$, $\hat{a}^\dagger_\mathrm{p}$ \eqref{eq:uniform_bose_gas_second_order_hamiltonian}. Grey and black lines denote condensate and excited particles, respectively. \cite{shi_finite-temperature_1998}
			(a) condensate self-interaction, 
			(b): Hartree interaction between a particle with momentum $\mathbf{p}$ and a particle in the condensate.
			(c): Fock interaction: condensate particle scatters into a state with momentum $\mathbf{p}$ and vice versa.
			(d), (e): Scattering of two condensate particles into $\mathbf{p}$ states and vice versa.
		}
	}
\end{figure}

Since terms $\mathcal{O} \left( a_\mathbf{p} \right) $ would violate momentum conservation, the next order in the approximation is quadratic in creation and annihilation operators with nonvanishing momentum:

\begin{IEEEeqnarray}{rCl} \label{eq:uniform_bose_gas_second_order_hamiltonian}
	\hat{H}^{\left( 2 \right)}
	&=&
	\overbrace{
		\frac{U_0}{2V} \hat{a}^\dagger_0 \hat{a}^\dagger_0 \hat{a}_0 \hat{a}_0
	}^{
		\mathrm{condensate \, self \, interaction, \, fig. \ref{fig:ground_state_interaction}}
	} 
	+ \overbrace{
		\sum_{\mathbf{p}} \frac{p^2}{2m} \hat{a}^\dagger_{\mathbf{p}} \hat{a}_{\mathbf{p}}
	}^{
		\mathrm{kinetic \, energy}
	}
	\\
	&+& 
	\frac{1}{2V}
	 \sum_{\mathbf{p} \neq 0} 
	\underbrace{
		U_\mathbf{p} \hat{a}^\dagger_0 \hat{a}^\dagger_0 \hat{a}_\mathbf{p} \hat{a}_{-\mathbf{p}}
	}_{
		\mathrm{fig.} \ref{fig:symmetric_scattering_interaction_1}
	} 
	+ \underbrace{
		U_\mathbf{p} \hat{a}^\dagger_\mathbf{p} \hat{a}^\dagger_{-\mathbf{p}} \hat{a}_0 \hat{a}_0
	}_{
	\mathrm{fig.} \ref{fig:symmetric_scattering_interaction_2}	
	}
	 +
	 \overbrace{
		2 \left(
		\underbrace{
			U_0
		}_{
			\mathrm{fig.} \ref{fig:hartree_interaction}
		}
	 + 
	\underbrace{
		U_\mathbf{p}
	}_{
		\mathrm{fig.} \ref{fig:fock_interaction}
	} \right) \hat{a}^\dagger_0 \hat{a}^\dagger_\mathbf{p} \hat{a}_0 \hat{a}_\mathbf{p}
	}^{
		\mathrm{Hartree \, and \, Fock \, interaction}	
	} \nonumber
\end{IEEEeqnarray}

While the first line in \eqref{eq:uniform_bose_gas_second_order_hamiltonian} reflects the condensate self-interactions and the kinetic energy, the second line containing interactions between excited states and the ground state deserves a closer look before the low-energy limit is taken and $U_\mathbf{p}$ is replaced by $U_0$ \cite{pethick_bose-einstein_2008}. The first two terms in the sum correspond to the scattering of two excited states into the condensate and vice versa (fig. \ref{fig:symmetric_scattering_interaction_1}, \ref{fig:symmetric_scattering_interaction_2}). The term mediated by $U_0$ is the Hartree scattering of an excited state with a condensate state (see fig. \ref{fig:hartree_interaction}) and the one including $U_\mathbf{p}$ is responsible for Fock interaction, where an excited state scatters into a zero momentum state and vice versa (fig. \ref{fig:fock_interaction}).

At this order, some corrections have to be made to the condensate self-energy term and the correspondence between the $U_0$ and $a_s$ \cite{pitaevskij_bose-einstein_2010}. The second-order perturbative expansion for the s-wave scattering length is \cite{landau_quantum_2013}

\begin{equation}
	a_s = \frac{m}{4 \pi \hbar^2} U_0 \left(
		1 - \frac{U_0}{V} \sum_{\mathbf{p} \neq 0} \frac{m}{p^2}
	\right)
\end{equation}

or, equivalently

\begin{equation}
	U_0 = g \left(
		1 - \frac{g}{V} \sum_{\mathbf{p} \neq 0} \frac{m}{p^2}
		\right) \; .
\end{equation}

For the ground state self-interaction, a correction to second order in the noncondensate ladder operators comes from the normalization

\begin{IEEEeqnarray}{RrCl}
	& \hat{a}^\dagger_0 \hat{a}_0 + \sum_{\mathbf{p} \neq 0} \hat{a}^\dagger_\mathbf{p} \hat{a}_\mathbf{p}
	&=&
	N
	\\
	\Rightarrow &
	\hat{a}^\dagger_0 \hat{a}^\dagger_0 \hat{a}_0 \hat{a}_0
	&=&
	 N^2 - 2 N \sum_{\mathbf{p} \neq 0} \hat{a}^\dagger_\mathbf{p} \hat{a}_\mathbf{p} + \mathcal{O} \left( \{\hat{a}^\dagger_\mathbf{p} , \, \hat{a}_\mathbf{p} \}^3 \right) \; .
\end{IEEEeqnarray}

With the above corrections and after the discarding the terms of cubic or higher order in the excited states, the Hamiltonian \eqref{eq:uniform_bose_gas_second_order_hamiltonian} becomes

\begin{equation} \label{eq:uniform_bose_gas_second_order_hamiltonian_2}
	\hat{H} = \frac{N^2 g}{2m} + \sum_{\mathbf{p} \neq 0}
		\left(
			\frac{p^2}{2m} + gn
		\right)
		\hat{a}^\dagger_\mathbf{p} \hat{a}_\mathbf{p} + \frac{g n}{2} \left(
			\hat{a}^\dagger_\mathbf{p} \hat{a}^\dagger_{- \mathbf{p}} + \hat{a}_\mathbf{p} \hat{a}_{- \mathbf{p}} + \frac{mgn}{p^2}
		\right) \; .
\end{equation}

The above Hamiltonian can be diagonalized through a transformation of the creation and annihilation operators known as a Bogoliubov transformation \cite{bogoliubov_theory_1947}. First used in the context of liquid helium, this technique has since found applications in the theoretical description of magnetism and superconductivity as well as (in a generalized form see appendix \ref{sec:bogoliubov_trafos}) quantum field theory in curved spacetime \cite{birrell_quantum_1984} or relativistic quantum metrology \cite{ahmadi_quantum_2014, ahmadi_relativistic_2015, sabin_phonon_2014}.

The transformation

\begin{IEEEeqnarray}{rClCl}
	\IEEEyesnumber \label{eq:uniform_bose_gas_bogo_trafo} \IEEEyessubnumber*
	\hat{a}_{ \mathbf{p} }
	& \equiv &
	u_{ \mathbf{p} } \hat{b}_{ \mathbf{p} }
	&+&
	v^*_{ -\mathbf{p} } \hat{b}^\dagger_{ -\mathbf{p} }
	\\
	\hat{a}^\dagger_{ \mathbf{p} }
	& \equiv &
	u^*_{ \mathbf{p} } \hat{b}^\dagger_{ \mathbf{p} }
	&+&
	v_{ -\mathbf{p} } \hat{b}_{ -\mathbf{p} }
\end{IEEEeqnarray}

and the requirement that $\hat{b}_{ \mathbf{p} }$ and $\hat{b}^\dagger_{ \mathbf{p'} }$ satisfy the same bosonic commutation relations as $\hat{a}_{ \mathbf{p} }$ and $\hat{a}^\dagger_{ \mathbf{p'} }$ defines the new creation and annihilation operators in the quasi-particle picture.

Demanding that, in this picture, the Hamiltonian \eqref{eq:uniform_bose_gas_second_order_hamiltonian_2} shall take the form

\begin{equation} \label{eq:uniform_bose_gas_noninteracting_bogo_hamiltonian}
	\hat{H}^{\left( 2 \right)} = E_0 + \sum_{ \mathbf{p} \neq 0} \epsilon_p \hat{b}^\dagger_{ \mathbf{p} } \hat{b}_{ \mathbf{p} }
\end{equation}

leads to the quasi-particle mode solutions \cite{pitaevskij_bose-einstein_2010}

\begin{equation} \label{eq:uniform_bose_gas_bogo_mode_solutions}
	u_{ \mathbf{p} }, \, v_{ -\mathbf{p} } = \pm \sqrt{\frac{p^2/2m + gn}{2 \epsilon_p} \pm \frac{1}{2}}
\end{equation}

with the ground state energy

\begin{equation} \label{eq:uniform_bose_gas_2nd_order_energy}
	E_0  = \frac{gN^2}{2 V} + \frac{1}{2} \sum_{ \mathbf{p} \neq 0 } \left( \epsilon_p - gn - \frac{p^2}{2m} + \frac{m g^2 n^2}{p^2} \right)
\end{equation}

and the \textit{Bogoliubov dispersion law}

\begin{equation} \label{eq:bogo_dispersion_law}
	\epsilon_p^2 =  \underbrace{\frac{gn}{m}}_{c_s^2} p^2 + \left( \frac{p^2}{2m} \right)^2 \; .
\end{equation}

Thus, we have transformed from a Hamiltonian of interacting particles in a Bose gas to a picture of quasi-particles with energies $\epsilon_p$ that is, up to quadratic order in the creation and annihilation operators, interaction-free.

At this point, a few remarks and consequences of results (\ref{eq:uniform_bose_gas_noninteracting_bogo_hamiltonian} - \ref{eq:bogo_dispersion_law}) are at hand:

\paragraph{Ground state and chemical potential:} what was called the ground state in the description above is in fact a metastable state with a finite life time for most physical systems, whereas the actual ground state, after also taking three-body collisions into account, would be a solid \cite{pitaevskij_bose-einstein_2010}. Nonetheless, $E_0$ can be computed by taking the continuum limit $\sum_{ \mathbf{p} } \rightarrow V/(2 \pi \hbar^3) \int \ud \mathbf{p}$: \cite{lee_eigenvalues_1957, lee_many-body_1957}

\begin{equation}
	E_0 = \frac{g N^2}{2V} \left( 1 + \frac{128}{15 \sqrt{\pi}} \left(n a_s^3 \right)^\frac{1}{2} \right) \; .
\end{equation}

With that, the chemical potential becomes

\begin{equation}
 \mu = \frac{\partial E_0}{\partial N} = g n \left( 1 + \frac{32}{3 \sqrt{\pi}} \left( n a_s^3 \right)^\frac{1}{2} \right) \; .
\end{equation}

\paragraph{Thermodynamic behaviour of the quasi-particles:} because there are no zero-momentum quasi-particle modes defined, it is to be expected that their average occupation number $N_{ \mathbf{p} } \equiv \langle \hat{b}^\dagger_{ \mathbf{p} } \hat{b}_{ \mathbf{p} } \rangle$ vanishes at zero temperature. As we are dealing with (in this approximation) independent bosonic excitations in a box, their chemical potential can be set to zero in analogy to the considerations in the previous section and the thermal occupation number is

\begin{equation} \label{eq:uniform_bose_gas_thermal_phonon_occupation_number}
	N_{ \mathbf{p} } = \frac{1}{\mathrm{e}^{\beta \epsilon_p} - 1} \; .
\end{equation}

\paragraph{Depletion of the condensate:} the average occupation number $n_{ \mathbf{p} } \equiv \langle \hat{a}^\dagger_{ \mathbf{p} } \hat{a}_{ \mathbf{p} } \rangle $ of states with nonzero momentum $\mathbf{p}$ can be calculated by means of \eqref{eq:uniform_bose_gas_bogo_trafo} to be \cite{pitaevskij_bose-einstein_2010}

\begin{equation}
	n_{ \mathbf{p} } = |u_{ \mathbf{p} }|^2 \langle \hat{b}^\dagger_{ \mathbf{p} } \hat{b}_{ \mathbf{p} } \rangle + |v_{ - \mathbf{p} } |^2 \langle \hat{b}^\dagger_{  - \mathbf{p} } \hat{b}_{ - \mathbf{p} } \rangle + |v_{ - \mathbf{p} } |^2 \; .
\end{equation}

The last term in the above equation is particularly remarkable:
it tells us that even at $T=0$, there is a nonvanishing depletion of the condensate that is a result of accounting for interactions. With \eqref{eq:uniform_bose_gas_bogo_mode_solutions}, 

\begin{equation}
	n_{ \mathbf{p} } = \frac{p^2/2m + gn}{2\epsilon_p} - \frac{1}{2} \; .
\end{equation}

\paragraph{Phonons and free particles:} to gain more intuition on the description of the quasi-particles, it is helpful to consider the asymptotic behaviour of the Bogoliubov dispersion relation \eqref{eq:bogo_dispersion_law} in the limit of small and large momenta.

For $p \ll mc_s$, coefficients $|v_{-p}| \approx |u_p|$ are of the same order and the energy of a mode

\begin{equation} \label{eq:bogo_dispersion_relation_low-energy_limit}
	\epsilon_p \approx c_s p
\end{equation}

becomes the energy of a (massless) sound wave. Thus, the low-energy excitations are referred to as phonons.

In the high-energy limit $ p \gg m c_s^2$, on the other hand, $\epsilon_p$ becomes the energy of a massive free particle

\begin{equation}
	\epsilon_p = \frac{p^2}{2m} + gn \; ,
\end{equation}

corresponding to only a weak Bogoliubov transformation, $|u_p| \approx 1 \gg |v_{-p}|$.

From this point on, we will only consider the phonon limit of the quasi-particles unless explicitly stated otherwise.

\subsection{Third Order: Phonon-phonon Interactions and Damping \label{sec:uniform_bose_gas_phonon-phonon_interactions}}

Up to second order in the noncondensate particle creation and annihilation operators, the quasi-particles are non-interacting and thus do not decay \cite{howl_quantum_2018}. If, however, the third-order terms in $\hat{a}_{ \mathbf{p} }$, $\hat{a}^\dagger_{ \mathbf{p} }$ in the Hamiltonian \eqref{eq:uniform_interacting_hamiltonian}

\begin{equation}
	\hat{H}^{(3)} = \frac{1}{V} \sum_{ \mathbf{p}, \mathbf{q} \neq 0 } U_{ \mathbf{p} } \left(
		\hat{a}^\dagger_0 \hat{a}^\dagger_{ \mathbf{p} + \mathbf{q} } \hat{a}_{ \mathbf{p} } \hat{a}_{ \mathbf{q} } + \hat{a}^\dagger_{ \mathbf{p} } \hat{a}^\dagger_{ \mathbf{q} } \hat{a}_0 \hat{a}_{ \mathbf{p}  + \mathbf{q}}
		\right)
\end{equation}

are taken into account, the resulting phonon-phonon interaction terms cause the phonons to have a finite life time. In the quasi-particle picture, the terms corresponding to a particular momentum $\mathbf{p}$ in $\hat{H}^{(3)}$ are \cite{howl_quantum_2018, pitaevskii_landau_1997, giorgini_damping_1998, bramati_spatial_2013}

\begin{equation}
	\hat{h}^{(3)}_{ \mathbf{p} } = \hat{b}_{ \mathbf{p} } \hat{E}^\dagger_{ \mathbf{p} } + \hat{b}^\dagger_{ \mathbf{p} } \hat{E}_{ \mathbf{p} }
\end{equation}

where

\begin{IEEEeqnarray}{rCl}
	\IEEEyesnumber \IEEEyessubnumber*
	\hat{E}_{ \mathbf{p} } 
	& \equiv &
	\hat{A}_{ \mathbf{p} } + \hat{B}_{ \mathbf{p} } + \hat{L}_{ \mathbf{p} }
	\\
	\hat{A}^\dagger_{ \mathbf{p} }
	& \equiv &
	g \sqrt{\frac{n}{V}} \sum_{ \mathbf{q}, \mathbf{q}' \neq \{ 0, \mathbf{p} \} }
		\mathcal{A}_{q, q'} \hat{b}_{ \mathbf{q} } \hat{b}_{ \mathbf{q'} } \delta_{- \mathbf{p}, \mathbf{q} + \mathbf{q}'}
	\\
	\hat{B}^\dagger_{ \mathbf{p} }
	& \equiv &
	g \sqrt{\frac{n}{V}} \sum_{ \mathbf{q}, \mathbf{q}' \neq \{ 0, \mathbf{p} \} }
		\mathcal{B}_{q, q'} \hat{b}^\dagger_{ \mathbf{q} } \hat{b}^\dagger_{ \mathbf{q'} } \delta_{ \mathbf{p}, \mathbf{q} + \mathbf{q}'}
		\label{eq:uniform_bose_gas_beliaev_damping operator}
	\\
	\hat{L}^\dagger_{ \mathbf{p} }
	& \equiv &
	g \sqrt{\frac{n}{V}} \sum_{ \mathbf{q}, \mathbf{q}' \neq \{ 0, \mathbf{p} \} }
		\mathcal{L}_{q, q'} \hat{b}_{ \mathbf{q} } \hat{b}^\dagger_{ \mathbf{q'} } \delta_{ \mathbf{p}, \mathbf{q}' - \mathbf{q}}
		\label{eq:uniform_bose_gas_landau_damping_operator}
\end{IEEEeqnarray}

and $\mathcal{A}_{q, q'}, \, \mathcal{B}_{q, q'}, \, \mathcal{L}_{q, q'}$ are determined by the Bogoliubov mode solutions $u, \, v$, such as for example \eqref{eq:uniform_bose_gas_bogo_mode_solutions}.\footnote{For the exact expressions for $\mathcal{A}_{q, q'}, \, \mathcal{B}_{q, q'}, \, \mathcal{L}_{q, q'}$, see \cite{howl_quantum_2018}.}

The interaction terms $\hat{b}_{ \mathrm{p} } \hat{L}^\dagger_{ \mathrm{p} }$ and $\hat{b}_{ \mathrm{p} } \hat{B}^\dagger_{ \mathrm{p} }$ correspond to the Landau and the Beliaev process. In the Beliaev process, a phonon in mode $\mathbf{p}$ spontaneously decays into two lower-energy modes,\footnote{This process is analogous to the quantum optics process of spontaneous parametric down-conversion.}
 whereas the Landau process consists of two phonons in modes $\mathbf{p}$ and $\mathbf{q}$ colliding and forming a higher-energy mode $\mathbf{p}' = \mathbf{p} + \mathbf{q}$. Both mechanisms lead to thermalization and therefore loss of coherence in phonon modes that are out of equilibrium. This puts an upper limit on the phonon life times.

\begin{figure}[h!pbt]
	\begin{subfigure}[b]{0.35\textwidth}
		\centering
		\includegraphics[width=\textwidth]{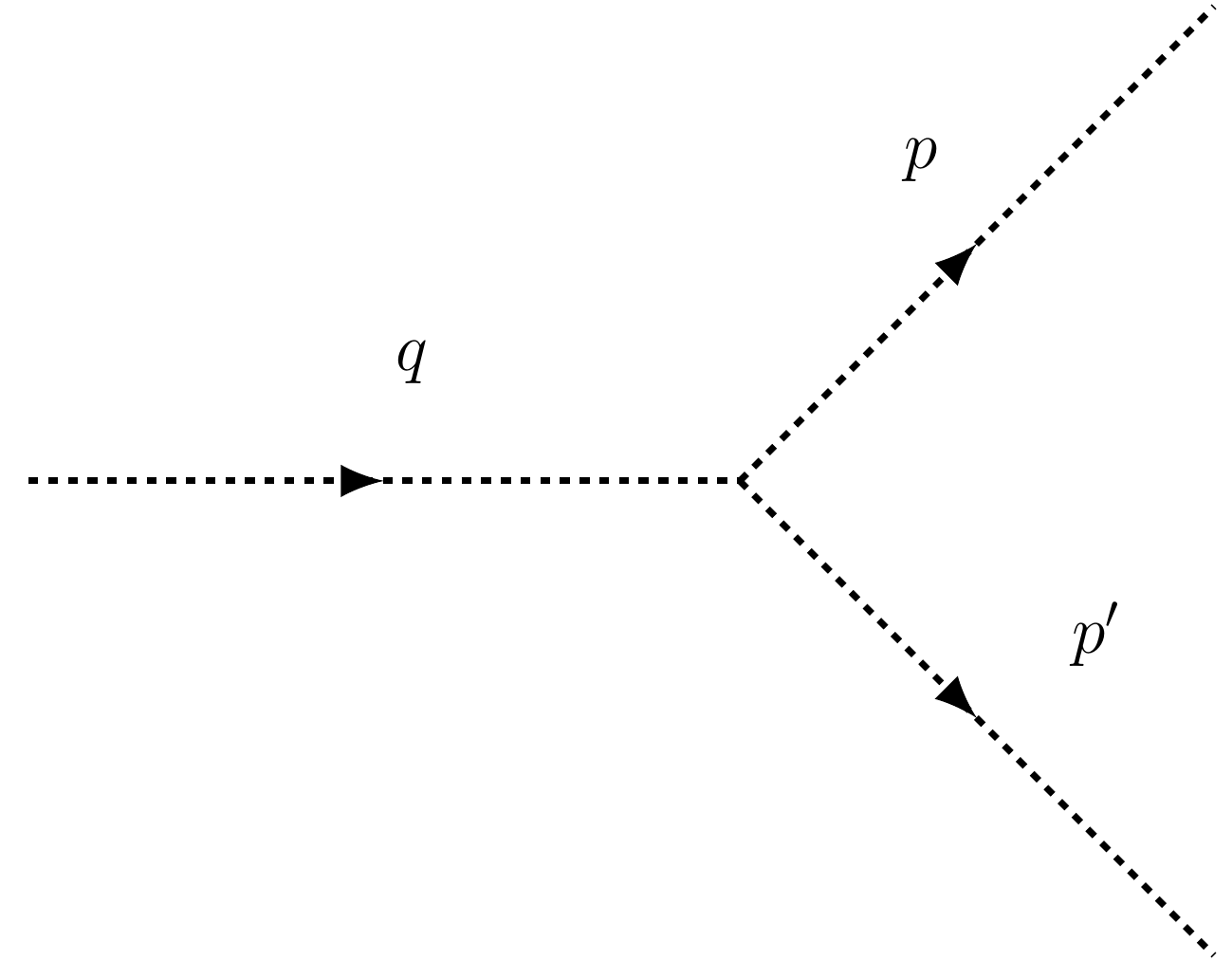}
		\caption{}
		\label{fig:beliaev_damping}
	\end{subfigure}
	\hfill
	\begin{subfigure}[b]{0.35\textwidth}
		\centering
		\includegraphics[width=\textwidth]{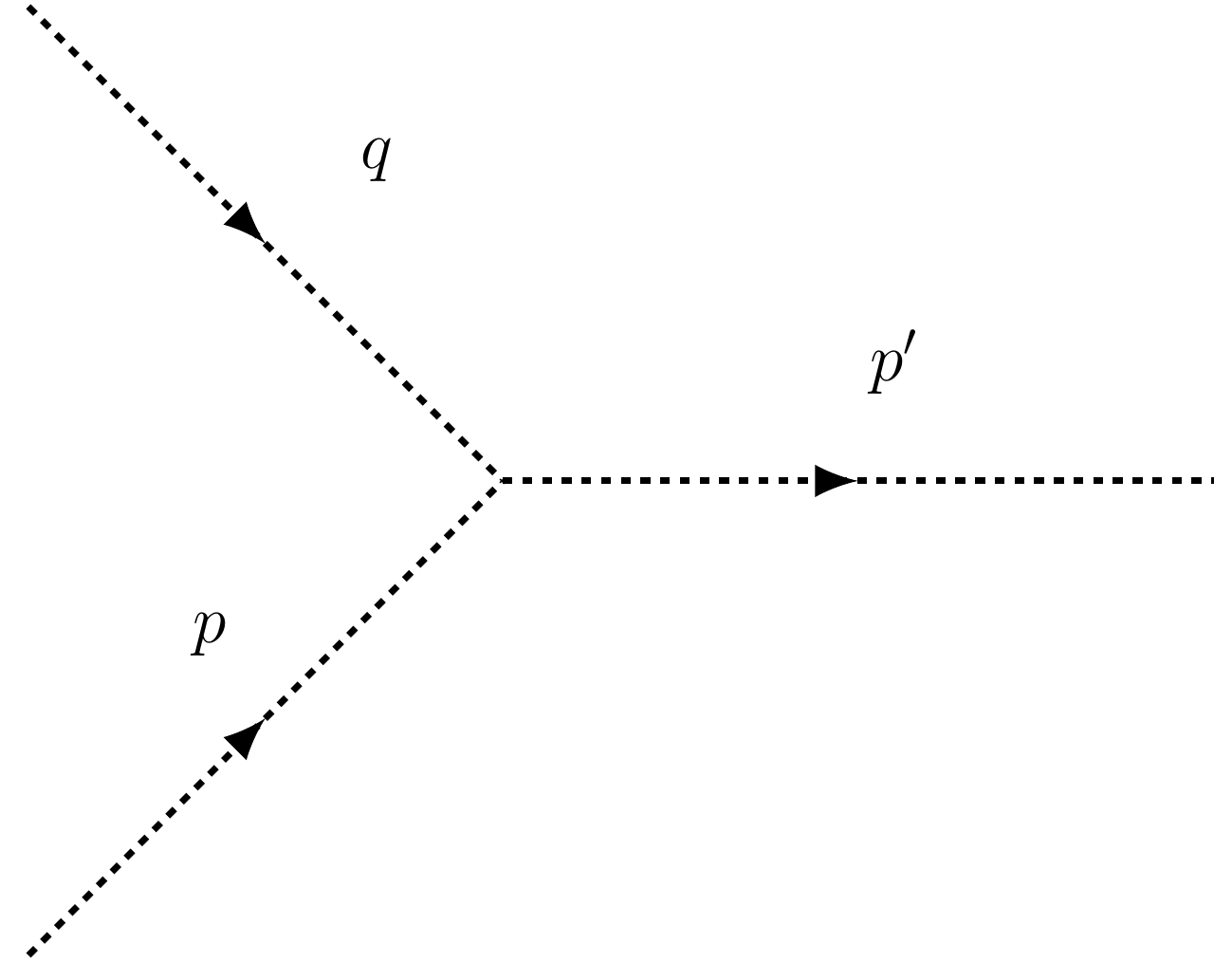}
		\caption{}
		\label{fig:landau_damping}
	\end{subfigure}
	\caption{
		{\footnotesize	
			Phonon-phonon interactions up to third order in $\hat{a}^\dagger_{ \mathbf{p}}$, $\hat{a}_{ \mathbf{p}}$ \cite{howl_decoherence_2016}.
			(a) Beliaev damping.
			(b) Landau damping.
		}
	}
\end{figure}

As already mentioned, the Beliaev decay can occur spontaneously and therefore also at $T=0$. The Landau process, in contrast, requires the thermal occupation of another mode. As can be seen from the expression for the thermal phonon occupation number \eqref{eq:uniform_bose_gas_thermal_phonon_occupation_number}, this means that the Landau process cannot happen at zero temperature. Thus, Landau and Beliaev damping can be expected to dominate at high and at low temperatures, respectively.

For the case of the uniform, weakly interacting Bose gas, the resulting Landau and Beliaev damping rates have been given in \cite{pitaevskii_landau_1997, giorgini_damping_1998} as:

\begin{IEEEeqnarray}{rCl} \label{eq:landau_and_beliaev_damping}
	\gamma^{\mathrm{La}}_p
	&=&
	\frac{4 m c_s a_s \omega_p}{\hbar} \int_0^\infty \ud x
		\left(
			\mathrm{e}^x - \mathrm{e}^{-x}
		\right)^{-2}
		\left(
			1 - \frac{1}{2u} - \frac{1}{2u^2}
		\right)^2
		\IEEEyesnumber		
		\IEEEyessubnumber
		\\
		\gamma^{\mathrm{Be}}_p
		&=&
		\gamma^{\mathrm{Be}, \, 0}_p
		\left(
			1 + 60 \int_0^1 \ud x
				\frac{x^2 \left(x-1 \right)^2}{\mathrm{e}^{\beta x c_s p}-1}
		\right) \; , \IEEEyessubnumber
\end{IEEEeqnarray}

where $\omega_p$ is the frequency of a phonon mode $\mathbf{p}$, $u \equiv \sqrt{1+ \left( 2x / \beta \mu \right)^2}$, the zero temperature Beliaev damping rate is

\begin{equation} \label{eq:zero_temperature_beliaev_damping_rate}
	\gamma^{\mathrm{Be}, \, 0}_p = \frac{3}{640 \pi \hbar^4} \frac{p^5}{m n_0}
\end{equation}

and $n_0 \equiv N_0/V$ is the condensate particle density. To gain a better intuition for the behaviour of the Landau damping rate, consider its high and low temperature limits

\begin{IEEEeqnarray}{rCl} \label{eq:landau_damping_limits}
	\gamma^{\mathrm{La}}_p 
	& \xrightarrow[k_B T \gg \mu]{} &
	\frac{3 \pi}{8 \hbar} \frac{a_s \omega_p}{\beta c_s}
	\IEEEyesnumber
	\IEEEyessubnumber
	\\
	\gamma^{\mathrm{La}}_p 
	& \xrightarrow[k_B T \ll \mu]{} &
	\frac{3 \pi^3}{40 \hbar^3} \frac{\omega_p}{m n_0 \beta^4 c_s^5} \; .
	\IEEEyessubnumber
\end{IEEEeqnarray}

For small temperatures, $\gamma^{\mathrm{La}}_p$ scales as $T^4$, vanishing at $T=0$, whereas $\gamma^{\mathrm{Be}}_p$ remains finite. In the regime $k_B T \ll \mu$, Landau processes will therefore make only a minor contribution to the overall phonon damping.

\section{Ground States of Nonuniform Condensates \label{sec:GPE}}

The considerations so far have been made under the assumption of a uniform trap with cyclic boundary conditions, resulting in a uniform ground state. However, because of the abundance of harmonic traps and also because of possible irregularities at the boundaries of a hard-wall box trap, it is important to have a theory describing the ground state in nonuniform Bose gases. This section is therefore centered around the Gross-Pitaevskii equation (GPE), which was independently discovered in 1961 by Gross and Pitaevskii \cite{pitaevskii_vortex_1961, gross_structure_1961}.

\subsection{The Gross-Pitaevskii Equation}

There are two possible paths for the derivation of the GPE, one of which is a variational approach employing the Hartree-Fock approximation \cite{pitaevskij_bose-einstein_2010,pethick_bose-einstein_2008}.\footnote{In the Hartree-Fock approximation, the many-body wave function of the ground state is written as a symmetrized product of single-particle wave functions.}
In the spirit of the previous sections, however, it comes more naturally to follow the arguments made in \cite{pitaevskij_bose-einstein_2010} and generalize the Bogoliubov theory to include nonuniform external potentials.

As a first step, let us recount the regime in which this generalized theory should be applicable:

\begin{itemize}
	\item \textit{Sufficiently large particle number} such that the term Bose-Einstein condensation can be used in a meaningful way.
	
	\item \textit{Weak interactions \eqref{eq:weak_interaction_approximation} and low temperatures} $k_B T \ll \mu$. With this, the interaction potential can again be replaced by an effective potential of which only the zero-momentum Fourier component is considered \eqref{eq:low_energy_limit_effective_potential}.
	
	\item \textit{No phenomena over small distances}: the GPE can only describe the behaviour of the ground state over distances much larger than the scattering length.
\end{itemize}

With the Hamiltonian that was used at the beginning of the derivation of the Bogoliubov theory \eqref{eq:full_interacting_hamiltonian}, the field operator $\hat{\psi} \left( \mathbf{r}, t \right)$ evolves as

\begin{IEEEeqnarray}{rCl}
	i \hbar \frac{\partial}{\partial t} \hat{\psi} \left( \mathbf{r}, t \right)
	&=& - \left[ \hat{H}, \, \hat{\psi} \left( \mathbf{r}, t \right) \right]
	\nonumber
	\\
	&=&
	\Bigg( - \frac{\hbar^2}{2m} \nabla^2 + \mathcal{V} \left( \mathbf{r}, t \right)
	\\
	& &
	+ \int \ud \mathbf{r} \hat{\psi}^\dagger \left( \mathbf{r}', t \right) U \left( \mathbf{r}' - \mathbf{r} \right) \hat{\psi} \left( \mathbf{r}', t \right)
	\Bigg)
	 \hat{\psi} \left( \mathbf{r}, t \right) \; .
	\nonumber
\end{IEEEeqnarray}

Under the above assumptions, the field operator $\hat{\psi} \left( \mathbf{r}, t \right)$ can be replaced by a classical field $\psi \left( \mathbf{r}, t \right)$ as was done in the zeroth-order approximation in the previous section. This step bears a strong analogy to the classical limit of quantum electrodynamics resulting in Maxwell's equations. Its result in this case is the time-dependent Gross-Pitaevskii equation

\begin{equation} \label{eq:gross-pita_time-dependent}
	i \hbar \frac{\partial}{\partial t} \psi \left( \mathbf{r}, t \right) = \left(
		- \frac{\hbar^2}{2m} \nabla^2 + \mathcal{V} \left( \mathbf{r}, t \right) + g |  \psi \left( \mathbf{r}, t \right) |^2 
		\right)
	\psi \left( \mathbf{r}, t \right) \; ,
\end{equation}

where $g$ is, again, the interaction strength \eqref{eq:zeroth_order_interaction_strength}. For stationary solutions

\begin{equation}
	\psi \left( \mathbf{r}, t \right) = \psi_0 \left( \mathbf{r} \right) \mathrm{e}^{-\frac{i}{\hbar} \mu t}
\end{equation}

and time-independent external potentials, the GPE reduces to an eigenvalue equation with the chemical potential $\mu$

\begin{equation} \label{eq:stationary_GPE}
	\left( - \frac{\hbar^2}{2m} \nabla^2 + \mathcal{V} \left( \mathbf{r} \right) + g | \psi_0 \left( \mathbf{r} \right)  |^2 \right) \psi \left( \mathbf{r} \right) = \mu \psi_0 \left( \mathbf{r} \right) \; .
\end{equation}

This resembles a nonlinear Schr\"odinger equation where, due to the self-interaction term, the chemical potential replaces the energy per particle as the eigenvalue.

\subsection{Harmonic Traps and the Thomas-Fermi Approximation \label{sec:thomas-fermi}}

As an attempt to approximate the ground state in a harmonic trap

\begin{equation} \label{eq:harmonic_trap_potential}
	\mathcal{V} \left( \mathbf{r} \right) = \frac{m}{2} \sum_{i=1}^3 \omega_i^2 r_i^2 \; ,
\end{equation}

a variational approach is taken in \cite{pethick_bose-einstein_2008}, using a solution $\psi_0$ of the stationary GPE without interactions, which is Gaussian, as a trial function. This trial function is then perturbed in the presence of interactions and it is found that the interaction energy per particle outgrows their kinetic energy by far for sufficiently large condensates.

This motivates the \textit{Thomas-Fermi approximation}, which can be taken for condensates with repulsive interactions and sufficiently large particle numbers. In this approximation, the kinetic energy term in the stationary GPE \eqref{eq:stationary_GPE} is neglected right away such that

\begin{equation}
	\left( \mathcal{V} \left( \mathbf{r} \right) + g | \psi_0 \left( \mathbf{r} \right) |^2 \right) \psi_0 \left( \mathbf{r} \right) = \mu \psi_0 \left( \mathbf{r} \right) \; .
\end{equation}

A solution $\psi_0 \left( \mathbf{r} \right)$ is easily found to be 

\begin{equation} \label{eq:thomas-fermi_solution}
	| \psi_0 \left( \mathbf{r} \right) |^2 =
	\begin{cases}	
		\frac{\mu - \mathcal{V} \left( \mathbf{r} \right)}{g} & \text{where} \, \mathcal{V} \left( \mathbf{r} \right) < \mu
		\\
		0 & \text{else.}
	\end{cases}
\end{equation}

For a harmonic potential \eqref{eq:harmonic_trap_potential}, the condensate density therefore has the shape of an inverse parabola with a maximum $\mu/g$ in $\mathbf{r}=0$ and falls to zero at $\mathcal{V} \left( \mathbf{r} \right) = \mu$. These boundaries are called the Thomas-Fermi radii

\begin{equation} \label{eq:thomas-fermi_radius}
	R_{TF, i}^2 = \frac{2 \mu}{m \omega_i^2} \; .
\end{equation}

The chemical potential is calculated using the normalization condition on $\psi_0 \left( \mathbf{r} \right)$ \cite{pethick_bose-einstein_2008}:

\begin{equation}
	\mu = \frac{15^{\frac{2}{5}}}{2} \left( \frac{N a_s}{\bar{a}_{ho}} \right)^\frac{2}{5} \hbar \bar{\omega}
\end{equation}

with the geometric mean of the trapping frequencies $\bar{\omega} \equiv \left( \omega_x \omega_y \omega_z \right)^{1/3}$ and the corresponding typical harmonic oscillator length $\bar{a}_{ho} \equiv \sqrt{\hbar/\left(m \bar{\omega} \right)}$. The Thomas-Fermi radii are then

\begin{equation}
	R_{TF, i} = \bar{R}_{TF} \frac{\bar{\omega}}{\omega_i} \; ,
\end{equation}

where the geometric mean of the the individual radii is

\begin{equation}
	\bar{R}_{TF}
	\equiv \left( R_{TF, x} R_{TF, y} R_{TF, z} \right)^\frac{1}{3}
	= 15^\frac{1}{5} \left( \frac{N a_s}{\bar{a}_{ho}} \right)^\frac{1}{5} \bar{a}_{ho} \; .
\end{equation}

\cite{pitaevskij_bose-einstein_2010} gives an estimate for the applicability of the Thomas-Fermi approximation: the kinetic energy term in \eqref{eq:stationary_GPE} becomes negligible if the typical distance $D$ over which variations in the condensate density extend becomes much larger than the \textit{healing length}

\begin{equation} \label{eq:healing_length_definition}
	\xi = \frac{\hbar}{\sqrt{2mgn}} \; .
\end{equation}

$D$ scales as $D \sim \left( \nabla^2 |\psi_0|/|\psi_0|\right)^{-1/2}$ and thus the condition for the Thomas-Fermi approximation becomes

\begin{equation}
	\frac{\nabla^2 |\psi_0|}{|\psi_0|} \ll \xi^{-2} \; .
\end{equation}

Another example in \cite{pethick_bose-einstein_2008} helps gain further intuition on the physical content of the healing length $\xi$. Imagine a hard-wall box trap such that $\psi_0$ must vanish at its boundaries. Away from the walls, on the other hand, the condensate is close to uniform, $|\psi_0| \approx \sqrt{N/V}$.

For an infinite potential wall at $x=0$, the solution of the one-dimensional stationary GPE is then shown to be

\begin{equation}
	\psi_0 (x) = \sqrt{\frac{N}{V}} \tanh \left( \frac{x}{\sqrt{2} \xi} \right).
\end{equation}

This clarifies that the perturbation to the condensate density caused by the box potential ``heals'' over a distance $\mathcal{O} (\xi)$.

\chapter{Feshbach Resonances \label{sec:feshbach_resonances}}

The phenomenon of Feshbach resonances has made a great impact in the field of ultracold gases and is of high relevance for the purposes of this thesis because it provides a tool to manipulate the interaction strength between the particles.

This section, being largely based on a review paper by Chin et al. \cite{chin_feshbach_2010} and the corresponding sections in Pethick and Smith's book \cite{pethick_bose-einstein_2008}, will give a few historical landmarks in the work on Feshbach resonances, a brief overview of the theoretical considerations behind them and finally some examples for various atomic species, with particular focus on magnetically tunable resonances and alkali atoms.

To get a basic picture of the underlying mechanism, consider two particles scattering in the low-energy limit $E \rightarrow 0$ at relative position $\left( \mathbf{R} \right)$ in the \textit{entrance} or \textit{open channel} governed by the background potential $U_{bg} \left( \mathbf{R} \right)$. Now imagine a \textit{closed channel}\footnote{corresponding to some different internal state of the scattering particles} represented by the potential $U_c \left( \mathbf{R} \right)$ which supports a bound state at energy $E_c$. If $E_c$ comes close to the threshold of the open channel, there is strong interference between the intermediate bound state in the closed channel and the free scattering state, even if the coupling between the channels is weak. This, in turn, results in a rapid change in the interaction strength even for small variations of $E_c$. Because the influence of the closed channel is only in terms of intermediate states in the scattering process, this is a second-order process. From perturbation theory, one would thus expect a correction to the scattering length to behave as \cite{pethick_bose-einstein_2008}

\begin{equation}
	\delta a \sim \frac{C}{E-E_c}
\end{equation}

with some constant $C$. If the magnetic moment of the bound state differs from that of the open channel, the relative energy gap between $E_c$ and the asymptotic value of $U_{bg}$ can be tuned by applying a homogeneous magnetic field. This is called a magnetically tuned Feshbach resonance. An optical Feshbach resonance, in contrast, is tuned by varying the frequency of an external optical field.

\begin{figure}[hbtp]
		\begin{centering}
		\includegraphics[width=0.8\textwidth]{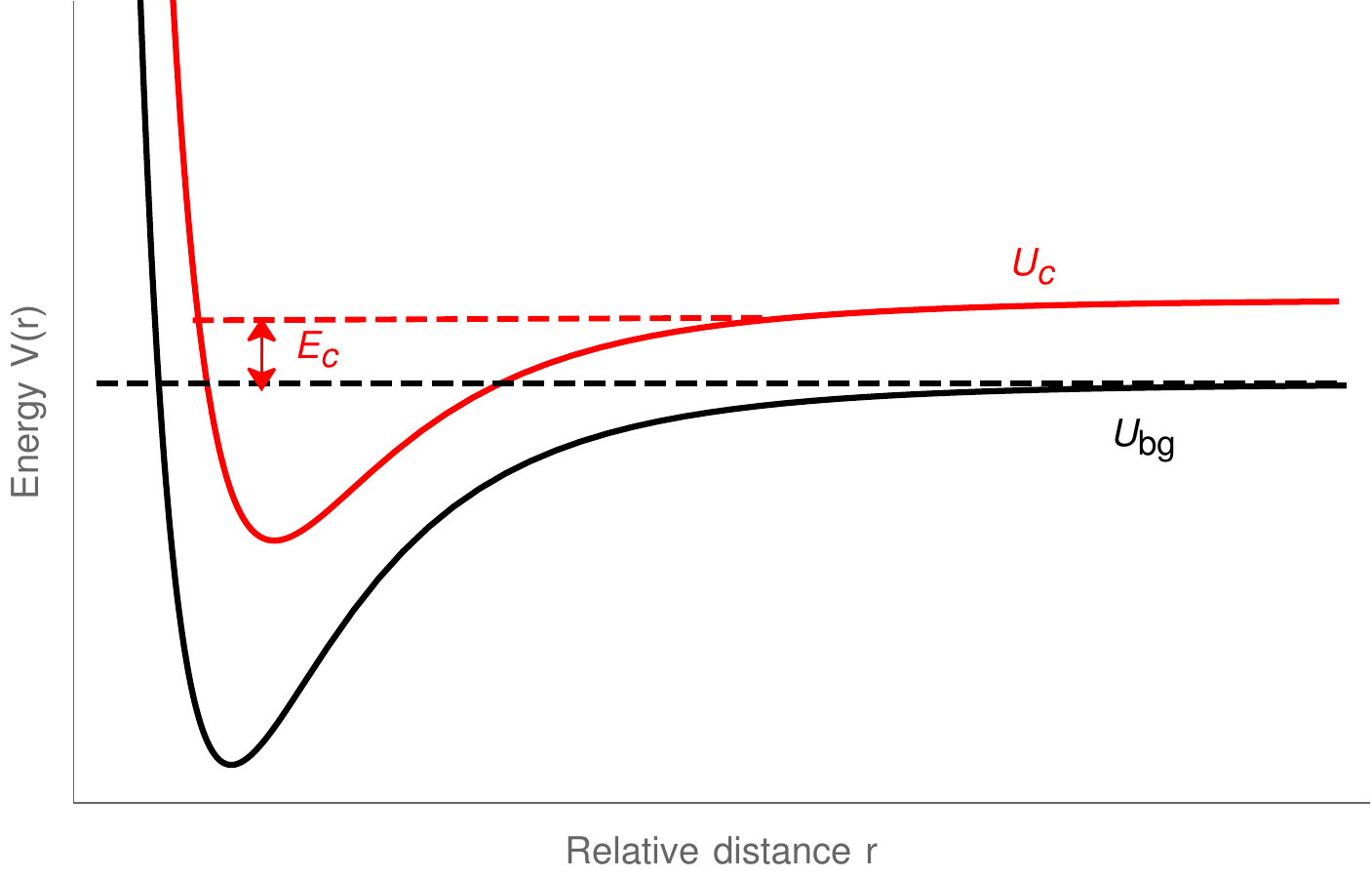}
		\caption{
			{\footnotesize
				Sketch the background and the closed channel potentials $U_{bg}$ (black) and $U_c$ (red, supporting a bound state) for a Feshbach resonance. The black and dashed lines mark the threshold energy of the open channel and the bound state energy of the background channel. Once the gap $E_c$ between the weakly bound state energy in the closed channel and the threshold energy in the open channel are tuned to zero, interference between the intermediate closed channel bound state and the scattering state in the open channel leads to a drastic variation in the interaction strength in the open channel.
			}
		\label{fig:feshbach_potential_plot}}
		\par\end{centering}
\end{figure}

While first steps in the theoretical description of the effects of coupling between bound states and the continuum date back to the early days of quantum mechanics \cite{rice_predissociation_1933, fano_sullo_1935}, the decisive works by Feshbach and Fano \cite{feshbach_unified_1958, feshbach_unified_1962, fano_effects_1961}, who approached the same problem from the different backgrounds of nuclear and atomic physics, followed around 1960. Because of their independent contributions to the same phenomenon, it is sometimes also referred to as Fano-Feshbach resonance.

The first experimental observation of a Feshbach resonance was made in 1977 \cite{bryant_observation_1977}. Although it had found applications in various fields soon afterwards, it was only in 1976 that it was first considered to play a role in quantum gases, where it was primarily thought of as a problem in the sense that it would enhance inelastic decay \cite{stwalley_stability_1976}. It was not until 1993 that their possible use as a tool to tune interaction strengths in ultracold atoms was predicted \cite{tiesinga_threshold_1993} and finally demonstrated experimentally in 1998 in a $^{23}$Na \cite{inouye_observation_1998} and a $^{85}$Rb BEC \cite{courteille_observation_1998}.

To gain a deeper theoretical understanding of Feshbach resonances, one has to take a minor detour and make some elementary observations from scattering theory. For more details, the reader is referred to \cite{pethick_bose-einstein_2008, landau_quantum_2013}.

\subsection{Single-channel Scattering and the $T$ Matrix \label{subsec:single-chanel_scattering}}

The first step is to determine the correspondence between the scattering potential, the \textit{scattering phase} and the scattering length in the low-energy limit. For two particles that are initially prepared in plane waves with reduced mass $\mu_r$, relative position $\mathbf{r}$ and relative momentum $\mathbf{k}$ colliding under the influence of a spherically symmetric potential $U(r)$, the wave function describing the incoming and the scattered wave can be expanded in its angular momentum components \cite{pethick_bose-einstein_2008}

\begin{equation} \label{eq:feshbach_partial_wave_expansion}
	\psi \left( \mathbf{r} \right) = \sum_{l=0}^\infty A_l P_l \left( \cos \theta \right) \frac{\phi_{kl} (r)}{r} \; ,
\end{equation}

where $P_l \left( \cos \theta \right)$ are Legendre polynomials, $A_l = i^l (2l+1)\mathrm{e}^{i \delta_l}$ and $l$ denotes the relative angular momentum, corresponding to $s, \, p, \, d, \ldots$ partial waves for $l=1, \, 2, \, 3, \ldots$. Due to the symmetry of $U(r)$, there is no coupling between partial waves of different relative angular momentum. The radial function $\phi_{kl} (r)$ then depends only on the magnitude of the relative position and solves the radial part of the Schr\"odinger equation \cite{chin_feshbach_2010}

\begin{equation} \label{eq:radial_SEQ_feshbach_resonances}
	\left( - \frac{\hbar^2}{2 \mu_r} \frac{\ud}{\ud r^2} + U_l (r) \right) \phi_{kl} (r) = E \phi_{kl} (r)
\end{equation}

with the pseudopotential $U_l (r) = U (r) + \hbar^2 l (l+1) / (2 \mu_r r^2)$ and the relative kinetic energy $E = \hbar^2 k^2 / (2 \mu_r)$. If $U(r)$ vanishes in the limit of large separations, the solution to \eqref{eq:radial_SEQ_feshbach_resonances} in the same limit $r \rightarrow \infty$ is given by \cite{pethick_bose-einstein_2008}

\begin{equation} \label{eq:feshbach_radial_wave_solution}
	\phi_{kl} (r) \approx \frac{\sin \left(k r - l \pi + \delta_l \right)}{k} \; .
\end{equation}

The scattering behaviour for given values of $l, k$ is thus encoded in the \textit{scattering phase} $\delta_l$. It has been shown \cite{sadeghpour_collisions_2000} that for potentials that behave as $U(r) \rightarrow r^{-6}$ in the limit of large $r$ and\footnote{In fact, the law found in \cite{sadeghpour_collisions_2000} works for general asymptotic behaviour $U(r) \rightarrow r^{-s}$. The particular exponent $s=6$ was picked here because it resembles the van der Waals potential.} for small relative momenta $k \rightarrow 0$, the proportionality between $\delta_l$ $k$ is well approximated by

\begin{equation}
	\delta_l \propto 
	\begin{cases}
		k & \text{for } l=0 \; (s \text{ waves})
		\\
		k^3 & \text{for } l=1 \; (p \text{ waves})
		\\
		k^4 & \text{for } l>1 \; (\text{all other partial waves}) \; .
	\end{cases}
\end{equation}

Thus, all partial waves with $l\neq 0$ can be safely neglected.

On the other hand, in the same limit of large separation and small momenta, the full wave function can be written as \cite{pethick_bose-einstein_2008}

\begin{equation} \label{eq:feshbach_wave_function_scattering_length}
	\psi (r) = 1 - \frac{a_s}{r} \; .
\end{equation}

Comparison with \eqref{eq:feshbach_partial_wave_expansion} and \eqref{eq:feshbach_radial_wave_solution} yields the desired relation

\begin{equation}
	\delta_0 = - k a_s \; .
\end{equation}

The momentum space wave function consisting of the initial plane wave of momentum $\mathbf{k}$ and the scattered wave $\psi_{sc}$, is

\begin{equation}
	\tilde{\psi} ( \mathbf{k}' ) = \left( 2 \pi \right)^3 \delta^{(3)} \left( \mathbf{k}' - \mathbf{k} \right) + \tilde{\psi}_{sc} \left( \mathbf{k}' \right) \; ,
\end{equation}

where $\tilde{\cdot}$ denotes a Fourier transform. The corresponding Schr\"odinger equation in $\mathbf{k}$ space is then \cite{pethick_bose-einstein_2008}

\begin{equation}
	\left( \frac{\hbar^2 k^2}{m} - \frac{\hbar^2 k'^2}{m} \right) \tilde{\psi}_{sc} \left( \mathbf{k}' \right) = \tilde{U} \left( \mathbf{k} - \mathbf{k}' \right)
	+ \frac{1}{V} \sum_{\mathbf{k}''}
		\tilde{U} \left( \mathbf{k}' - \mathbf{k}'' \right) \tilde{\psi}_{sc} \left( \mathbf{k}'' \right)
\end{equation}

and thus

\begin{IEEEeqnarray}{rCl}
	\tilde{\psi}_{sc} \left( \mathbf{k}' \right)
	&=&
	\left( \frac{\hbar^2 k^2}{m} - \frac{\hbar^2 k'^2}{m} + i \delta \right)^{-1} \cdot
	\nonumber
	\\
	\label{eq:feshbach_k-space_scattering_solution}
	& & \cdot
	\left(
		\tilde{U} \left( \mathbf{k} - \mathbf{k}' \right)
	+ \frac{1}{V} \sum_{\mathbf{k}''}
		\tilde{U} \left( \mathbf{k}' - \mathbf{k}'' \right) \tilde{\psi}_{sc} \left( \mathbf{k}'' \right)
	\right)
	\IEEEyesnumber
	\IEEEyessubnumber
	\\
	\label{eq:feshbach_T-matrix_definition}
	& \equiv &
	\left( \frac{\hbar^2 k^2}{m} - \frac{\hbar^2 k'^2}{m} + i \delta \right)^{-1}
	T \left( \mathbf{k}', \, \mathbf{k}; \, E=\frac{\hbar^2 k^2}{m} \right) \; .
	\IEEEyessubnumber
\end{IEEEeqnarray}

In \eqref{eq:feshbach_k-space_scattering_solution}, $i \delta, \, \delta \in \mathbb{R}$ denotes the condition that the scattered wave should be outgoing. In the solution, the limit $\delta \rightarrow 0$ is then taken. The more important point in the equations above, however, is the definition of the \textit{scattering matrix} $T$ whose elements in the plane wave basis lead to $\tilde{\psi}_{sc}$ and which solves the Lippmann-Schwinger equation

\begin{IEEEeqnarray}{rCl} \label{eq:feshbach_lippmann-schwinger}
	T \left( \mathbf{k}', \, \mathbf{k}; \, E \right)
	&=&
	\tilde{U} \left( \mathbf{k}' - \mathbf{k} \right)
	\nonumber
	\\
	& &
	+ \frac{1}{V} \sum_{\mathbf{k}''}
		\tilde{U} \left( \mathbf{k}' - \mathbf{k}'' \right)
		\left(
			E - \frac{\hbar^2 k''^2}{m} + i \delta
		\right)^{-1}
		T \left( \mathbf{k}'', \, \mathbf{k}; \, E \right) .
\end{IEEEeqnarray}

Taking our familiar limit $k \rightarrow 0, \, r \rightarrow \infty$, the scattered wave is determined by the zero momentum elements of $T$ and comparison with the previous result for the s-wave scattering length \eqref{eq:feshbach_wave_function_scattering_length} leads to  \cite{pethick_bose-einstein_2008}

\begin{equation} \label{eq:as-T-matrix_correspondence}
	a_s = \frac{m}{4 \pi \hbar^2} T \left(0, 0; \, 0 \right) \; .
\end{equation}

The relation \eqref{eq:zeroth_order_interaction_strength} between the zero momentum component of $\tilde{U} ( \mathbf{k} )$ that was applied in section \ref{sec:uniform_weakly_interacting} on the uniform, weakly interacting Bose gas, is then recovered by taking the Born approximation in which the terms in the last line of \eqref{eq:feshbach_lippmann-schwinger} are neglected:

\begin{equation} \label{eq:feshbach_scattering_length_from_U0}
	a_{s, Born} = \frac{m}{4 \pi \hbar^2} \underbrace{\tilde{U} (0)}_{= U_0} = \frac{m}{4 \pi \hbar^2} \int \ud \mathbf{r} \, U \left( \mathbf{r} \right) \; .
\end{equation}

In addition to finding a relation between the $T$ matrix elements and the scattering length, \eqref{eq:feshbach_scattering_length_from_U0} emphasizes a point that was already used in section \ref{sec:uniform_weakly_interacting}: for the purposes of s-wave scattering in the low energy limit, the interaction potential can be replaced by any effective potential that leads to the same scattering length.


\subsection{Scattering Between Multiple Channels and the Feshbach Resonance}

When considering scattering between open and closed channels, it is helpful to split up the Hilbert space for the internal degrees of freedom into the subspaces corresponding to open and closed channels $\mathcal{H} = \mathcal{O} \otimes \mathcal{C}$ and use projections onto the respective subspaces $\hat{O}$ and $\hat{C}$ to divide the state

\begin{equation}
	| \psi \rangle = \underbrace{ \hat{O} | \psi \rangle}_{\equiv | \psi_O \rangle} + \underbrace{\hat{C} | \psi \rangle}_{\equiv | \psi_C \rangle} \;
\end{equation}

and the state vectors in the subspaces can then be denoted by their internal state and their relative momentum

\begin{IEEEeqnarray}{rCl}
	| \psi_O \rangle &=& |o, \mathbf{k} \rangle
	\IEEEyesnumber
	\IEEEyessubnumber
	\\
	| \psi_C \rangle &=& |c, \mathbf{k}' \rangle \; .
	\IEEEyessubnumber
\end{IEEEeqnarray}

After projecting the Schr\"odinger equation $\hat{H} | \psi \rangle = E | \psi \rangle$ onto the open and closed subspaces and recombining the resulting coupled equations, one gets for $| \psi_O \rangle$ \cite{pethick_bose-einstein_2008}

\begin{equation} \label{eq:feshbach_SEQ}
	\left( E - \hat{H}_{OO}  - \hat{H}_{OO}' \right) | \psi_O \rangle = 0 \; ,
\end{equation}

where $\hat{H}_{IJ}$ denotes the projection of the Hamiltonian onto the subspace $\mathcal{I}$ from the left and $\mathcal{J}$ from the right and

\begin{equation}
	\hat{H}_{OO}' = \hat{H}_{OC} \left( E - \hat{H}_{CC} + i \delta \right)^{-1} \hat{H}_{CO} \; .
\end{equation}

The infinitesimal imaginary addition $i \delta $ in the denominator above again denotes outgoing scattered waves. $\hat{H}_{OO}'$ lives on the $\mathcal{O}$ subspace, but it contains all the information on interactions between the open and the closed channels and within the closed channels. It is thus the term that is responsible for Feshbach resonances.

If the open channel Hamiltonian is divided into the interaction-free term $\hat{H}_0$ and the potential $\hat{U}_O$, the total effective interaction in $\mathcal{O}$ is expressed as

\begin{equation}
	\hat{U} = \hat{U}_O + \underbrace{\hat{H}_{OO}'}_{\equiv \hat{U}_C}
\end{equation}

and the Schr\"odinger equation \eqref{eq:feshbach_SEQ} becomes

\begin{equation}
	\left( E - \hat{H}_0 - \hat{U} \right) | \psi_O \rangle = 0
\end{equation}

Deriving the s-wave scattering length for the open channel is then again a matter of finding the $T$ matrix elements in the low-energy limit, with $\hat{T}$ fulfilling

\begin{equation}
	\hat{T} = \hat{U} + \hat{U} \hat{G}_0 \hat{T}
\end{equation}

and

\begin{equation}
	\hat{G}_0 = \left( E - \hat{H}_0 + i \delta \right)^{-1} \; ,
\end{equation}

in analogy to the single-channel scattering case. Up to first order in $\hat{U}_C$, $\hat{T}$ is approximated by \cite{pethick_bose-einstein_2008}

\begin{equation}
	\hat{T} = \hat{T}_O + \left( 1 - \hat{U}_O G_0 \right)^{-1} \hat{U_C} \left( 1 - G_0 \hat{U} \right)^{-1}
\end{equation}

with $\hat{T}_O = \hat{U}_O + \hat{U}_O G_0 \hat{T}_O$. Thus, the matrix elements are expanded as\footnote{Indices denoting the internal states are suppressed here.}

\begin{equation} \label{eq:feshbach_multi-channel_T-matrix_elements}
	\langle \mathbf{k}' | \hat{T} | \mathbf{k} \rangle = \langle \mathbf{k}' | \hat{T}_O | \mathbf{k} \rangle +
	\underbrace{
	\langle \mathbf{k}'	| 	\left( 1 - \hat{U}_O G_0 \right)^{-1}
	}_{\equiv \langle \mathbf{k}' ; \, \hat{U}_O, - |}
	\hat{U}_C
	\underbrace{
			\left( 1 - G_0 \hat{U} \right)^{-1} | \mathbf{k} \rangle
	}_{\equiv | \mathbf{k} ; \, \hat{U}, + \rangle} \; .
\end{equation}

For $r \rightarrow \infty$, the states $| \mathbf{k} ; \, \hat{U}_{(O)}, +/- \rangle$ defined above describe a state consisting of a plane wave and an incoming/outgoing spherical wave. The first term on the r.h.s. of  \eqref{eq:feshbach_multi-channel_T-matrix_elements} alone would lead to the scattering length without any interactions with closed channels.

If we consider only one particular open channel and take the low-energy limit, the distinction between incoming and outgoing spherical waves breaks down and we can take $| \mathbf{k} ; \, \hat{U}_{(O)}, +/- \rangle \rightarrow | \psi_0 \rangle $. By inserting a complete set of states, one can show

\begin{IEEEeqnarray}{rCl} \label{eq:feshbach_low-energy_matrix_elements}
	\lim_{\mathbf{k}, \mathbf{k}' \rightarrow 0} \langle \mathbf{k}' ; \, \hat{U}_O, - | \hat{U}_C | \mathbf{k} ; \, \hat{U}, + \rangle
	&=&
	\langle \psi_0 | \hat{H}_{OC} \left( E_{th} - \hat{H}_{CC} + i \delta \right)^{-1} \hat{H}_{CO} | \psi_0 \rangle \nonumber
	\\
	&=&
	\sum_n \langle \psi_0 | \hat{H}_{OC}
	\left( E_{th} - \hat{H}_{CC} + i \delta \right)^{-1} | n \rangle
	\langle n | \hat{H}_{CO} | \psi_0 \rangle \nonumber
	\\
	&=& \sum_n \frac{| \langle \psi_0 | \hat{H}_{OC} | n \rangle |^2}{E_{th} - E_n}
\end{IEEEeqnarray}

in the limit $\delta \rightarrow 0$. Remember that in the case of the Feshbach resonance, we have $E_c \approx E_{th}$ for one particular bound state energy. The other terms in the sum in \eqref{eq:feshbach_low-energy_matrix_elements} are then approximately constant in $E$ and contribute to the \textit{background scattering length} $a_{bg}$. Using the correspondence \eqref{eq:as-T-matrix_correspondence} between $\hat{T} (0, 0; 0)$ and $a_s$, one therefore gets

\begin{equation}
	a_s = a_{bg} + \frac{m}{4 \pi \hbar^2} \frac{| \langle \psi_0 | \hat{H}_{OC} | c \rangle |^2}{E_{th} - E_c} \; .
\end{equation}

So far, we have not made any assumptions about the mechanism used to tune $E_{th} - E_c$: for a difference in magnetic moments $\delta \mu$ between the open and the resonant closed channel, their energy difference can be manipulated by applying an external magnetic field:

\begin{equation}
	E_{th} - E_c = \delta \mu \left( B - B_0 \right)
\end{equation}

and the scattering length becomes

\begin{equation} \label{eq:feshbach_resonance_formula}
	a_s = a_{bg} \left( 1 - \frac{\Delta B}{B - B_0} \right) \; .
\end{equation}

\begin{figure}[hbtp]
	\begin{centering}
	\includegraphics[width=0.8\textwidth]{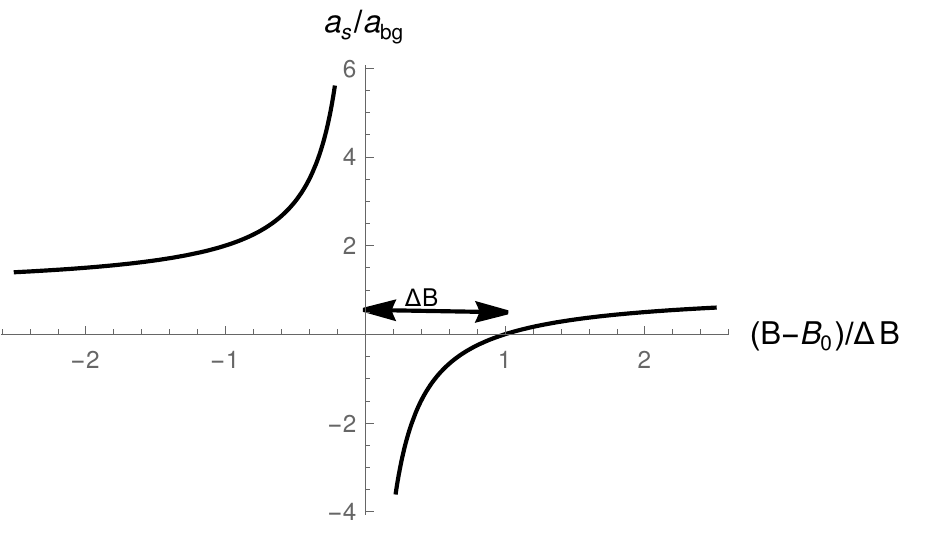}
	\caption{
		{\footnotesize		
			Behaviour of the s-wave scattering length (in terms of its background value) near a Feshbach resonance centered at $B_0$.
			\label{fig:feshbach_resonance_shape}
		}
	}
	\par\end{centering}
\end{figure}

This is the well-known expression for a Feshbach resonance with width $\Delta B$

\begin{equation}
	\Delta B = - \frac{m}{4 \pi \hbar^2 a_{bg}} \frac{| \langle \psi_0 | \hat{H}_{OC} | c \rangle |^2}{\delta \mu} \; ,
\end{equation}

centered around $B_0$ (see fig. \ref{fig:feshbach_resonance_shape}). Note that the scattering length can not only be tuned to arbitrarily high (positive and negative) values, but also

\begin{equation}
	a_s \big|_{B = B_0 + \Delta B} = 0 \; ,
\end{equation}

yielding non-interacting particles.

While the possibility to achieve negative values of the interaction strength opens the door to interesting physics, such as the formation of ultracold Feshbach molecules \cite{chin_feshbach_2010}, the positive range of $a_s$ is the one relevant for the squeezing mechanism this thesis is centered around.

Although \eqref{eq:feshbach_resonance_formula} would theoretically allow to reach any desired scattering length, given perfect precision in the strength of the external magnetic field, there are some obstacles to this. Firstly, $B$ can not be controlled with infinite precision and Feshbach resonances in cold atoms are typically narrow. However, there are major differences between the resonance widths for different atomic species and even for single species between multiple resonances.\footnote{For an overview of discovered Feshbach resonances, see \cite{chin_feshbach_2010}.} Thus, when considering different species for an experiment that requires controlling $a_s$, it is prudent to take the values of $\Delta B$ of their respective resonances into account.
Secondly and most importantly, high interaction strengths lead to strongly enhanced three-body recombination (see fig. \ref{fig:feshbach_three-body_losses})\footnote{In fact, the occurrence of rapidly increased three-body losses was used to first demonstrate the ability to manipulate $a_s$ in a BEC \cite{inouye_observation_1998}.}, a more detailed description of which is given in the subsequent chapter \ref{sec:three-body collisions}. Among the consequences of three-body recombination are particle loss.\footnote{
	and possibly condensate heating \cite{dziarmaga_bose-einstein-condensate_2003}, although it can be argued \cite{roy_test_2013} that the increase in temperature in harmonic traps is due to the scaling of three-body losses with the third power of the particle density, resulting in the preferential ejection of particles at the trap center where the density is maximized. Hence, it would not apply uniform BECs.}
Hence, they put an upper bound on both the condensate and the phonon life time.

\begin{figure}[h!ptb]
	\begin{centering}
	\includegraphics[width=0.8\textwidth]{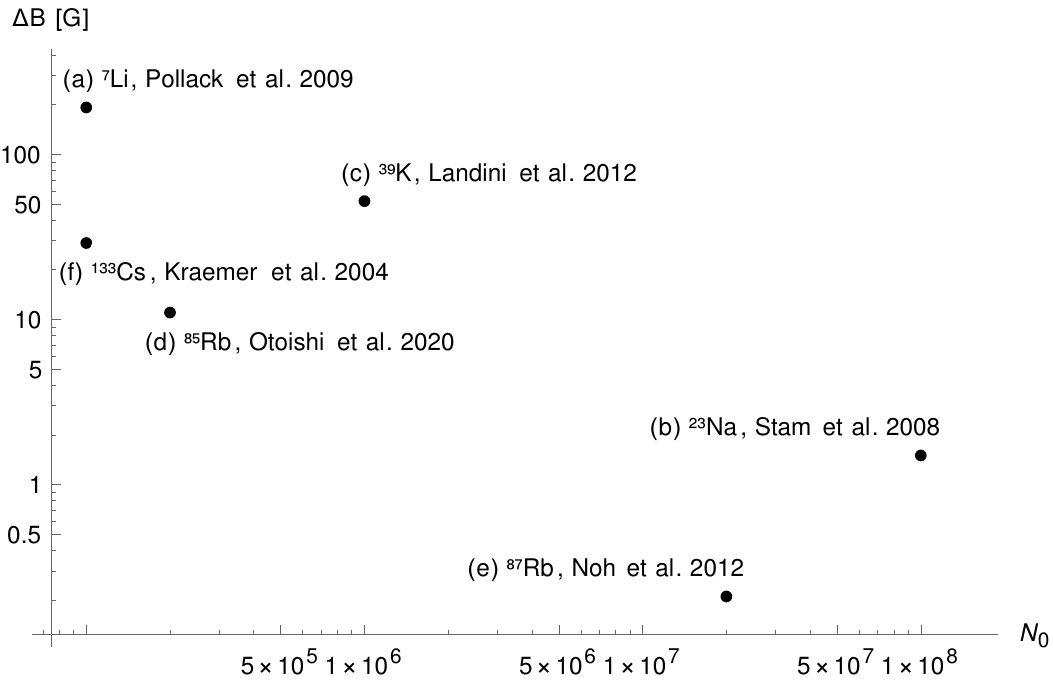}
	\caption{
		{\footnotesize			
			Double logarithmic plot of achieved condensate particle numbers $N_0$ and the width of the broadest Feshbach resonance of the atomic species in the condensate.
			(a) $^7$Li, \cite{pollack_extreme_2009},
			(b) $^{23}$Na, \cite{van_der_stam_large_2007},
			(c) $^{39}$K, \cite{landini_direct_2012},
			(d) $^{85}$Rb, \cite{otoishi_rapid_2020},
			(e) $^{87}$Rb, \cite{noh_high-performance_2012},
			(f) $^{133}$Cs, \cite{kraemer_optimized_2004}
			\label{fig:feshbach_species_scatter_plot}
		}
	}
	\par\end{centering}
\end{figure}

\begin{figure}[hpbt]
	\begin{centering}
	\includegraphics[width=0.5\textwidth]{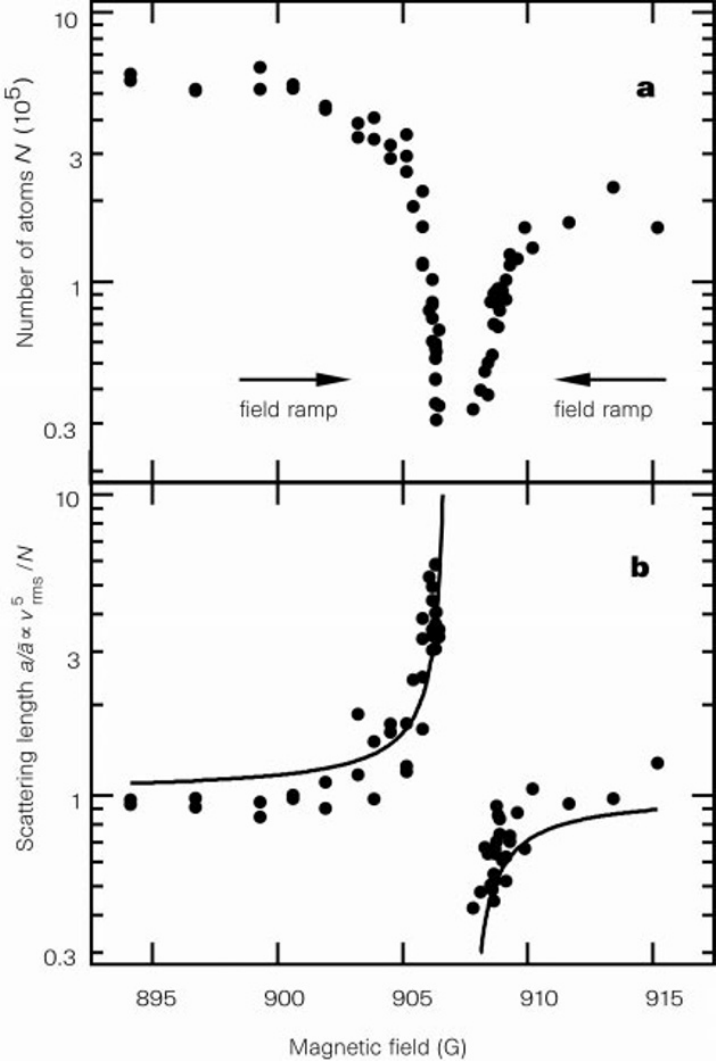}
	\caption{
		{\footnotesize			
			Drastically enhanced three-body losses near a Feshbach resonance are shown in the upper panel. The lower panel pictures the corresponding scattering lengths in terms of $a_{bg}$. \cite{inouye_observation_1998}
			\label{fig:feshbach_three-body_losses}
		}
	}
	\par\end{centering}
\end{figure}

Because the phonon squeezing parameter in a BEC is limited by the number of particles in the condensate (once the number of excitations becomes the same order of magnitude as $N_0$, it does not make sense to speak of a BEC anymore), there is a possible tradeoff between the atom numbers that can be achieved in an experiment and the tunability of the interactions of the considered species. To give the reader an impression of the vast range of these parameters, figure \ref{fig:feshbach_species_scatter_plot} depicts the particle numbers for a few exemplary alkali atom BEC experiments in correspondence with the width of the broadest Feshbach resonance of the species in question.

\newpage


\chapter{Three-Body Losses \label{sec:three-body collisions}}

The task of estimating limitations of the condensate life time due to three-particle recombination has presented itself as one of the most puzzling challenges to the author in the context of this work.

\begin{figure}[hpbt]
	\begin{centering}
	\includegraphics[width=0.5\textwidth]{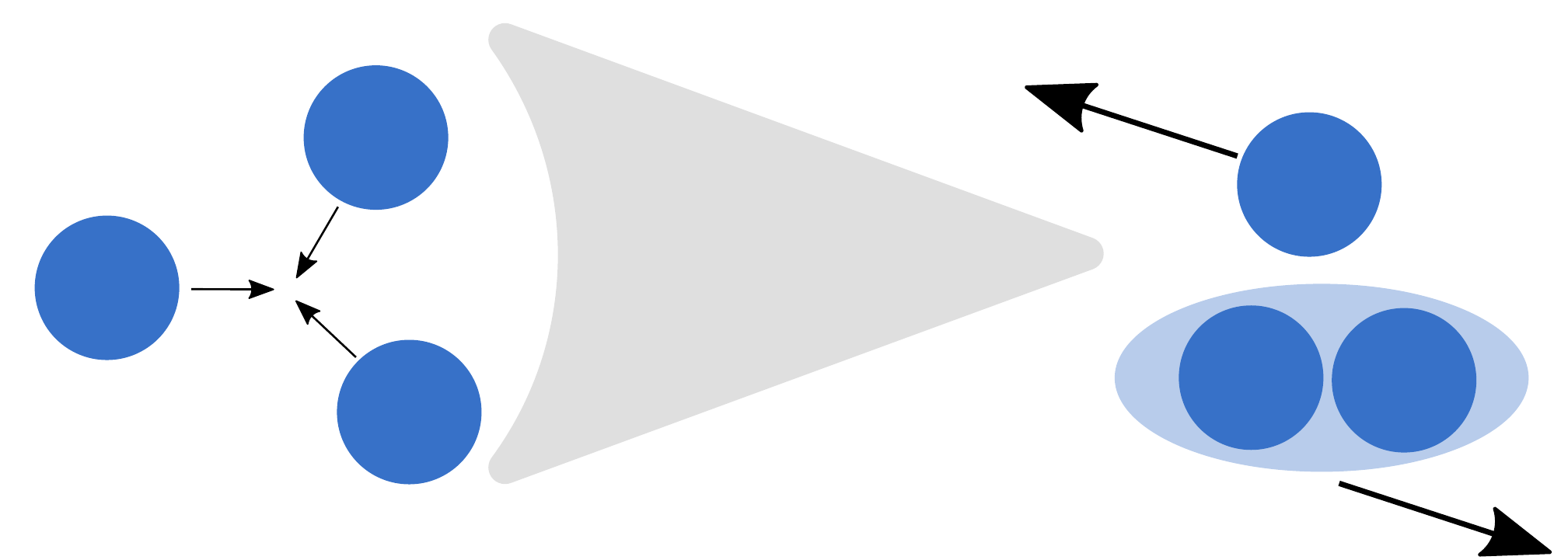}
	\caption{
		{\footnotesize			
			Sketch of a three-body recombination process: three particles collide, two of them form a molecule and the redistribution of the binding energy as kinetic energy ejects the particle and the newly formed molecule from the trap.
			\label{fig:three_body_sketch}
		}
	}
	\par\end{centering}
\end{figure}

The basic process, though, is easily sketched (figure \ref{fig:three_body_sketch}): imagine three identical interacting particles where any two-particle subsystem allows for a bound state or \textit{dimer} with binding energy $\epsilon_{dim}$. If the three particles then collide, two of them can form a dimer and the $\epsilon_{dim}$ is distributed between the newly formed dimer and the remaining atom as kinetic energy. Since, in most BEC experiments, the binding energy is much larger than the trap depth, all three particles are then lost from the trap \cite{weber_three-body_2003}. Thus, the particle number of the condensate decreases in time.

A reader who is new to the field of BECs might first encounter this phenomenon in textbooks, such as Pethick and Smith's \cite{pethick_bose-einstein_2008}, where the dependence of the loss rate on the particle density is intuitively explained. In order for a three-body process to occur, three particles need to be in close proximity to each other, at a range where the two-body interaction becomes relevant. The loss rate must therefore scale as

\begin{equation}
	\frac{ \ud n}{\ud t} = - L_3 n^3 \; ,
\end{equation}

with the \textit{rate coefficient} $L_3$. The particle density (and with it the number of particles in the trap) is then time-dependent and the solution to the differential equation above for a constant rate coefficient is

\begin{equation}
	n (t) = \frac{1}{\sqrt{n^{-2}(0) + 2 L_3 t}}
\end{equation}

and we can deduce a half-life time of the condensate due to three-particle losses

\begin{equation} \label{eq:three-body_half-life}
	t_{1/2} = \frac{3}{2 L_3 n^2(0)}
\end{equation}

as an estimate for an upper bound on the condensate life time.

While it is mentioned in \cite{pethick_bose-einstein_2008} that $L_3$ depends strongly on the atomic species and some values are given for alkali atoms, the reader already familiar with Feshbach resonances could already speculate that this is not the entire truth. After all, this is an interaction problem and one could therefore expect that the rate coefficient would have to depend on the scattering length. Indeed, even from dimensional analysis, one can already show a rough proportionality \cite{sorensen_three-body_2013}: when only considering s-wave scattering in the low energy limit, $a_s$ is the only characteristic length scale of the system. The rate coefficient, by virtue of its definition above, must have the dimension of length$^6/$time. The only combination of $\hbar$, the particle mass $m$ and $a_s$ that yields such a dimension is $\hbar a_s^4/m$ and thus one can write

\begin{equation} \label{eq:three-body_a^4_scaling}
	L_3 = C_3 \frac{\hbar a_s^4}{m}
\end{equation}

with a dimensionless factor $C_3$. Scientific interest in the exact $a_s$ dependence of $L_3$ spiked with the experimental realization of Feshbach resonances. Early theoretical works \cite{fedichev_three-body_1996, nielsen_low-energy_1999, bedaque_three-body_2000} came to the same conclusion \eqref{eq:three-body_a^4_scaling} and their results were roughly supported by experiments, albeit not in close proximity to a Feshbach resonance \cite{stenger_strongly_1999}.

Notably, while \cite{fedichev_three-body_1996, nielsen_low-energy_1999} found only upper and lower bounds $0 < C_3 \lesssim 65$, Bedaque et al. already \cite{bedaque_three-body_2000} found the oscillatory behaviour of $C_3$ with values between zero and $67.9$. These oscillations are also a signature of Efimov physics, a framework in which Braaten and Hammer have found analytical expressions for $C_3(a_s)$ which are in good agreement with experiments \cite{pollack_universality_2009}. For a more in-depth review, the reader is referred to the excellent review paper by Naidon and Endo \cite{naidon_efimov_2017}, on which the following section is largely based.

\section{Efimov Physics}

Three-body losses in ultracold gases are, of course, an instance of the quantum mechanical three-body problem for which no general solution is at hand. However, some of its aspects are still known. Results of particular relevance to this thesis are based on the theoretical work of Vitaly Efimov \cite{efimov_energy_1970, efimov_weakly-bound_1971} on particles interacting via a short-range potential\footnote{That is a potential which decays faster than $r^{-3}$. Therefore, the van der Waals potential falls into this class.}
near a resonance and hence in the limit of large scattering lengths. The most striking features of what was later named the Efimov effect are the appearance of an effective three-body attractive force and with it an \textit{infinite spectrum of three-body bound states} (\textit{Efimov trimers}), connected by a \textit{discrete scale invariance}.

As was the case for Feshbach resonances, the original motivation for Efimov's came from nuclear physics. For a long time after its publication in 1970, tough, it was thought to be almost impossible to experimentally observed \cite{naidon_efimov_2017}, even more so because the appearance of more than one Efimov trimer necessitates the two-body interaction to be extremely close to resonance, a condition that is rarely fulfilled in nature. The observation of a single three-particle bound state by itself might not be uniquely attributed to the Efimov effect because it lacks its signature discrete scaling invariance. The first measurement of an Efimov trimer was achieved in 1994 using $^4$He \cite{schollkopf_nondestructive_1994}.

The great breakthrough in the experimental work on Efimov physics came with the realization of Feshbach resonances and thus tunable interactions in BECs, starting with the 2006 paper in the observation of an Efimov trimer in an ultracold $^{133}$Cs gas by Grimm and N\"agerl \cite{kraemer_evidence_2006}. The first measurements of a second Efimov trimer and thus the most important confirmation of the discrete scaling invariance, were achieved in 2014 by the same group \cite{huang_observation_2014} as well as the Chin group in Chicago and the Weidemüller group in Heidelberg \cite{tung_geometric_2014, pires_observation_2014}.

The following paragraphs give an outline of the derivation of the essential features of Efimov physics in \cite{naidon_efimov_2017} and finally present the results by Braaten and Hammer \cite{braaten_universality_2006} that are going to be used in chapter \ref{sec:implementation} to estimate three-body losses in the experiments considered.

The considerations on two-body scattering in the section \ref{sec:uniform_weakly_interacting} and chapter \ref{sec:feshbach_resonances} were made in the domain of \textit{universality}, where two-body interactions between particles are short-ranged and energies are sufficiently low such that the inner workings of the interaction can be neglected and the scattering behaviour is determined entirely by the s-wave scattering length. For interactions that are also near resonance, one can make use of \textit{zero-range theory}, of which there are several implementations. What they all have in common is that they reproduce the asymptotic shape of the two-body wave function \eqref{eq:feshbach_wave_function_scattering_length}.\footnote{
	The maybe most intuitive example of a zero-range theory is to replace the interaction potential $U ( \mathbf{r} )$ by an effective Dirac delta function reproducing the same scattering length.
}
The one chosen in \cite{naidon_efimov_2017} is to have the system fulfill the free Schr\"odinger equation along with Bethe-Peierls boundary conditions\footnote{
	For a separation $r$ between any pair of particles, the many-particle wave function must satisfy $-(r \psi)^{-1} \partial_r (r \psi) \xrightarrow[r \rightarrow 0]{} 1/a$ \cite{bethe_quantum_1935}.
}
The three-body wave function then solves \cite{naidon_efimov_2017}

\begin{equation} \label{eq:efimov_SEQ_jacobi_coordinates}
	\left( - \nabla^2_{r_ {12}} - \nabla^2_{\rho_{12}} - k^2 \right) \psi ( r_{12}, \rho_{12} ) = 0 \; ,
\end{equation}

where the wave number $k$ encodes the total energy $E= \hbar^2 k^2 /m$ and $ \{ r_{12}, \rho_{12} \}$ are a set of Jacobi coordinates

\begin{IEEEeqnarray}{rCl}
	\IEEEyesnumber \IEEEyessubnumber*
	\mathbf{r}_{ij} &=& \mathbf{r}_i - \mathbf{r}_j
	\\
	\boldsymbol{\rho}_{ij} &=& \frac{2}{\sqrt{3}} \left( \mathbf{r}_k - \frac{\mathbf{r}_i + \mathbf{r}_j}{2} \right)
\end{IEEEeqnarray}

with the individual particle positions $\mathbf{r}_i$ and $\{i, j, k\}$ being permutations of $\{1, 2, 3\}$. When dealing with bosons, $\psi$ must be symmetric under particle exchange and can therefore be decomposed as

\begin{equation}
	\psi = \chi ( \mathbf{r}_{12}, \boldsymbol{\rho}_{12} )
		+ \chi ( \mathbf{r}_{23}, \boldsymbol{\rho}_{23} )
		+ \chi ( \mathbf{r}_{31}, \boldsymbol{\rho}_{31} ) \; .
\end{equation}

Considering only the partial wave with vanishing total angular momentum, $\chi$ becomes ``hyperspherically symmetric'' in $( \mathbf{r}, \boldsymbol{\rho} )$ and can be decomposed into

\begin{equation}
	\chi ( \mathbf{r}, \boldsymbol{\rho} ) = \frac{\chi_0 ( r, \rho )}{r \rho}
\end{equation}

and transformed into \textit{hyperspherical coordinates}

\begin{IEEEeqnarray}{rCl}
	\IEEEyesnumber \label{eq:efimov_hyper-spherical_coords_definition} \IEEEyessubnumber*
	r &=& R \sin \alpha
	\\
	\rho &=& R \cos \alpha  \; .
\end{IEEEeqnarray}

In the limit of large scattering lengths, the problem of the free Schr\"odinger equation with Bethe-Peierl boundary conditions becomes separable in $(R, \alpha)$ \cite{naidon_efimov_2017}:

\begin{IEEEeqnarray}{rCl}
	\IEEEyesnumber \IEEEyessubnumber*
	\chi_0 (R, \alpha ) &=& F (R) \phi (\alpha)
	\\
	\IEEEeqnarraymulticol{3}{l}{
		\mathrm{with}
	} \nonumber
	\\
	\phi_n (\alpha) &=& \sin \left( s_n ( \frac{\pi}{2} - \alpha ) \right)
	\\
	\left(
		- \frac{\partial^2}{\partial_R^2} + V_n (R) - k^2
	\right) \sqrt{R} F_n (R) &=& 0
	\label{eq:efimov_one-dim_SEQ}
	\\
	V_n (R) &=& \frac{s_n^2 - 1/4}{R^2}
	\label{eq:efimov_hyper-radial_potential}
\end{IEEEeqnarray}

\eqref{eq:efimov_one-dim_SEQ} and \eqref{eq:efimov_hyper-radial_potential} thus resemble a one-dimensional Schr\"odinger equation with a hyper-radial potential for each channel $n$ of hyper-radial motion. The boundary conditions are encoded into $s_n$ which solves

\begin{equation} \label{eq:efimov_s_n_equation}
	-s_n \cos \left( s_n \frac{\pi}{2} \right) + \frac{8}{\sqrt{3}} \sin \left( s_n \frac{\pi}{6} \right) = 0 \; .
\end{equation}

For the $n=0$ channel, the solution $s_0 \approx \pm 1.00624 \, i$ of \eqref{eq:efimov_s_n_equation} is purely imaginary. From the definition \eqref{eq:efimov_hyper-spherical_coords_definition}, one can show that

\begin{equation}
	R^2 = \frac{2}{3} \left( r_{12}^2 + r_{23}^2 + r_{31}^2 \right)
\end{equation}

and thus $s_0$ defines an effective long-range ($\propto -R^{-2}$), attractive three-body potential

\begin{equation}
	V_0 (R) = - \frac{|s_0|^2 - 1/4}{R^2} \; .
\end{equation}

This \textit{Efimov attraction} is the first one of Efimov's results that we set out to illustrate at the beginning of this section. The origin of the Efimov trimers and the discrete scale invariance then lies in the solution to an inherent problem in the hyper-radial SEQ \eqref{eq:efimov_one-dim_SEQ}: $V_n (R)$ scales the same way in $R$ as the kinetic energy operator. Hence, the hyper-radius can be arbitrarily rescaled $R \rightarrow \lambda R$, yielding a rescaled energy $\lambda^2 E$. Therefore, the energy spectrum has now lower bound and the Efimov trimer would be subject to a phenomenon called \textit{Thomas collapse} \cite{thomas_interaction_1935}. This is a consequence of the usage of zero-range theory which, by its nature, cannot describe interactions at inter-particle distances of the order of the range of the physical potential.

Efimov addressed this problem by imposing an additional boundary condition in the region of some small hyper-radius $R_0$. In doing so, he accounted for the short-distance behaviour of the two-body potential and introduced a new length scale, the \textit{three-body parameter}.

Independently of what kind of boundary conditions are chosen at $R \approx R_0$, their effect is the same: for a low-energy solution of the SEQ \eqref{eq:efimov_one-dim_SEQ} written as \cite{naidon_efimov_2017}

\begin{equation}
	F_0 ( R \gtrsim R_0) = \alpha R^{i |s_0|} + \beta R^{-i |s_0|} \; ,
\end{equation}

they fix the ratio $\beta / \alpha$. Because $R$ has the dimension of length, we can express this ratio as $\Lambda^{-2i |s_0|}$ with some inverse length $\Lambda$. We then have

\begin{IEEEeqnarray}{rCrCl} \label{eq:efimov_logarithmic_periodicity}
	F_0 ( R \gtrsim R_0) & \propto &
	(R \Lambda )^{i |s_0|} &+& (R \Lambda )^{-i |s_0|}
	\nonumber
	\\*[3pt]
	& & =
	\mathrm{e}^{i |s_0| \ln R \Lambda} &+& \mathrm{e}^{-i |s_0| \ln R \Lambda}
	\nonumber
	\\
	& & \IEEEeqnarraymulticol{3}{l}{
		=
		\cos \left(
			|s_0| ( \ln R + \ln \Lambda )
		\right)
	} \; .
\end{IEEEeqnarray}

The phase arising from $\Lambda$ is only one of multiple equivalent representations of the three-body parameter. The arbitrary scale invariance of \eqref{eq:efimov_one-dim_SEQ} has become a discrete one, resembled in the logarithmic periodicity in \eqref{eq:efimov_logarithmic_periodicity}, which is invariant under transformations $R \rightarrow \lambda_0^n R$ with the \textit{Efimov scaling factor}

\begin{equation}
	\lambda_0 = \mathrm{e}^{\pi/|s_0|} \approx 22.7 \; .
\end{equation}

The energy of the system, on the other hand, is rescaled by $E \rightarrow \lambda_0^{-2n} E$ under the same transformation. Thus, we have found the missing two signatures of the Efimov effect, the discrete scale invariance and the infinite series of Efimov trimers.

As mentioned before, there are various ways to characterize the three-body parameter. In figure \ref{fig:efimov_plot}, the dissociation scattering length $a_-$, the scattering length $a_*$ at the particle-dimer scattering threshold and the $a \rightarrow \pm \infty$ binding wave number $\kappa_*$ are depicted. The one that is most relevant for this work, however, is $a_+$, the position of a local minimum of the coefficient $C_3$ in the three-body loss rate $L_3$ \eqref{eq:three-body_a^4_scaling}. $a_+$ marks the merging point between a trimer state and the particle-dimer scattering continuum. At this point, there are two allowed channels for recombination to dimers exhibiting destructive interference \cite{naidon_efimov_2017, pollack_universality_2009}.

Because they are mutually equivalent (at least in zero-range theory), different instances of the three-body parameter are connected to one another by universal relations, for instance \cite{braaten_universality_2006}

\begin{equation}
	a_+ \approx 0.32 \kappa_*^{-1}
\end{equation}

Experimental findings on the various ratios connecting the three-body parameters \cite{roy_test_2013, dyke_finite-range_2013, gross_observation_2009} have been in fair agreement with these predictions.

Zero-range theory does, however, not make predictions about the absolute values of the three-body parameters. For this, the short-range details of the two-body interaction need to be taken into account. Measurements of Efimov trimers in ultracold atoms hinted at a strong correlation with the van der Waals length across several species (for an overview, see figure 11.2 in \cite{naidon_efimov_2017}). This correlation, in turn, sparked theoretical considerations on this connection, which have found to be in reasonably good agreement with the experimental data. However, in this thesis, we will choose experimental values over theoretical predictions wherever possible.

Another subtlety Efimov's original work could not account for is the possible decay of a trimer into another atom-dimer state at lower energy. Such a state might exist, for example, due to a neighboring Feshbach resonance. This was accounted for by Hammer and Braaten \cite{braaten_universality_2006}, who introduced an \textit{inelasticity parameter} $\eta$ which encodes the finite life time of an Efimov trimer. The introduction of $\eta$ finally leads us to another finding presented in the same review paper, which was the initial motivation for the author to delve into the field of Efimov physics: Hammer and Braaten's analytic expression for the $a_s$-dependence of the coefficient $C_3$ in the three-body loss rate \eqref{eq:three-body_a^4_scaling} at positive scattering lengths:

\begin{equation} \label{eq:efimov_C3-expression}
	C_3 (a_s) \approx 67.12 \mathrm{e}^{-2 \eta_+}
	\left[
		\sin^2 \left( s_0 \ln \frac{a}{a_+} \right) + \sinh^2 \eta_+
	\right]
	+ 16.84 \left( 1 - \mathrm{e}^{-4 \eta_+} \right)
\end{equation}

\begin{figure}[hbtp]
	\begin{centering}
	\includegraphics[width=0.6\textwidth]{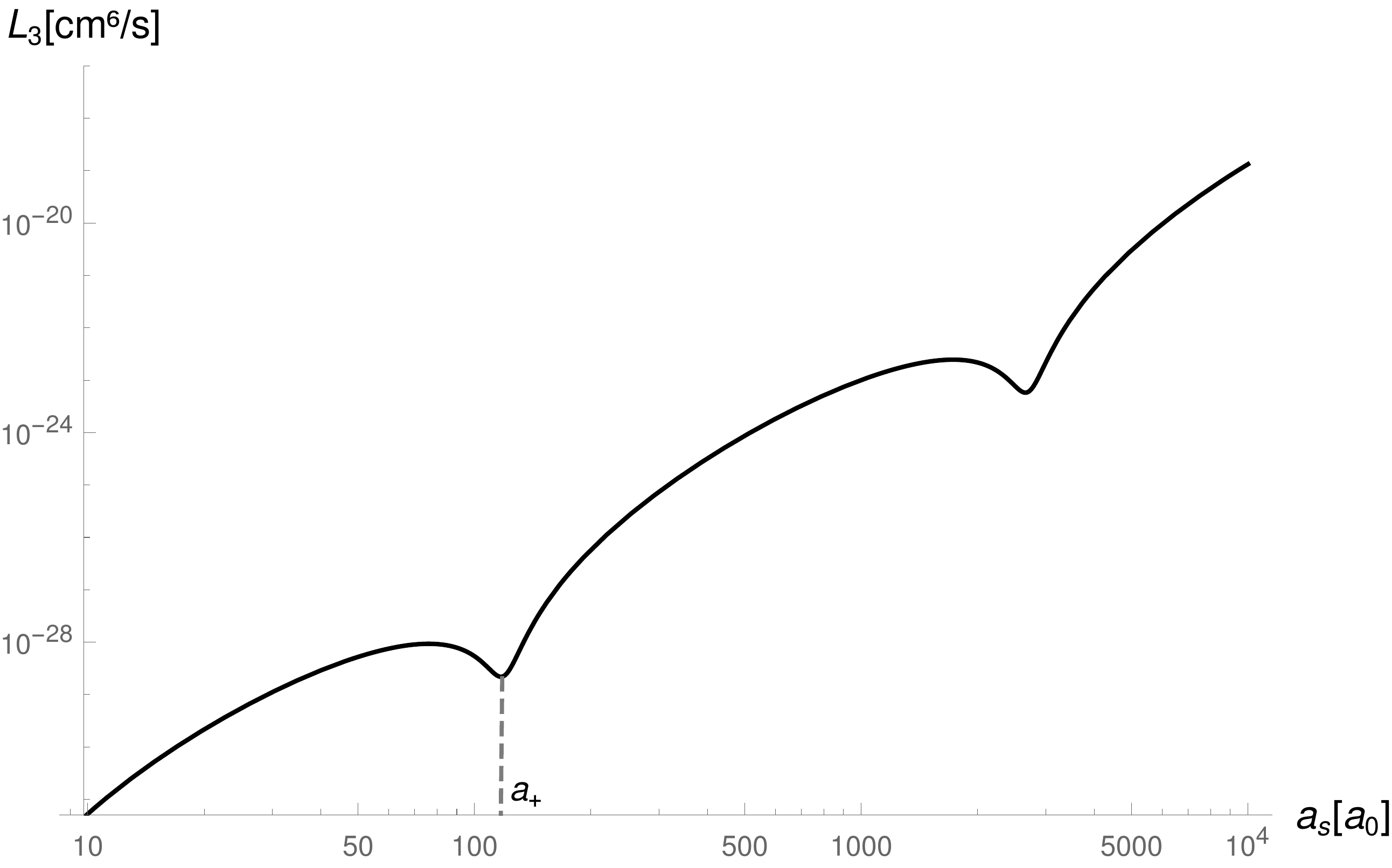}
	\caption{
		{\footnotesize
			Three-body losses as described by equations (\ref{eq:three-body_a^4_scaling}, \ref{eq:efimov_C3-expression}) with the values for $a_+$ and $\eta_+$ as measured in \cite{pollack_universality_2009}. The first dip in the three-body loss rate is found near $a_s = a_+ \approx 119 a_0$.
			\label{fig:efimov_dip_hulet}
		}
	}
	\par\end{centering}
\end{figure}

\begin{figure}[phbt]
	\begin{centering}
	\includegraphics[width=0.8\textwidth]{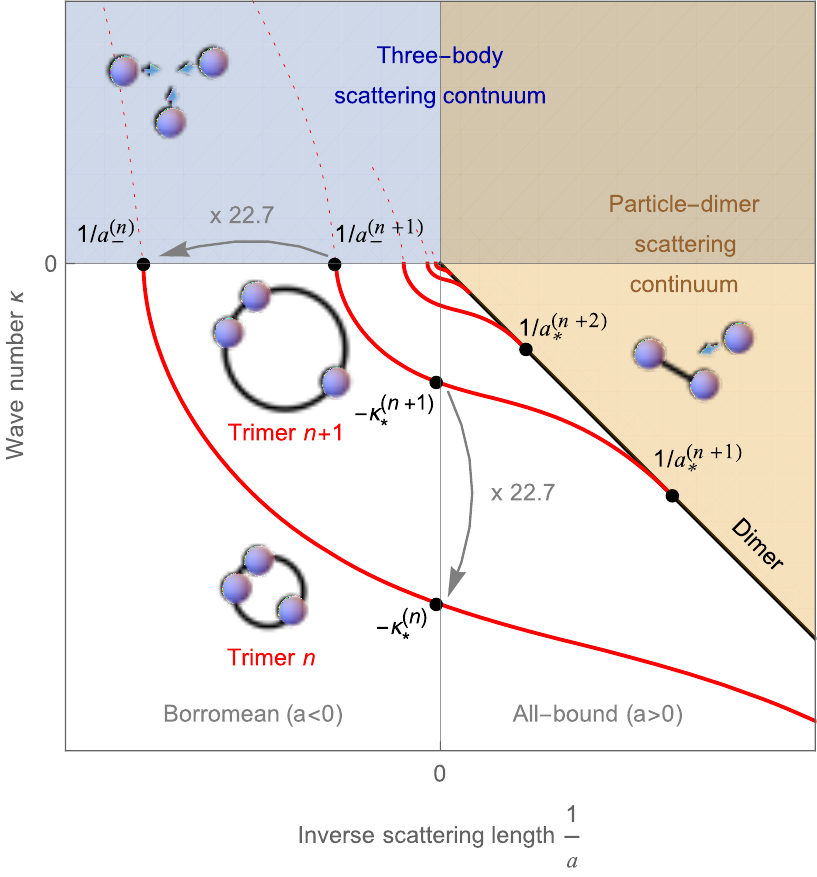}
	\caption{
		{\footnotesize
			From \cite{naidon_efimov_2017}: Efimov plot illustrating the discrete scaling invariance of Efimov trimers: wave numbers $\kappa$ corresponding to the binding energy $E=- \hbar^2 \kappa^2/m$ of the states are shown in dependence of a given inverse scattering length. The region of Efimov trimers is bounded by the three-body scattering continuum where all three particles are free in the positive energy half plane and the particle-dimer scattering continuum, where $E+\hbar^2/(ma_s^2) >0$ and $\hbar^2/(ma_s^2)$ is the binding energy of a dimer. $a_-^{(n)}$, $a_*^{(n)}$ and $\kappa_*^{(n)}$ are all equivalent representations of the three-body parameter. $a_-^{(n)}$ is the scattering length at which the $n$th Efimov trimer reaches the scattering threshold for three free particles and can be observed in three-body loss peaks in ultracold gases with negative scattering lengths \cite{pollack_universality_2009}. $a_*^{(n)}$ marks the particle-dimer scattering threshold and $\kappa_*^{(n)}$ is the binding energy of the trimer right on resonance, i.e. in the limit $a_s \rightarrow \pm \infty$.
			\newline Observe how for every trimer $n$, there exists a trimer $n+1$ of $ \sim 22.7$ times its size, an s-wave scattering length $ \sim 22.7$ times larger and with $\sim 1/515$ times its energy.
			\label{fig:efimov_plot}
		}
	}
	\par\end{centering}
\end{figure}

The first and the second term in \eqref{eq:efimov_C3-expression} correspond to coupling to weakly and to deeply bound dimer states, respectively. This expression will be used in the maximization of the two-mode phonon squeezing in section \ref{sec:implementation} to estimate an upper bound to the condensate life time. An example for the $a_s$-dependence of the three-body losses applying \eqref{eq:efimov_C3-expression} with the parameters measured by the Hulet group's $^7$Li experiment \cite{pollack_universality_2009} can be seen in FIGURE \ref{fig:efimov_dip_hulet}.

\newpage


\chapter{Squeezed States \label{sec:squeezed_states}}

\begin{framed}
	\begin{quote}
		Squeezed states belong to the class of ``non-classical'' states, which are considered to be at the heart of quantum mechanics. These states are defined as those that cannot be described as a mixture of coherent states. \cite{schnabel_squeezed_2017}
	\end{quote}
\end{framed}

\begin{framed}
	\begin{quote}
		Heisenberg (sic) uncertainty principle imposes that $\Delta  x \Delta p \geq 1 $ for all possible quantum states. Single-mode states that saturate this inequality are called the \textit{single-mode squeezed states}. (...) as for general squeezed states $\Delta x \neq \Delta p$, the noise in one of the quadratures can be made smaller than in the other. \cite{demkowicz-dobrzanski_quantum_2015}
	\end{quote}
\end{framed}

\begin{framed}
	\begin{quotation}
		Here we go, folks. A coherent state, very simply, is a state in a harmonic oscillator potential which, at $t=0$, i) is a Gaussian that is displaced from the origin by $x_0$, ii) has a phase which is proportional to the position $x$, and iii) the width of the Gaussian is that of the ground state, $\sigma_0$. (...)
		
		A squeezed state is the same Gaussian but the width is different than that of the ground state,
		
		\begin{equation*}
			\sigma_0 \rightarrow \sigma = s \sigma_0
		\end{equation*}
		
		That's all it is, honest to God. You can go out eat lunch in Santa Fe. \cite{moore_what_1986}
	\end{quotation}
\end{framed}

\begin{framed}
	\begin{quotation}
		As coherent states, the more general squeezed states can be defined in three equivalent ways:
		
		1) \textit{Displacement-Operator Method}. To obtain squeezed states, one applies both the squeeze and displacement operators to the ground state. (...)
		
		2) \textit{Ladder- (Annihilation-) Operator Method}. Using a Holstein-Primakoff/Bogoliubov transformation (...)
		
		3) \textit{Minimum-Uncertainty Method}. (...)
		\cite{nieto_discovery_1997}
	\end{quotation}
\end{framed}

As the reader can see, there are multiple ways that have been used to define \textit{squeezed states}. What most of their descriptions have in common, though, is that they put them in contrast with \textit{coherent states}. The latter were found by Schr\"odinger in 1926 \cite{schrodinger_stetige_1926} in an attempt to find states that reproduce the motion of a classical system and are thus often referred to as \textit{classical states}.

In the next year (and after Heisenberg had presented his uncertainty relation), Kennard studied the time evolution of a Gaussian in a harmonic oscillator potential \cite{kennard_zur_1927}. What he discovered was a set of states with uncertainties in position and momentum $\Delta x$ and $\Delta p$ whose sum remained constant, but whose product varied in time. In addition, he found that coherent states were a subset of these newly found states with $\Delta x = \Delta p$.

Interest in this new set of states, however, would remain small until the 1970s. The term squeezed states originates from a 1979 paper where they were discussed in the context of Weber bar-type gravitational wave detectors \cite{hollenhorst_quantum_1979} (see chapter \ref{sec:detector_proposal}). Possibly due to the multiplicity of their definitions,\footnote{Or possibly the origin of this multiplicity.} they were rediscovered in several instances in the following years, being referred to as \textit{correlated coherent states} \cite{dodonov_generalized_1980}, \textit{pulsating states} \cite{fujiwara_pulsating_1980}, \textit{new coherent states} \cite{rajagopal_new_1982} or \textit{twisted states} \cite{yuen_contractive_1983}, each publication emphasizing different properties.

The most prominent application of squeezed states, finally, is happening these days, as they are being used to push the limits of interferometric gravitational wave detectors (see for example \cite{the_ligo_scientific_collaboration_enhanced_2013}).

Due to their comparably simple generation in photons, most of the experimental and theoretical work on squeezed states has been done in the field of light-based interferometry. Because this is a rather intuitive example for more general parameter estimation problems, we will stick with this approach for the rest of this chapter whenever a specific system is considered. Some remarks on the generalization to \textit{frequency interferometry} will be given in section \ref{sec:pumped-up_SU_11} on the pumped-up $SU(1,1)$ scheme.

To see the significance of non-classical states in general and squeezed states in particular in parameter estimation problems, imagine a Mach-Zehnder interferometer where we aim to estimate the relative phase $\varphi$ between the two arms by measuring the light intensity at one of the outputs. In a classical picture, the precision of $\varphi$ is not bounded by any fundamental limits. This situation changes when we adopt a quantum description, where the intensity at the detector is measured in terms of the number $N$ of detected photons. The error in the parameter estimation is then $\Delta \varphi \propto \Delta \hat{N} / \langle \hat{N} \rangle$ \cite{demkowicz-dobrzanski_quantum_2015}. If the interferometer is run with coherent states, it is subjected to their Poisson statistics with $\Delta \hat{N} = \sqrt{\langle \hat{N} \rangle}$ and the uncertainty in the phase scales as $\Delta \varphi \propto 1 / \sqrt{\langle \hat{N} \rangle}$.

While this \textit{shot noise limit} is a hard bound for the precision achievable using classical states, it is not a fundamental restriction. This was demonstrated in an early proposal for the combined usage of squeezed and coherent states in a Mach-Zehnder interferometer \cite{caves_quantum-mechanical_1981}, where a $1/ \langle \hat{N} \rangle ^{2/3}$ scaling was found, and pushed further by the consideration of \textit{two-mode squeezed states} to reach \cite{bondurant_squeezed_1984}

\begin{equation} \label{eq:heiseinberg_limit}
	\Delta \varphi \propto 1/ \langle \hat{N} \rangle \; .
\end{equation}

What these early proposals lacked was generality in that they were considering specific measurement strategies. This generalization came with the arrival of quantum metrology and the quantum Fisher information, which allowed to put a lower bound on the error of an estimated parameter for an optimal measurement strategy (see appendix \ref{sec:RQM_new}). With these tools, it was shown that for arbitrary unitary parameter estimations,\footnote{
	That is for unitary single-particle transformations that encode a parameter $\varphi$ as $U_\varphi = \mathrm{exp} (-i \hat{H} \varphi)$ and $N$-particle transformations $U_\varphi^{\otimes N}$ \cite{demkowicz-dobrzanski_quantum_2015}
}
\eqref{eq:heiseinberg_limit} is indeed a fundamental bound \cite{giovannetti_quantum-enhanced_2004, giovannetti_quantum_2006} and it was called the \textit{Heisenberg limit}.

In the following discussion of coherent and squeezed states, we will stick to the first definition mentioned in \cite{nieto_discovery_1997} and describe them using the operators they are generated by.

\paragraph{Coherent states,} as mentioned before, are the Gaussian states that behave most similar to classical light. A single-mode coherent state can be created from the vacuum state by the displacement operator

\begin{equation} \label{eq:displacement_operator}
	\hat{D} ( \alpha ) = \mathrm{e}^{\alpha \hat{a} - \alpha^* \hat{a}}
\end{equation}

such that

\begin{equation}
	| \alpha \rangle \equiv \hat{D} ( \alpha ) |0 \rangle =
	\mathrm{e}^{
		-| \alpha|^2/2
	}
	\sum_{n=0}^\infty
		\frac{\alpha^n}{\sqrt{n!}} | n \rangle \; .
\end{equation}

Let us define the quadratures as linear combinations of a mode's creation and annihilation operator at a relative angle\footnote{
	This is a generalization of the quadratures in the real picture of the covariance matrix formalism displayed in appendix \ref{sec:CM_real_complex} and will come in handy in the definition of squeezed states.
}
$\gamma$ such that $\hat{x}_\gamma \equiv \frac{1}{\sqrt{2}} ( \hat{a} \mathrm{e}^{-i \gamma} + \hat{a}^\dagger \mathrm{e}^{i \gamma})$ and $\hat{p}_\gamma \equiv \hat{x}_{\gamma + \pi/2} = \frac{1}{\sqrt{2}i} ( \hat{a} \mathrm{e}^{-i \gamma} - \hat{a}^\dagger \mathrm{e}^{i \gamma})$. It is then easy to show that coherent states saturate the uncertainty relation and that their uncertainties are balanced.


\begin{IEEEeqnarray}{rCl}
	\IEEEyesnumber \IEEEyessubnumber*
	\Delta \hat{x}_\gamma \Delta \hat{p}_\gamma &=& \frac{1}{2}
	\\
	\Delta \hat{x}_\gamma = \Delta \hat{p}_\gamma &=& \frac{1}{\sqrt{2}} \qquad \forall \, \gamma
\end{IEEEeqnarray}

\paragraph{Single-mode squeezed states} are Gaussian states that saturate Heisenberg's uncertainty relation. They can be created from the vacuum by applying a combination of displacement and the squeezing operator \cite{demkowicz-dobrzanski_quantum_2015}

\begin{equation}
	| \alpha, \zeta \rangle = \hat{D} (\alpha) \hat{S} (\zeta) |0\rangle
\end{equation}

with

\begin{equation} \label{eq:single-mode_squeezing_operator}
	\hat{S} (\zeta) \equiv \mathrm{e}^{\frac{1}{2} ( \zeta^* \hat{a}^2 - \zeta \hat{a}^{\dagger 2})} \; .
\end{equation}

The complex squeezing parameter can be decomposed as $\zeta = r \mathrm{e}^{i \theta}$ with $r = |\zeta|$. $r$ and $\theta$ are often referred to as the \textit{squeeze factor} and the \textit{squeeze angle}. One can show \cite{drummond_quantum_2004} that $\theta$ is the angle at which the uncertainty in the quadrature $\hat{x}_\theta$ is minimized to $\Delta \hat{x}_\theta = \mathrm{e}^{-r}/ \sqrt{2}$ (and, due to the uncertainty relation, $\Delta \hat{p}_\theta = \mathrm{e}^{r}/ \sqrt{2}$). This is the essential feature of squeezed states: the uncertainty in one of their quadratures can be pushed arbitrarily low (while the other one grows exponentially) and as a consequence, the fluctuations in the number operator can also be pushed below the shot-noise limit \cite{demkowicz-dobrzanski_quantum_2015}. In the literature, $r$ is often given in terms of the ratio between the uncertainty of the squeezed and the vacuum state in $dB$ and thus defined as \cite{schnabel_squeezed_2017}

\begin{equation}
	r [dB] \equiv - 10 \log_{10} \left( \frac{\Delta^2 \hat{x}_{\theta}}{ \Delta^2 \hat{x}_{\theta, vac}} \right) \approx 8.69  \, r \; .
\end{equation}

Note that, because we are free to set $\alpha$ and $\zeta$ to zero, both the vacuum state $|0\rangle$ and the coherent state $| \alpha \rangle$ are special instances of squeezed states, as is the \textit{squeezed vacuum state}

\begin{equation}
	| \zeta \rangle = \hat{S} (\zeta) | 0 \rangle =
	\frac{1}{\sqrt{\cosh r}} \sum_{n=0}^{\infty}
		\frac{\big[ (2n) ! \big]^{1/2}}{2^n n!} \left( \mathrm{e}^{i \theta} \tanh r \right)^n |2n \rangle \; .
\end{equation}

The $|2n \rangle$ states in the sum reveal a distinctive feature of the single-mode squeezed vacuum: it only contains even particle number states. Demkowicz-Dobrzański et al. \cite{demkowicz-dobrzanski_quantum_2015} argue that only entangled states can beat the shot-noise limit, but on the other hand products of coherent and squeezed states $|\alpha \rangle \otimes |\zeta \rangle$ have been proposed to achieve exactly this. They resolve this issue by clarifying the notions of \textit{mode} and \textit{particle entanglement}.\footnote{
	Elsewhere in this thesis, the term entanglement will be used synonymously with mode entanglement if not stated otherwise.
}
While a product state is clearly separable in its modes, it can be particle entangled, which they illustrate by considering the two-particle sector of the two-mode Fock space

\begin{IEEEeqnarray}{rCl}
	\left( | \alpha \rangle_a | \zeta \rangle_b \right)^{N=2}
	&\propto &
	\alpha^2 |2 \rangle_a |0 \rangle_b + \tanh r |0 \rangle_a |2 \rangle_b
	\nonumber
	\\
	& &
	= \alpha^2 |a \rangle_1 |a \rangle_2 + \tanh r |b \rangle_1 |b \rangle_2 \; ,
\end{IEEEeqnarray}

where the last line indicates \textit{particle} $1$ or $2$ being populated by \textit{mode} $a$ or $b$. This particle entanglement is enabled by the $|2n\rangle$ feature of single-mode squeezed states.

Although the term ``squeezed vacuum'' might suggest otherwise, the average particle number of $| \zeta \rangle$ is $\langle \hat{N} \rangle_\zeta =\sinh^2 r$.

\paragraph{Two-mode squeezed vacuum states,} which will be the ones relevant in part \ref{part:2} of this thesis, are states that can be created from the vacuum by the two-mode squeezing operator

\begin{IEEEeqnarray}{rCl}
	\IEEEyesnumber \label{eq:two-mode_squeezed_vacuum} \IEEEyessubnumber*
	| \zeta \rangle_2 &\equiv & \hat{S}_2 ( \zeta ) |0, 0 \rangle
	\\
	&=&
	\frac{1}{\cosh r} \sum_{n=0}^{\infty}
		(-1)^n \mathrm{e}^{i n \theta} \tanh^n r \, | n, n \rangle
\end{IEEEeqnarray}

with

\begin{equation} \label{eq:squeezed_states_two-mode_squeezing_operator}
	\hat{S}_2 ( \zeta ) = \mathrm{e}^{\zeta^* \hat{a} \hat{b} - \zeta \hat{a}^\dagger \hat{b}^\dagger} \; .
\end{equation}

There are several ways to see that the state \eqref{eq:two-mode_squeezed_vacuum} is, in the terminology of Demkowicz et al., mode entangled. One way is to take the partial trace over one of its subsystems and recognize it to be a thermal and thus mixed state. This also means that the two-mode squeezed state is not a product of single-mode squeezed states.

As was the case for the single-mode squeezed vacuum, the particle number $ \langle \hat{N}^A \rangle_{\zeta_2} = \langle \hat{N}^B \rangle_{\zeta_2} = \sinh^2 r$ is nonzero and is now evenly distributed between the modes.

The squeezedness of the two-mode squeezed vacuum becomes apparent not in the observables of a single subspace, but in the variance of \textit{joint} observables.\footnote{
	For a more detailed example and derivation, see appendix \ref{sec:appendix_two-mode_squeezed_variance}.
} Consider, for example, the sum of quadratures $( \hat{x}_\gamma^A + \hat{x}_\gamma^B )$, where the upper index denotes the subspace of the operator. The variance of this joint quadrature is $\cosh (2r) - \cos (\theta - 2 \gamma) \sinh (2r)$. It is thus minimized at the squeezing angle $\theta = 2 \gamma$ and can be pushed arbitrarily low in the limit of infinite squeezing:

\begin{equation}
\Delta^2 ( \hat{x}_{\theta/2}^A + \hat{x}_{\theta/2}^B ) = \mathrm{e}^{- 2r}
\end{equation}

and, correspondingly, 

\begin{equation}
\Delta^2 ( \hat{p}_{\theta/2}^A + \hat{p}_{\theta/2}^B ) = \mathrm{e}^{ 2r} \; .
\end{equation}

As can be intuitively seen in \eqref{eq:two-mode_squeezed_vacuum}, the particle numbers in the two modes are perfectly correlated. Therefore, the variance of the \textit{difference} in particle numbers of the squeezed vacuum $\Delta(\hat{N}^A - \hat{N}^B)$ must vanish.

In the limit of infinitely large squeeze factors $r \rightarrow \infty$, $| \zeta \rangle_2$ approaches the infinitely entangled Einstein-Podolsky-Rosen (EPR) state \cite{adesso_entanglement_2007}. Such a state, however, is nonnormalizable and therefore unphysical. Two-mode squeezed states can be used as an arbitrarily good approximation of EPR states in experiments. The first EPR-type experiment making use of this approximation was performed by Ou et al. in 1992 \cite{ou_realization_1992}, where the photon states were produced by a non-linear process called type-II parametric down conversion (PDC) \cite{bachor_guide_2004}.

As for Bose-Einstein condensates, there have been multiple considerations on squeezing in various degrees of freedom, for instance number squeezing in systems with two internal states (think of double well traps) or spin squeezing (see eg. \cite{johnsson_squeezing_2013, sorensen_many-particle_2001}). When it comes to two-mode phonon squeezing, however, the field is still lacking the standard technique that parametric down conversion has become for quantum optics.


\chapter{Gravitational Waves and the TT Gauge \label{sec:gravitational_waves}}

Since the central motivation for this thesis was the proposal for a gravitational wave detector, which is outlined in chapter \ref{sec:detector_proposal}, a basic discussion of gravitational waves might be appropriate at this point. The following chapter is largely based on Rindler's \cite{rindler_relativity_2006} and Misner, Thorne and Wheeler's \cite{misner_gravitation_2017} textbooks.

When describing gravitational waves as localized vacuum solutions of the linearized Einstein field equations, the transversal traceless gauge presents itself particularly handy in eliminating nonphysical degrees of freedom with only two polarizations of the wave remaining.

Consider the Einstein field equations

\begin{equation} \label{eq:Einstein_field_eqns}
	G_{ik} := R_{ik} - \frac{1}{2} g_{ik} R = - \kappa T_{ik} \; ,
\end{equation}

where $G_{ik}$ is the called the Einstein Tensor, $R_{ik}$ is the Ricci curvature tensor, $g_{ik}$ is the metric and $R \equiv {R^i}_i$ is the Ricci or curvature scalar. In order to linearize \eqref{eq:Einstein_field_eqns}, take the metric to be Minkowskian with a small perturbation

\begin{equation}
	g_{ik} = \eta_{ik} + 2 h_{ik} \; ,
\end{equation}

such that second and higher order terms in $h_{ik}$ and all its derivatives can be neglected. Defining the trace reverse $\psi_{ik} := h_{ik} - 1/2 \, \eta_{ik} {h^j}_j$ and applying the harmonic gauge\footnote{This is an analogy to the Lorenz gauge in electromagnetism and can always be realized locally.}

\begin{equation} \label{eq:harmonic_gauge}
	{\psi_{ik}}^{, k} = {h_{ik}}^{,k}  - \frac{1}{2} {h^k}_{k, i} = 0
\end{equation}

simplifies the Ricci tensor to $R_{ik} = \Box h_{ik}$ and the field equations \eqref{eq:Einstein_field_eqns} to

\begin{equation}
	\Box h_{ik} = - \kappa \left( T_{ik} - \frac{1}{2} \eta_{ik} {T^j}_j \right)
\end{equation}

or, for the trace inverse $\psi_{ik}$

\begin{equation}
	\Box \psi_{ik} = - \kappa T_{ik} \; .
\end{equation}

For a plane wave solution of the vacuum field equations propagating in the $z$-direction $\psi_{ik} = \psi_{ik}(t-z)$ s.t. $\Box \psi_{ik}=0$, \eqref{eq:harmonic_gauge} results in four gauge conditions

\begin{IEEEeqnarray}{rClL}
	\IEEEyesnumber \IEEEyessubnumber*
	\dot{\psi}_{i0} &=& -\dot{\psi}_{i3} & \mathrm{or}
	\\
	\psi_{i0} &=& - \psi_{i3} + \mathrm{const.} &
\end{IEEEeqnarray}

The integrational constants do not have any effect on $R_{ik}$ and can be discarded immediately, leaving six degrees of freedom. However, for the solutions considered here, four of these degrees of freedom are due to disturbances to the coordinate system that result in zero curvature as well and are thus unphysical. Exploiting the remaining gauge freedom, functions $\Lambda_i (t-z)$ can be found such that after the transformation

\begin{equation} \label{eq:TT_gauge_trafo}
	\psi_{ik} \rightarrow \psi_{ik} + \Lambda_{i, k} + \Lambda_{k, i} - \eta_{ik} {\Lambda^j}_j \quad \mathrm{with} \quad \Box \Lambda_i = 0 \; ,
\end{equation}

we have 

\begin{equation}
	\psi_{10} = \psi_{20} = \psi_{30} = {\psi_i}^i \quad \mathrm{and} \quad \psi_{ik}=h_{ik}
\end{equation}

and the trace of the perturbation vanishes and only two independent components, the transversal components, remain. This is called the \textit{transverse traceless} (TT) gauge. In matrix notation, $h$ can be written as

\begin{equation}
	\left( h_{ij} \right) = 
	 \begin{pmatrix}
		0 & 0 & 0 & 0
		\\
		0 & h_+ & h_\times & 0
		\\
		0 & h_\times & -h_+ & 0
		\\
		0 & 0 & 0 & 0
	\end{pmatrix} \; .
\end{equation}

Note that the transformation \eqref{eq:TT_gauge_trafo} leaves $h_+$ and $h_\times$ invariant. As a consequence, plane wave solutions in the harmonic gauge must already have the same entries up to some constants and there is no need to find a suitable transformation $\Lambda_i$. Instead, all components except $h_{22}, \, h_{23}, \, h_{32}, \, h_{33}$ can simply be dropped and a constant subtracted from $h_{22}$ and $h_{33}$ such that the trace vanishes.

\part{Measuring Gravitational Waves and Creating Squeezed Phonons \label{part:2}}

\chapter{Gravitational Wave Detector Proposal \label{sec:detector_proposal}}

Only a short time after the publication of his theory of general relativity, Einstein predicted the existence of gravitational waves (GWs) \cite{einstein_naherungsweise_1916, einstein_uber_1918}. He did, however, find that it would be virtually impossible to measure gravitational waves that were produced artificially in a laboratory \cite{aufmuth_gravitational_2005}. Even for astronomic sources, the expected amplitudes were extremely small and the quest to detect their signals thus became one of the great challenges in physics in the following century, culminating in the observation of a merger event of two $\sim 30$ solar mass black holes on the 14 September 2015 at the Laser Interferometer Gravitational-Wave Observatory (LIGO), which was published in February 2016 \cite{the_ligo_scientific_collaboration_observation_2016, castelvecchi_einsteins_2016} and followed by a Nobel Prize awarded to Weiss, Thorne and Barish ``for decisive contributions to the LIGO detector and the observation of gravitational waves'' in 2017.

Almost sixty years earlier, Weber proposed the use of what later became known as \textit{Weber bars} \cite{weber_detection_1960}. These are resonant mass antennas, solid bodies that would theoretically pick up space time distortions at their resonant frequency and thus amplify them to a detectable level. However, scientific consensus remained that Weber's measurements \cite{weber_evidence_1969, weber_anisotropy_1970} were unsuccessful \cite{aufmuth_gravitational_2005}. The concept itself, though, is still subject of ongoing research, an example for which is MiniGRAIL.

More recently, proposals of GW detectors have been based on atom interferometry \cite{canuel_matter-wave_2014, chaibi_low_2016, dimopoulos_general_2008, graham_new_2013, wolf_quantum_2009}. Here, a BEC is brought into a macroscopic spatial superposition and recombined at a later time. If a GW passes during this process, then the relative phase between the two paths carries information about it.

Nonetheless, to this day, laser interferometers remain the only type of GW detector that have accomplished their goal.

The major motivation for this thesis and subject of the following section are the findings of Sabín et al. \cite{sabin_phonon_2014}, where it was shown that the amplitude of a passing gravitational wave can get resonantly encoded into the phonon state populating a BEC with Heisenberg scaling\footnote{
	See the introduction to chapter \ref{sec:squeezed_states} and particularly \eqref{eq:heiseinberg_limit}.}
and could, in principle, be observable. Their derivation employs techniques from relativistic quantum metrology \cite{ahmadi_quantum_2014} and analogue gravity \cite{barcelo_analogue_2011}. This scheme would allow to target frequency ranges of $10^3$Hz to $10^7$Hz --which is higher than the region of peak sensitivity of light-based interferometric detectors such as LIGO or VIRGO-- and look for persistent GW signals. Proposed sources include pulsars, spinning neutron stars, supernovae or small neutron star mergers. 

Ongoing work \cite{howl_active_2019, howl_quantum_2020-1, howl_quantum_2020} has found the possibility to improve the sensitivity of the original scheme \cite{sabin_phonon_2014} by several orders of magnitude and introduced two central notions: the \textit{quantum Weber bar} and \textit{frequency metrology}. The origin of the former lies in the fact that, in analogy to the classical Weber bar, phonon creation happens on resonance with the GW signal. In the present case however, they are created parametrically. Since the phonon states will be squeezed, the detector has earned the \textit{quantum} in its name. The term \textit{frequency interferometry} expresses that the modes that make up the ``arms'' of the interferometer are frequency modes that are all delocalized within the same trap. Hence, they can interact at any time, allowing for estimations not only of phase imprints, but also parameters that manifest themselves in entangling unitaries \cite{howl_quantum_2020-1}. This is in sharp contrast with \textit{spatial interferometry}, both light based (such as LIGO) or with matter waves, where the arms are actual arms\footnote{Well, not actual arms.} that are spatially separated and estimations can only be made from the imprints left by \textit{local} unitaries.


Quantum metrology exploits phenomena whose nature is completely quantum, such as entanglement, to improve the experimental estimation of parameters beyond limits that would be imposed by classical physics. When dealing with curved space-time, however, it has to be generalized to account for relativistic effects. This is done in relativistic quantum metrology \cite{ahmadi_quantum_2014} by means of quantum field theory in curved space-time \cite{birrell_quantum_1984}.

In the field of analogue gravity, experimentally more easily accessible systems and particularly Bose-Einstein condensates are employed to investigate general relativistic (GR) phenomena. One of the findings in this context that is most relevant to this thesis is the fact that the equations of motion for phonons on a BEC can be expressed as a Klein-Gordon equation with an \textit{acoustic metric} that is composed of the (potentially curved) space-time metric and (experimentally controllable) parameters of the BEC \cite{fagnocchi_relativistic_2010}. While in the analogue gravity community, this property is applied to simulate GR settings, the opposite approach was taken in \cite{sabin_phonon_2014} and the publications that followed, where the BEC parameters in the acoustic metric are kept uniform in order to study the effects of a gravitational wave (GW) on the phonon modes and propose a GW detector. 

Since the publication of \cite{sabin_phonon_2014}, there has been a debate on whether the predicted effect can be measured with present day or near-future technology \cite{schutzhold_interaction_2018, howl_comment_2018}. For the quantum Weber bar, it was found that the quantum Fisher information of the state transformation induced by the GW depends strongly on the squeezing of the initial state. It is therefore of high interest to find estimates on what squeeze factors one can hope to achieve in existing experiments and how setups that are best suited for two-mode phonon squeezing can be found. This is the contribution this thesis intends to make.

\section{Phonons on a Curved Background \label{sec:analogue_gravity}}

An extensive overview of the history of analogue gravity is given in a review paper by Barceló et al. \cite{barcelo_analogue_2011}. Here, we restrict ourselves to the ideas that are most relevant in the given context, the earliest of which were on the correspondence between electrodynamics and general relativity. A 1923 paper by Gordon \cite{gordon_zur_1923}, and a problem in Landau and Lifshitz' book ``The Classical Theory of Fields'' \cite{landau_classical_1994} followed similar approaches, but in opposite directions. While Gordon's intention was to have a gravitational system simulate a dielectric medium, the relation was reversed in Landau and Lifshitz' problem. Gordon found what he called an \textit{effective metric}\footnote{In the subsequent years, the term Gordon metric was adopted for it.} to describe the medium, consisting of two terms: the background space-time metric and the 4-velocity of the medium. However, the latter still had to be assumed to be constant. This restriction was overcome in 1975 by Anderson and Spiegel \cite{anderson_radiative_1975}.

The first steps into the realm of \textit{acoustic} analogue gravity were kicked off by White's 1973 hydrodynamical paper \cite{white_acoustic_1973}, followed\footnote{among others} by Moncrief in 1980 \cite{moncrief_stability_1980}, who showed that the linear perturbations on a perfect fluid living on the background of a Schwarzschild metric obey a relativistic wave equation on an effective metric, thus already allowing for a curved background. Montcrief's thoughts were therefore much closer to the spirit of the proposed GW detector \cite{sabin_phonon_2014} than the more recent work on in the field, where the the focus lies on the simulation of GR while controlling the parameters of the system in question.\footnote{To point out just two among the numerous impactful contributions to this train of thought, the reader is referred to the 1981 paper ``Experimental Black Hole Evaporation'' by Unruh \cite{unruh_experimental_1981} and the observation of an acoustic analogue of Hawking radiation by Steinhauer in 2014 \cite{steinhauer_observation_2014}.}

A derivation of the acoustic Klein-Gordon equation for the phonons on a BEC is given in \cite{fagnocchi_relativistic_2010} and sketched here. The Lagrangian density for a Bose gas on a background metric $g_{ij}$ with determinant $g$ is \cite{edward_bruschi_testing_2014}

\begin{equation} \label{eq:analogue_gravity_lagrangian}
	\mathcal{L} (x^i) = \sqrt{-g} g^{ab} \partial_a \hat{\psi}^\dagger \partial_b \hat{\psi}
	- \left(
		\frac{m^2 c^2}{\hbar^2} + \mathcal{V} \left( x^i \right)
	\right) \hat{\psi}^\dagger \hat{\psi}
	- U \left( \hat{\psi}^\dagger \hat{\psi}; \{ \lambda_i \} \right) \; .
\end{equation}

In our familiar notation, $\mathcal{V} (x^i)$ is an external potential and interactions are mediated by $U \left( \hat{\psi}^\dagger \hat{\psi}; \{ \lambda_i \} \right)$ which, in turn, depends on a set of coupling constants $\{ \lambda_i \}$\footnote{In \eqref{eq:analogue_gravity_metric_derivation_step}, $U$ will be absorbed into the newly defined quantity $c_0$ with no need arising to specify the interaction. However, the reader can imagine the interaction to be of the previously used nature $U \left( \hat{\psi}^\dagger \hat{\psi}; \{ \lambda_i \} \right) = U_0 \left( \hat{\psi}^\dagger \hat{\psi} \right)^2$.}
and $c$ is the speed of light.

To derive an equation of motion for the ground state in the ultracold limit, one can take the same step as in the derivation of the Gross-Pitaevskii equation in section \ref{sec:GPE} and, under the same restrictions, replace the field operator in \eqref{eq:analogue_gravity_lagrangian} by a classical field $\psi$. $\mathcal{L}$ then becomes

\begin{equation} \label{eq:analogue_gravity_relativistic_GPE}
	\mathcal{L} (x^i) =
	\sqrt{-g} g^{ab} \partial_a \psi^* \partial_b \psi
	- \left(
		\frac{m^2 c^2}{\hbar^2} + \mathcal{V} \left( x^i \right)
	\right) | \psi |^2
	- U \left( | \psi |^2; \{ \lambda_i \} \right)
\end{equation}

and thus the Euler-Lagrange equations are

\begin{equation}
	\Box_g \psi - \left( \frac{m c^2}{\hbar} + \mathcal{V} (x^i) \right) \psi - U' \left( n_0; \{ \lambda_i \} \right) = 0 \; ,
\end{equation}

where the prime denotes a partial derivative $U' \left( n_0; \{ \lambda_i \} \right) = \partial U \left( n_0; \{ \lambda_i \} \right) / \partial n_0$ and\footnote{To avoid possible confusion: in section \ref{sec:ideal_bose_gas}, $n_0$ was defined as the condensate particle density and in \ref{sec:uniform_bose_gas_phonon-phonon_interactions}, the overall particle density was denoted by $n$. If nonclassical perturbations to the field are neglected altogether, these two quantities coincide.}

\begin{equation}
	\Box_g \equiv \sqrt{-g}^{-1} \partial_a \left( \sqrt{-g} \partial^a \right)
\end{equation}

is the generalized d'Alembert operator corresponding to a metric $g_{ij}$. In the non-relativistic limit, \eqref{eq:analogue_gravity_relativistic_GPE} reduces to the Gross-Pitaevskii equation \eqref{eq:gross-pita_time-dependent} \cite{fagnocchi_relativistic_2010}.

For the quantum fluctuations over the condensate, recall the Bogoliubov approximation \eqref{eq:bogo_approx_field_operator}, although with a slight change in notation that will turn out more convenient in the subsequent calculations.

\begin{equation} \label{eq:analogue_gravity_bogo_approximation}
	\hat{\psi} \left( x^i \right)
	=
	\psi \left( x^i \right) + \delta \hat{\psi} \left( x^i \right)
	\equiv
	\psi \left( x^i \right) \left( 1 + \delta \hat{\Pi} \left( x^i \right) \right)
\end{equation}

With the same Lagrangian \eqref{eq:analogue_gravity_lagrangian} as before, the equation of motion for $\delta \hat{\Pi}$ can be found as \cite{fagnocchi_relativistic_2010}

\begin{equation}
	\Box_g \delta \hat{\Pi}
	+
	2 g^{ab} \left( \partial_a \ln \psi \right) \partial_b  \delta \hat{\Pi}
	-
	n_0 U'' \left( n_0, \{ \lambda_i \} \right) \left( \delta \hat{\Pi} + \delta \hat{\Pi}^\dagger \right) = 0 \; .
\end{equation}

With the help of a few redefinitions, this is cast into the shape

\begin{equation} \label{eq:analogue_gravity_metric_derivation_step}
	\left(
		\left(
			i \hbar u^a \partial_a + T_{n_0}
		\right)
		\frac{1}{c_0^2}
		\left(
			- i \hbar u^b \partial_b  + T_{n_0}
		\right)
		 - \frac{\hbar^2}{n_0} g^{cd} \partial_c n_0 \partial_d
	\right)
	\delta \hat{\Pi} = 0 \; .
\end{equation}

To elaborate on the new quantities occuring in \eqref{eq:analogue_gravity_metric_derivation_step}, let us start by factorizing $\psi$ into the (real) particle density and local phase

\begin{equation}
	\psi (x^i) = \sqrt{n_0 (x^i)} + \mathrm{e}^{i \vartheta ( x^i )} \; .
\end{equation}

$u^i$ is then defined as

\begin{equation}
	u^i (x^j) \equiv \frac{\hbar}{m} g^{ik} \partial_k \vartheta (x^j) = \frac{\hbar}{m n_0} j^i (x^j)
\end{equation}

and is proportional to the conserved\footnote{The continuity equation for $j^i$ can be shown by applying Noether's theorem to the global $U(1)$ symmetry of \eqref{eq:analogue_gravity_lagrangian}.} condensate particle density current $j^i$.

$c_0$ connects mass, condensate particle density and interactions and is given by

\begin{equation}
	c_0^2 \equiv \frac{\hbar^2}{2m^2} n_0 U'' (n_0, \{\lambda_i \}) \; .
\end{equation}

Its dimension is that of velocity.

Lastly, the generalization of the kinetic energy

\begin{equation}
	T_{n_0} \equiv -\frac{\hbar^2}{2m}
	\left(
		\Box_g + g^{ab} \partial_a \ln n_0 \partial_b \right)
\end{equation}

assumes the form of the nonrelativistic kinetic energy in the limit $c^2 \rightarrow \infty$.

If the perturbations $\delta \hat{\Pi}$ are then expanded into phonon modes of momentum $\mathbf{k}$ and energy $\hbar \omega$, one can take the low energy limit\footnote{Where also the phonon limit of the Bogoliubov dispersion relation \eqref{eq:bogo_dispersion_law} is recovered.} and what is dubbed the \textit{eikonal approximation} in \cite{fagnocchi_relativistic_2010} and is analogous to the Thomas-Fermi approximation in section \ref{sec:thomas-fermi} in the sense that it requires that all quantities determined by the condensate wave function vary slowly --here in four-dimensional space-time-- and results in the --now generalized-- kinetic energy term in \eqref{eq:analogue_gravity_metric_derivation_step} being discarded. The equation of motion for the quantum perturbations then becomes \cite{fagnocchi_relativistic_2010}

\begin{equation}
	\partial_a
	\underbrace{	
		\left(
			\frac{n_0}{c_0^2} u^a u^b - n_0 g^{ab}
		\right)
	}_{
		\equiv \sqrt{- \mathfrak{g}} \mathfrak{g}^{ab}
	}
	\partial_b \delta \hat{\Pi} = 0 \; .
\end{equation}

As already pointed out by the underbrace above, this has the shape of a massless Klein-Gordon equation with the acoustic metric

\begin{equation}
	\mathfrak{g_{ab}}
	\equiv
	\frac{n_0}{\sqrt{1 - u_c u^c / c_0^2}}
	\left(
		g_{ab}
		\left(
			1 - \frac{u_d u^d}{c_0^2}
		\right)
		+ \frac{u_a u_b}{c_0^2}
	\right) \; .
\end{equation}

To compare their findings with the results on relativistic barotropic
irrotational fluid flow in curved space-time \cite{visser_acoustic_2010}, Fagnocchi et al. then introduce the norm of $u^i$, the speed of sound $c_s$ and the four-velocity $v^i$

\begin{IEEEeqnarray}{rCl}
	||u|| & \equiv & \sqrt{-u_a u^a}
	\IEEEyesnumber
	\IEEEyessubnumber
	\\*[5pt]
	c_s^2
	& \equiv &
	\frac{c^2 c_0^2/||u||^2}{1 + c_0^2 / ||u||^2}
	\IEEEyessubnumber
	\\
	v^i & \equiv & \frac{c}{||u||} u^i
	\IEEEyessubnumber
\end{IEEEeqnarray}

to find \cite{fagnocchi_relativistic_2010, visser_acoustic_2010, edward_bruschi_testing_2014, sabin_phonon_2014}

%
%
%
%

\begin{equation} \label{eq:analogue_gravity_acoustic_metric_carlos}
	\mathfrak{g}_{ab} =
	\frac{1}{c_s}	
	\frac{n_0^2}{p_0 + \rho_0}
	\left(
		g_{ab} + \left( 1 - \frac{c_s^2}{c^2} \right) v_a v_b
	\right)
\end{equation}


\section{GW Detector Scheme \label{sec:GW_detector_setup}}

The situation considered in \cite{sabin_phonon_2014} is described by the combination of the results of the previous section on the acoustic metric for a Bose gas in the ultracold limit and the GW metric derived in section \ref{sec:gravitational_waves} with the hard-wall boundary conditions of an elongated box-type trap.

As mentioned in section \ref{sec:analogue_gravity}, a path in the opposite direction of the usual analogue gravity approach is chosen: the BEC is taken to be at rest and the four-velocities in \eqref{eq:analogue_gravity_acoustic_metric_carlos} are $v^i= (c, \, 0, \, 0, \, 0)$.

An interesting effect can already be seen in the absence of any space time distortions. One then has

\begin{IEEEeqnarray}{rCl}
	\IEEEyesnumber \IEEEyessubnumber*
	\mathfrak{g}_{ab}
	&=&
	\frac{1}{c_s}	
	\frac{n_0^2}{p_0 + \rho_0}
	\left(
		\eta_{ab} + \left( 1 - \frac{c_s^2}{c^2} \right)
		\begin{pmatrix}
			c^2 & 0 & 0 & 0
			\\
			0 & 0 & 0 & 0
			\\
			0 & 0 & 0 & 0
			\\
			0 & 0 & 0 & 0
		\end{pmatrix}
	\right)
	\\
	&=&
	\frac{1}{c_s}	
	\frac{n_0^2}{p_0 + \rho_0}
	\begin{pmatrix}
		-c_s^2 & 0 & 0 & 0
		\\
		0 & 1 & 0 & 0
		\\
		0 & 0 & 1 & 0
		\\
		0 & 0 & 0 & 1
	\end{pmatrix} \; .
\end{IEEEeqnarray}

Up to a conformal factor, the Minkowski metric is recovered. This means that in the comoving frame and for a flat background space time, the phonons live on an effective Minkowski metric, but with a \textit{rescaled time coordinate}. In a sense, time flows slower for the phonons \cite{sabin_phonon_2014}, which is an indicator for why resonance phenomena with GWs can be found for typical trap lengths as small as $L \sim 100 \mu$m.

For an elongated trap such that its length in the $x$ direction $L$ is reasonably longer than $L_y$ and $L_z$ and in the ultracold limit, it is sufficient to consider phonons whose momentum components along the $y$ and the $z$ axis vanish since non-negligible excitations of these modes would require too much energy to be populated thermally or induced by a GW.

One can then adopt a description in $1+1$ dimensions and make use of the phonon dispersion relation \eqref{eq:bogo_dispersion_relation_low-energy_limit} $\omega_k = c_s k$ and the hard-wall boundary conditions to find the expansion of the phonon field (see appendix \ref{sec:bogoliubov_trafos})

\begin{equation}
	\hat{\delta \Pi} (x, t) = \sum_n \varphi_n (x, t) \hat{a}_n + \varphi_n^* (x, t) \hat{a}_n^\dagger
\end{equation}

with the mode solutions

\begin{equation}
	\varphi_n (x, t) = \frac{1}{\sqrt{n \pi}} \sin
	\left(	
		\frac{n \pi \left(x - x_L \right)}{L}
	\right)		
		\mathrm{e}^{-i \omega_n t}
\end{equation}

and the frequencies $\omega_n = n \pi c_s/L$. $x_L$ denotes one boundary of the trap.

In the comoving frame, the acoustic metric for a sinusoidal gravitational wave with amplitude $h_{+, 0}$ and frequency $\Omega$ can be expressed in terms of the invariant line element

\begin{equation}
	\ud s^2 = - c_s^2 \ud t^2 + \big( 1 + \underbrace{h_+ (t)}_{\equiv h_{+, 0} \sin \Omega t} \big) \ud x^2 \; .
\end{equation}

It is then not surprising that the effects found in \cite{sabin_phonon_2014} are the same as for a cavity on a flat background whose length varies sinusoidally \cite{ji_production_1997}.

Under the coordinate change

\begin{equation} \label{eq:detector_proposal_coordinate_trafo}
	x' = x \left( 1 + \frac{h_{+, 0}}{2} \right) \, , \quad t' = t \; ,
\end{equation}

the Klein-Gordon equation for the transformed phonon field $\delta \hat{\Pi}' (x', t')$ assumes the same form\footnote{Up to $\mathcal{O}(h_{+, 0})$} as for $\hat{\delta \Pi} (x, t)$ on a flat background and the mode solutions $\varphi'_n (x', t')$ are therefore

\begin{equation}
	\varphi'_n (x', t') = \frac{1}{\sqrt{n \pi}} \sin
	\left(	
		\frac{n \pi \left(x' - x'_L \right)}{L'}
	\right)		
		\mathrm{e}^{-i \omega'_n t'} \; .
\end{equation}

In \cite{sabin_phonon_2014}, the trap is then assumed to have a constant proper length. Under this assumption, the mode solutions $\varphi'_n$ can be expressed to $\mathcal{O} (h_{+, 0})$ in the coordinates $x, t$ as

\begin{equation}
	\varphi'_n (x, t) = \frac{1}{\sqrt{n \pi}} \sin
	\left(	
		\frac{n \pi \left(x (1+ h_{+, 0}/2) - x_L \right)}{L}
	\right)		
		\mathrm{e}^{-i \omega_n t} \; .
\end{equation}

$\delta \hat{\Pi}' (x', t')$ and $\hat{\delta \Pi} (x, t)$ describe the same field. They are therefore related by a Bogoliubov transformation (see appendix \ref{sec:bogoliubov_trafos})

\begin{equation}
	\delta \hat{\Pi}' = \sum_l \varphi_l' \hat{b}_l + \varphi_l'^* \hat{b}_l^\dagger = \sum_n \varphi_n \hat{a}_n + \varphi_n^* \hat{a}_n ^\dagger = \delta \hat{\Pi}
\end{equation}

with

\begin{equation}
	\hat{b}_l = \sum_n \alpha_{ln, 0}^* \hat{a}_n + \beta_{ln, 0}^* \hat{a}_n^\dagger
\end{equation}

and the Bogoliubov coefficients $\alpha_{ln, 0}$ and $\beta_{ln, 0}$ are defined by the Klein-Gordon inner products between the mode functions $\varphi_l'$, $\varphi_n$. For the transformation in question, with $x_L=0$ and up to $\mathcal{O} (h_{+, 0})$, they have been found in \cite{sabin_phonon_2014}:

\begin{IEEEeqnarray}{rCl}
	\IEEEyesnumber \label{eq:detector_bogo_coefficients_instantaneous} \IEEEyessubnumber*
	\beta_{ln, 0}
	&=&
	\begin{cases}
		- \frac{(-1)^{l+n} \sqrt{ln}}{2 (l+n)} h_{+, 0} & \text{for } l \neq n
		\\
		0 & \text{for } l = n
	\end{cases}
	\\
	\alpha_{ln, 0} 
	&=&
	\begin{cases}
		 \frac{(-1)^{l+n} \sqrt{ln}}{2 (l-n)} h_{+, 0} & \quad \text{for } l \neq n
		\\
		1 & \quad \text{for } l = n
	\end{cases}
\end{IEEEeqnarray}

The nonvanishing coefficients $\beta_{ln, 0}$ already indicate particle creation by the passing GW. However, the \eqref{eq:detector_proposal_coordinate_trafo} represents only an instantaneous transformation from an initially flat space-time to one whose spatial component is distorted by a factor $(1+h_{+, 0} / 2)$. In order do calculate the final state after the interaction with a sinusoidal GW, a technique first presented in \cite{bruschi_mode-mixing_2013} has to be applied, where the entire transformation is pieced together from discrete $h_{+, 0}$-perturbations followed by intervals of free evolution and finally the continuous limit is taken \cite{sabin_phonon_2014}. With \eqref{eq:detector_bogo_coefficients_instantaneous}, and $h_+(t) = h_{+, 0} \sin \Omega t$, the coefficients become

\begin{IEEEeqnarray}{rCLl}
		\IEEEyesnumber \label{eq:detector_bogo_continuous_both} \IEEEyessubnumber*
		\beta_{ln} (t)
		&=&
		i \beta_{ln, 0} \left( \omega_l + \omega_n \right)
		&
		\int \ud t' \mathrm{e}^{
			-i (\omega_l + \omega_n ) t'
		}
		\sin \Omega t'
		\label{eq:detector_bogo_continuous_creation}
		\\
		\alpha_{ln} (t)
		&=&
		i \alpha_{ln, 0} \left( \omega_l - \omega_n \right)
		&
		\int \ud t' \mathrm{e}^{
			-i (\omega_l - \omega_n ) t'
		}
		\sin \Omega t'
\end{IEEEeqnarray}

Hence, the phonon creation reflected in \eqref{eq:detector_bogo_continuous_creation} is a resonance phenomenon. If the GW is on resonance with pair of modes $\Omega = \omega_l + \omega_n$ and the interaction time is sufficiently long ($\omega_1 t \gg 1$), the rotating wave approximation can be taken and (up to $\mathcal{O}(h_{+, 0})$) the Bogoliubov coefficients for particle creation are calculated as \cite{sabin_phonon_2014}

\begin{equation}
	\beta_{ij} (t) = \frac{\omega_m}{4} \sqrt{\frac{n}{l}} \delta_{i+j, l+n} \, h_{+, 0} \, t
\end{equation}

for an odd sum of mode numbers $l+n$.

\section{Phonon State Transformation and Quantum Fisher Information}

To see how the transformation induced by the interaction with the GW gets imprinted onto the phonon state and how information about the GW can be extracted from this imprint, the \textit{covariance matrix} (CM) \textit{formalism}\footnote{
	See appendix \ref{sec:CM_formalism_new} and, for a comprehensive overview, \cite{adesso_entanglement_2007}	
} and the framework of relativistic quantum information\footnote{
	See appendix \ref{sec:RQM_new} and \cite{ahmadi_quantum_2014, ahmadi_relativistic_2015}
} provide some useful tools.

Gaussian states with vanishing first moments are uniquely described by the covariance matrix $\sigma_{ij} \equiv  \langle \hat{X}_i \hat{X}_j + \hat{X}_j \hat{X}_i \rangle - \langle \hat{X}_i \rangle \langle \hat{X}_j \rangle $ with the quadrature operators $\hat{X}_{i, j}$. Transformations preserving the Gaussian character of a state are given by the group of symplectic transformations (\eqref{eq:CM_symplectic_trafo_definition_new}, \eqref{eq:CM_symplectic_form_invariance_new}). For Bogoliubov transformations, the corresponding transformation can be calculated directly from the set of coefficients $(\alpha_{mn}, \beta_{mn})$ using \eqref{eq:CM_symplectic_trafo_from_bogo_new}.\footnote{
	See also table \ref{table:CM_formalism_real_complex}.
}

A system that is initially in the state $\sigma_0$ and then interacts with a GW with amplitude $h_{+, 0}$ is then \cite{sabin_phonon_2014}

\begin{equation}
	\sigma_0 \rightarrow \sigma_h = S_h \sigma_0 S_h^T \; ,
\end{equation}

where $S_h$ is the symplectic transformation calculated from the coefficients \eqref{eq:detector_bogo_continuous_both} with help of \eqref{eq:CM_symplectic_trafo_from_bogo_new} under the above assumptions for mode numbers and the interaction time.

Because the intention in \cite{sabin_phonon_2014} is the detection of gravitational waves, the interest lies in the optimum precision for an estimate on $h_{+, 0}$, which is limited by the quantum Cramér-Rao bound \cite{braunstein_statistical_1994}.

\begin{equation} \label{eq:original_detector_proposal_sensitivity}
	\left\langle \left( \Delta h_{+, 0} \right)^2 \right\rangle \geq \frac{1}{M H_h}
\end{equation}

In the equation above, $M$ is the number of independent measurements. $H_h$ is the quantum Fisher information for the GW amplitude. In the CM formalism, it can be computed from the Uhlmann fidelity (\eqref{eq:RQM_QFI_from_uhlmann_fidelity_new}, \eqref{eq:RQM_uhlmann_fidelity_new}, \cite{marian_uhlmann_2012})  between two states $\sigma_h$ and $\sigma_{h + \delta h}$ whose transformation differs by an infinitesimal change in the GW amplitude.

The quantum Fisher information for $h_{+, 0}$ was found in \cite{sabin_phonon_2014} to be

\begin{figure}[hptb]
	\begin{centering}
	\includegraphics[width=0.7\textwidth]{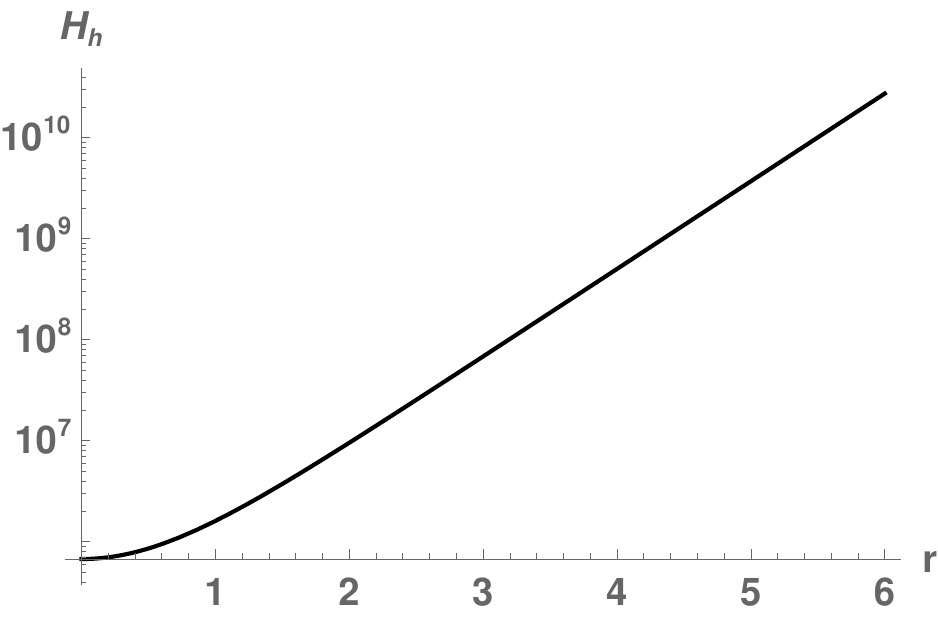}
	\caption{
		{\footnotesize			
			Logarithmic plot of the quantum Fisher information as a function of the initial squeezing $r$ for a set of parameters in analogy to the $^7$Li BEC in \cite{pollack_universality_2009} with $t = 10$s, $\Omega \approx 1$kHz, $n=8$, $l=9$.
			Note the exponential growth of $H_h(r)$ for $r \gtrsim 0.75$.
			\label{fig:detector_QFI_plot}
		}
	}
	\par\end{centering}
\end{figure}

\begin{IEEEeqnarray}{rCl} \label{eq:detector_proposal_QFI}
	H_h &=& \frac{n}{4l} \omega_l^2 t^2 \left[ \left( 2 \sinh^2 r \right)^2 + 4 \right]
	\nonumber
	\\
	&=& \frac{n}{4l} \omega_l^2 t^2 \left( N_{phon}^2 + 4 \right)
\end{IEEEeqnarray}



for an initial squeezed vacuum state with squeeze factor $r$ and phonon number $N_{phon}$. This shows the Heisenberg scaling of the proposed detector, as was described in chapter \ref{sec:squeezed_states}. Because of its strong dependence on the $r$, equation \eqref{eq:detector_proposal_QFI} was the initial motivation for this thesis. For $r \gtrsim 0.7$, the quantum Fisher information scales exponentially with the initial squeezing (fig. \ref{fig:detector_QFI_plot}) and quadratically with the phonon number in the initial state. When trying to measure the imprint a GW leaves on the phonon states of a BEC, it is therefore crucial to know the upper bound for $r$.

\newpage

\section{Improvements: the Pumped-Up $SU(1, 1)$ Scheme \label{sec:pumped-up_SU_11}}

Although the original idea in \cite{sabin_phonon_2014} does allow for Heisenberg scaling, it scales with the phonon number which is much smaller than the number of particles $N_0$ in the bulk of the condensate. As a consequence, it would be outperformed by spatial BEC-based atom interferometric experiments \cite{canuel_matter-wave_2014, chaibi_low_2016, graham_new_2013} that make use of the larger resource of the condensate ground state, even though they cannot surpass the standard quantum limit.

This shortcoming inspired the ongoing work \cite{howl_active_2019, howl_quantum_2020-1, howl_quantum_2020} on the generalization of the \textit{pumped-up} $SU(1, 1)$ scheme. In its essence, it is the most recent link in a historical chain of generalizations of the original idea of an interferometer: from the extension to fields other than light to nonclassical input states, over the 1986 $SU(1, 1)$ scheme \cite{yurke_su2_1986}, which includes active elements and particularly parametric amplifiers rather than only beam splitters. This allows for the generation of squeezing and entanglement \textit{inside} the interferometer. Its name stems from its mathematical description which distinguishes it from traditional interferometers: in the latter, all operators that resemble the building elements obey an $SU(2)$ algebra, whereas here, the algebra that defines them is $SU(1, 1)$ \cite{howl_quantum_2020-1}.

The next link in our chain and the step decisive to exploiting the large particle number of the BEC is the pumped $SU(1, 1)$ scheme \cite{szigeti_pumped-up_2017}. Here, a tritter (a three-way beam splitter) is placed after the parametric amplification stage, linearly mixing the previously squeezed side modes (the phononic modes in our case) and the classical pump mode (the condensate). The tritter angle $\theta$ then controls the mode mixing and can increase the population of the side modes, albeit at the cost of them no longer being in a squeezed vacuum state. For the estimation of a parameter that gets imprinted onto the phase $\varphi$ of the arms, this leads to $\Delta \varphi \propto 1/\sqrt{N_0 N_{phon}}$, thereby potentially improving the sensitivity from \cite{sabin_phonon_2014} by multiple orders of magnitude.

To add some more clarity, let us go through the individual stages:

\begin{description}
	\item[The input state] (the pump mode) is considered a large coherent beam, that is a classical state.\footnote{
		Note that this is well in agreement with our description of a BEC ground state in the Bogoliubov approximation, where it becomes an eigenstate of the annihilation operator.}
	\item[Parametric amplification] or up-conversion, populating the (phonon) side modes with a two-mode squeezed state. This is the element that this thesis is dedicated to.
	\item[Tritter] mixing the pump and side modes and thus letting all of them participate in the parameter estimation. The tritter angle $\theta$ decides how the interferometer behaves between the limiting cases of traditional $SU(2)$ and $SU(1, 1)$ \cite{howl_active_2019}.
	\item[Phase imprint] of the parameter that is to be estimated.
	\item[Tritter$^{-1}$]
	\item[Parametric amplification$^{-1}$] or down-conversion. The sequence of the inverse tritter and this element recombines the side modes.
	\item[Number-sum measurement]
\end{description}

The final link in the chain so far takes us one step further in terms of generalization. Up until this point, the pumped-up $SU(1, 1)$ was designed only for spatial interferometry and thus restricted to local channels. If, however, we are to consider frequency metrology on a BEC, there is no spatial separation between the modes. This allows for non-local channels in the imprint stage, which has been incorporated into the pumped-up $SU(1, 1)$ scheme in recent work by Howl, Fuentes et al. \cite{howl_active_2019, howl_quantum_2020-1, howl_quantum_2020}. In particular, they have managed to consider not only phase imprints, but also squeezing and mode mixing and therefore a complete set of two-mode Gaussian unitary operators. In \cite{howl_quantum_2020}, they have found an expression in terms of BEC parameters for what would be the sensitivity for the GW amplitude in the quasi- $(1+1)$-dimensional case of the previous section \eqref{eq:original_detector_proposal_sensitivity}

\begin{equation} \label{eq:pumped-up_SU11_sensitivity}
	\Delta h_{+, 0} \geq
	\frac{m \pi}{4 \sqrt{2} \hbar}
	\frac{1}{\sqrt{\tau N_{phon} N_d}}
	\frac{\sqrt{l n} \left(l-n \right)^2}{l^2+n^2}
	\frac{\alpha^3}{\theta N_0^2}
	\sqrt{\frac{L^7}{a_s^3 t}} \; ,
\end{equation}

where $m$ is the particle mass, $N_d$ the number of detectors, $(l, n)$ the phonon mode numbers. $L$ is the length of the trap $\alpha = R/L$ the ratio between its length and width, $\tau$ the integration time and $t$ the duration of a single experiment. Equation \eqref{eq:pumped-up_SU11_sensitivity} will be used at the end of chapter \ref{sec:implementation} to estimate the usefulness of the squeeze factor and the parameters derived there.

\chapter{Squeezing Mechanism \label{sec:squeezing mechanism}}

Once the question of why two-mode phonon squeezing is desirable, one might naturally ask how to achieve it. This is the subject of the following section. One possible answer has its origins again in quantum metrology, albeit this time in a non-relativistic context. More specifically, the ansatz comes from a recent paper by R\"atzel et al. \cite{ratzel_dynamical_2018}, who investigated whether it would be possible to detect the influence of a small oscillating mass on a BEC placed next to it. While this was motivated by the potential use of the BEC as a sensing device for other experiments, for instance the superposition of macroscopic quantum systems \cite{romero-isart_large_2011}, it was found that one of the effects of the oscillating Newtonian potential sourced by the mass on the phonon states populating the BEC is two-mode squeezing. In the case considered in \cite{ratzel_dynamical_2018}, the degree of squeezing is, of course, limited by the magnitude of the gravitational interaction. If the experimenter's interest, however, lies not in the detection of said potential but in its exploitation, it does not need to be gravitational in nature. Electromagnetic fields, which are already crucial to the trapping mechanism, can be used to create perturbations of the same character as that of an oscillating nearby mass, but orders of magnitude larger, such that the created two-mode squeezing becomes not only detectable, but could possibly even be used to maximize the sensitivity of the GW detector described in \cite{sabin_phonon_2014}.

The following chapter will follow the arguments made in \cite{ratzel_dynamical_2018} and start by extending the Hamiltonian for a weakly interacting Bose gas \eqref{eq:full_interacting_hamiltonian} to time-dependent fields while keeping the (also time-dependent) external potential general. In the now familiar Bogoliubov approximation, the Hamiltonian can then be expanded in its fluctuations around the mean field to derive mode equations for the phonons. These are then applied to the situation considered in \cite{ratzel_dynamical_2018}: a Bose gas in an elongated box trap with a time-dependent Newtonian potential. Finally, it is shown that the resulting time evolution for the phononic state contains a squeezing operator.

\section{Quantum Field Theory Ansatz}

Let us begin by generalizing the Hamiltonian for a weakly interacting Bose gas \eqref{eq:full_interacting_hamiltonian} to account for an external potential that includes a time-dependent perturbation $\mathcal{V} ( \mathbf{r}, t) = \mathcal{V}_0 (\mathbf{r}) + \delta \mathcal{V} ( \mathbf{r}, t)$. Taking only the zero-momentum component of the interaction potential \eqref{eq:low_energy_limit_effective_potential}, $\hat{H} ( t )$ becomes

\begin{IEEEeqnarray}{rCl}
	\hat{H} (t) &=&
	\int \ud \mathbf{r} \; \hat{\psi}^\dagger ( \mathbf{r}, t)
	\left(
		- \frac{\hbar^2}{2m} \nabla^2 + \mathcal{V}_0 (\mathbf{r})
	\right)
	\hat{\psi} ( \mathbf{r}, t)
	\nonumber
	\\
	& & +
	\int \ud \mathbf{r} \; \hat{\psi}^\dagger ( \mathbf{r}, t)
	\delta \mathcal{V} ( \mathbf{r}, t)
	\hat{\psi} ( \mathbf{r}, t)
	\nonumber
	\\
	& & + \frac{U_0}{2}
	\int \ud \mathbf{r} \;
	\hat{\psi}^\dagger ( \mathbf{r}, t) \hat{\psi}^\dagger ( \mathbf{r}, t)
	\hat{\psi} ( \mathbf{r}, t) \hat{\psi} ( \mathbf{r}, t) \; .
\end{IEEEeqnarray}

The aim here is again to take the Bogoliubov approximation and diagonalize the Hamiltonian up to second order in the fluctuations, but only after separating the time evolution due to $\delta \mathcal{V}$ out of the solution for the ground state. In order to do so, consider a solution $\psi_o (\mathbf{r})$ of the stationary Gross-Pitaevskii equation \eqref{eq:stationary_GPE} with only the constant part $\mathcal{V}$ of the external potential and with the interaction strength $g$ replaced by $U_0$.

\begin{equation} \label{eq:squeezing_mechanism_stationary_GPE}
	\left( - \frac{\hbar^2}{2m} \nabla^2 + \mathcal{V}_0 (\mathbf{r}) + U_0 | \psi_0 (\mathbf{r})|^2 \right) \psi_0 (\mathbf{r}) = \mu \psi_0 (\mathbf{r})
\end{equation}

Neglecting the perturbation to the external potential, the ground state would evolve as $\psi (\mathbf{r}, t) = \psi_0 (\mathbf{r}) \mathrm{e}^{-i \mu t /\hbar}$. However, $\delta \mathcal{V}$ adds a time-dependent change to the chemical potential:

\begin{equation} \label{eq:squeezing_mechanism_chemical_potential_perturbation}
	\delta \mu (t) = \frac{
			\int \ud \mathbf{r} \; \psi_0^* (\mathbf{r}) \delta \mathcal{V} (\mathbf{r}, t ) \psi_0 (\mathbf{r})
	}{
			\int \ud \mathbf{r} \; | \psi_0 (\mathbf{r}) |^2
	}
\end{equation}

This definition comes natural: the numerator is the additional energy of the condensate due to $\delta \mathcal{V}$, the denominator normalizes it to the energy per particle. With \eqref{eq:squeezing_mechanism_chemical_potential_perturbation}, the actual ground state of the condensate in the presence of a time-dependent perturbing potential evolves as

\begin{equation}
	i \hbar \partial_t \psi (\mathbf{r}, t ) =
	\left(
		- \frac{\hbar^2}{2m} \nabla^2 + \mathcal{V}_0 (\mathbf{r}) + \delta \mu (t) + U_0 | \psi_0 (\mathbf{r})|^2
	\right)
	\psi (\mathbf{r}, t )
\end{equation}

and therefore

\begin{equation}
	\psi (\mathbf{r}, t ) = \psi_0 (\mathbf{r})
	\mathrm{e}^{
		-\frac{i}{\hbar} \left( \mu t + \int_0^{t'} \ud t' \; \delta \mu (t') \right)
	} \; .
\end{equation}

The time evolution of $\psi$ can now be separated out of the full bosonic field to define

\begin{equation}
	\hat{\psi}' (\mathbf{r}, t) \equiv \hat{\psi} (\mathbf{r}, t )
	\mathrm{e}^{
		\frac{i}{\hbar} \left( \mu t + \int_0^{t'} \ud t' \; \delta \mu (t') \right)
	} \; .
\end{equation}

In the Heisenberg picture, $\hat{\psi}$ evolves as $-i \hbar \partial_t \hat{\psi} = [\hat{H}, \hat{\psi}]$. $\hat{\psi}$ and $\hat{\psi}'$ both obey the bosonic equal time commutation relations.\footnote{
$[\hat{\psi} (\mathbf{r}, t), \hat{\psi}^\dagger (\mathbf{r}', t)] = \delta^{(3)} (\mathbf{r} - \mathbf{r}')$, $[\hat{\psi} (\mathbf{r}, t), \hat{\psi} (\mathbf{r}', t)] = [\hat{\psi}^\dagger (\mathbf{r}, t), \hat{\psi}^\dagger (\mathbf{r}', t)] = 0$
}
Hence, the evolution of $\hat{\psi}'$ is governed by \cite{ratzel_dynamical_2018}

\begin{IEEEeqnarray}{rCl}
	-i \hbar \partial_t \hat{\psi}' (\mathbf{r}, t)
	&=&
	\left(
		\left( -i \hbar \partial_t + \mu + \delta \mu (t) \right)
		\hat{\psi} (\mathbf{r}, t )
	\right)
	\mathrm{e}^{
		\frac{i}{\hbar}
		\left(
			\mu t + \int_0^{t'} \ud t' \; \delta \mu (t')
		\right)
	}
	\nonumber
	\\
	&=&
	\left[
		\hat{H} (t), \hat{\psi}' (\mathbf{r}, t)
	\right]
	+ \left( \mu + \delta \mu(t) \right) \hat{\psi}' (\mathbf{r}, t)
	\nonumber
	\\
	&=&
	\left[
		\left(
			\underbrace{
				\hat{H} (t) - \left( \mu + \delta \mu(t) \right) \hat{N}(t)
			}_{
				\equiv \hat{H}' (t)
			}
		\right)
		, \hat{\psi}' (\mathbf{r}, t)
	\right] \; ,
\end{IEEEeqnarray}

where the number operator $\hat{N}$ was introduced in the second line and its equal time commutation relation with $\hat{\psi}'$ was used in the third line.\footnote{
	$\hat{N} (t) \equiv \int \ud \mathbf{r} \, \hat{\psi}'^\dagger (\mathbf{r}, t) \hat{\psi}' (\mathbf{r}, t), \qquad [\hat{N} (t), \hat{\psi}' (\mathbf{r}, t)] = -\hat{\psi}' (\mathbf{r}, t)$
}

$H' = \hat{H} - (\mu + \delta \mu) \hat{N}$ is the grand canonical Hamiltonian. Taking the Bogoliubov approximation $\hat{\psi}' = \psi_0 + \delta \hat{\psi}$, $\hat{H}'$ expands to\footnote{
For the sake of readability, we drop the variables ($\mathbf{r}, t)$ for now.	
}

\begin{IEEEeqnarray}{rCl}
	\hat{H}' [\hat{\psi}']
	&=&
	\int \ud \mathbf{r} \; 
		\psi_0^*
		\left(
			-\frac{\hbar^2}{2m} \nabla^2 + \mathcal{V}_0 + \frac{U_0}{2} |\psi_0|^2
		\right) \psi_0
		\nonumber
	\\
	& & +
	\int \ud \mathbf{r} \;
		\delta \hat{\psi}^\dagger
		\left(
			-\frac{\hbar^2}{2m} \nabla^2 + \mathcal{V}_0 + U_0 | \psi_0|^2
		\right)
		\psi_0
		+ \mathrm{h.c.}
		\nonumber
	\\
	& & +
	\int \ud \mathbf{r} \;
		\delta \hat{\psi}^\dagger
		\left(
			-\frac{\hbar^2}{2m} \nabla^2 + \mathcal{V}_0 + 2 U_0 | \psi_0|^2
		\right)
		\delta \hat{\psi}
		+ \frac{U_0}{2}
		\left(
			\psi_0^2 \delta \hat{\psi}^{\dagger 2} + \psi_0^{*2} \delta \hat{\psi}^2
		\right)
		\nonumber
	\\
	& & +
	\int \ud \mathbf{r} \;
		U_0 \left(
			\psi_0 \delta \hat{\psi}^{\dagger 2} \delta \hat{\psi} + \psi_0^* \delta \hat{\psi}^\dagger \delta \hat{\psi}^2
		\right)
		+ \frac{U_0}{2} \delta \hat{\psi}^{\dagger 2} \delta \hat{\psi}^2
		\nonumber
	\\
	& & +
	\int \ud \mathbf{r} \;
		|\psi_0|^2 \delta \mathcal{V}
		+ \left(
			\psi_0^* \delta \mathcal{V} \delta \hat{\psi} + \psi_0 \delta \mathcal{V} \delta \hat{\psi}^\dagger 
		\right)
		+ \delta \hat{\psi}^\dagger \delta \mathcal{V} \delta \hat{\psi}
		\nonumber
	\\
	& & -
	\int \ud \mathbf{r} \;
		\mu |\psi_0|^2
		+ \delta \mu |\psi_0|^2
		+ \mu \left(
			\psi_0^* \delta \hat{\psi} + \psi_0 \delta \hat{\psi}^\dagger
		\right)
		\nonumber
	\\
	& & -
	\int \ud \mathbf{r} \;
		\delta \mu \left(
			\psi_0^* \delta \hat{\psi} + \psi_0 \delta \hat{\psi}^\dagger
		\right)
		+ \mu \delta \hat{\psi}^\dagger \delta \hat{\psi}
		+ \delta \mu \delta \hat{\psi}^\dagger \delta \hat{\psi} \; ,
\end{IEEEeqnarray}

which looks rather complicated. Luckily, parts of the first and fifth and the entire second and sixth line can be shown to cancel out with help of the stationary GPE and the definition \eqref{eq:squeezing_mechanism_chemical_potential_perturbation} of $\delta \mu$ \cite{ratzel_dynamical_2018}. The remaining terms can be grouped by their order in $\delta \hat{\psi}$, beginning with the zeroth order term: the ground state energy

\begin{equation}
	H^{(0)} = - \frac{U_0}{2} \int \ud \mathbf{r} \, | \psi_0|^4 \; ,
\end{equation}

where the hat accent has been omitted due to its classicality. Once the trapping potential is specified, $H^{(0)}$ can be used to calculate the chemical potential $\mu = \partial H^{(0)}/ \partial N_0$ where $N_0$ is the ground state particle number.

The part of the Hamiltonian mediating the phonon-condensate interaction via the perturbing potential can be defined as

\begin{equation} \label{eq:squeezing_mechanism_interaction_hamiltonian}
	H^{(1)}_{int} = \int \ud \mathbf{r} \,
	\left(
		\delta \mathcal{V}- \delta \mu	
	\right)
	\left(
		\psi_0^* \delta \hat{\psi} + \psi_0 \delta \hat{\psi}^\dagger
	\right)
	+
	\delta \hat{\psi}^\dagger
	\left(
		\delta \mathcal{V} - \delta \mu	
	\right)
	\delta \hat{\psi}
\end{equation}

and will lead to the nontrivial part of the time evolution.

For the terms $\mathcal{O} (\delta \hat{\psi}^2)$, one can apply normal ordering to omit the vacuum energy \cite{ratzel_dynamical_2018} and form the Bogoliubov Hamiltonian

\begin{equation}
	\hat{H}^{(2)} = \, \vcentcolon \int \ud \mathbf{r} \,
		\delta \hat{\psi}^\dagger
		\left(
			-\frac{\hbar^2}{2m} \nabla^2 + \mathcal{V}_0 - \mu + 2 U_0 | \psi_0|^2
		\right)
		\delta \hat{\psi}
		+ \frac{U_0}{2}
		\left(
			\psi_0^2 \delta \hat{\psi}^{\dagger 2} + \psi_0^{*2} \delta \hat{\psi}^2
		\right) \vcentcolon
\end{equation}

Third and higher order terms

\begin{equation}
	\hat{H}^{(3+)}_{int} = U_0 \int \ud \mathbf{r} \,
		\psi_0 \delta \hat{\psi}^{\dagger 2} \delta \hat{\psi} + \psi_0^* \delta \hat{\psi}^\dagger \delta \hat{\psi}^2 + \frac{1}{2} \delta \hat{\psi}^{\dagger 2} \delta \hat{\psi}^2
\end{equation}

are neglected for the remainder of this chapter and accounted for in chapter \ref{sec:implementation} by considering phonon damping and three-particle collisions.

As was stated at the beginning of the calculations above, we want to diagonalize $\hat{H}^{(2)}$ and change to the quasi-particle picture by doing so. Hence, with a Bogoliubov transformation in analogy to \eqref{eq:uniform_bose_gas_bogo_trafo}, $\delta \hat{\psi}$ is expanded as \cite{ratzel_dynamical_2018}

\begin{equation}
	\delta \hat{\psi} ( \mathbf{r}, t) = \sum_n u_n ( \mathbf{r}) \hat{b}_n + v_n^* ( \mathbf{r}) \hat{b}_n^\dagger \; ,
\end{equation}

with the orthonormal mode functions satisfying

\begin{equation}
	\int \ud \mathbf{r} \,
		u_n^* ( \mathbf{r}) u_l ( \mathbf{r}) - v_n^* ( \mathbf{r}) v_l ( \mathbf{r}) = \delta_{ln} \; .
\end{equation}

Demanding for the Bogoliubov Hamiltonian to become diagonal

\begin{equation} \label{eq:squeezing_mechanism_diagonalized_bogo_hamiltonian}
	\hat{H}^{(2)} = \sum_n \hbar \omega_n \hat{b}^\dagger_n \hat{b}_n
\end{equation}

leads to the stationary Bogoliubov-de Gennes (BdG) equations \cite{ratzel_dynamical_2018, stoof_ultracold_2009}\footnote{
	There are some subtleties in the derivation of Bogoliubov equations and analogously in the BdG equations, which are well covered in section 8.2 of Pethick and Smith's book \cite{pethick_bose-einstein_2008} and \cite{lewenstein_quantum_1996}.
}

\begin{IEEEeqnarray}{rClCl}
		\IEEEyesnumber \label{eq:squeezing_mechanism_bdg_arbitrary_potential} \IEEEyessubnumber*
		\hbar \omega_n u_n &=&
		\left(
			-\frac{\hbar^2}{2m} \nabla^2 + \mathcal{V}_0 - \mu + 2 U_0 | \psi_0|^2
		\right) u_n
		&+&
		U_0 \psi_0^2 v_n
		\\
		- \hbar \omega_n v_n &=&
		\left(
			-\frac{\hbar^2}{2m} \nabla^2 + \mathcal{V}_0 - \mu + 2 U_0 | \psi_0|^2
		\right) v_n
		&+&
		U_0 \psi_0^{* 2} u_n
\end{IEEEeqnarray}

\section{Simplifications and Time Evolution in a Uniform Trap}

For the situation described in \cite{ratzel_dynamical_2018}, the above expressions and the BdG equations simplify quite substantially (for more detailed derivations, the reader is referred to appendix \ref{sec:appendix_box_trap}). Consider an elongated box type trap with length $L$ and cross section $A$ (and thus volume $V=L A$) and a sinusoidally time-dependent perturbing potential\footnote{
	In the original paper, a spatially uniform $\Phi_{0, \Omega}$ term is included in $\delta \mathcal{V}$ as well as a proportionality to the particle mass, since it is considered to be a gravitational potential. However, the spatially uniform term does not result in any squeezing and we are not specifying a particular type of force here. Thus, $\Phi_{0, \Omega}$ and the mass factor are omitted in the following description.
}

\begin{equation} \label{eq:squeezing_mechanism_perturbing_potential}
	\delta \mathcal{V} (t) = \left( a_\Omega x + \mathfrak{G}_\Omega \frac{x^2}{2} \right) \sin \Omega t \equiv \delta \bar{\mathcal{V}}  \sin \Omega t \; .
\end{equation}

The perturbation to the chemical potential \eqref{eq:squeezing_mechanism_chemical_potential_perturbation} must then share the same time dependence

\begin{equation}
	\delta \mu (t) = \frac{\mathfrak{G}_\Omega L^2}{24} \delta \bar{\mu} \equiv \delta \bar{\mu} \sin \Omega t \; .
\end{equation}

From section \ref{sec:uniform_weakly_interacting}, we know that for an interacting Bose gas in a box trap, the ground state becomes uniform $\psi_0 = \sqrt{n_0}$ with the condensate particle density $n_0$ and the chemical potential \eqref{eq:zeroth_order_chemical_potential} thus turns out as $\mu = U_0 n_0$. The stationary BdG equations for modes with nonvanishing momenta only in the direction of the potential reduce to

\begin{IEEEeqnarray}{rCl}
	\IEEEyesnumber \label{eq:squeezing_mechanism_BdG_eqns_box_potential_final} \IEEEyessubnumber*
	\omega_n u_n ( x) &=& \frac{\hbar}{2m} \left[ \left( -  \nabla^2 + \xi \right) u_n ( x) + \xi n_0 v_n ( x) \right]
	\label{eq:squeezing_mechanism_BdG_eqns_box_potential_un}
	\\
	- \omega_n v_n ( x) &=& \frac{\hbar}{2m} \left[ \left( -  \nabla^2 + \xi \right) v_n ( x) + \xi n_0 u_n ( x) \right]
\end{IEEEeqnarray}

with von Neumann boundary conditions at $x= \pm L/2$. The resulting mode solutions are

\begin{IEEEeqnarray}{rClrCl}
	\IEEEyesnumber \label{eq:squeezing_mechanism_BdG_mode_solutions} \IEEEyessubnumber*
	u_n &=& \alpha_n \varphi_n
	&
	v_n &=& \beta_n \varphi_n
	\\
	\alpha_n &=& \sqrt{\frac{1}{V}} \sqrt{\frac{1}{\sqrt{2}k_n \xi} + 1} \qquad
	&
	\beta_n &=& - \sqrt{\frac{1}{V}} \sqrt{\frac{1}{\sqrt{2}k_n \xi} - 1}
	\\
	\varphi_n &=& \cos \left( k_n (x+\frac{L}{2}) \right)
	&
	k_n &=& \frac{n \pi}{L}
\end{IEEEeqnarray}

and the phonon dispersion relation $\omega_n = c_s k_n$ \eqref{eq:bogo_dispersion_relation_low-energy_limit} is recovered in the limit $k_n \xi \ll 1$ \cite{ratzel_dynamical_2018}.

We are trying to show that the time evolution due to $\delta \mathcal{V}$ contains a squeezing operator. Therefore, we need the corresponding Hamiltonian $H^{(1)}_{int}$ \eqref{eq:squeezing_mechanism_interaction_hamiltonian} in the interaction picture (see appendix \ref{sec:appendix_matrix_elements_time_evolution_box_trap}, \cite{ratzel_dynamical_2018}), which can be composed in terms of matrix elements.

\begin{IEEEeqnarray}{rCl}
	H^{(1)}_{int, I}
	&=&
	\sum_n
		\mathcal{M}_{0n} (u_n + v_n)
		\left(
			\mathrm{e}^{-i ( \omega_n - \Omega ) t} \hat{b}_n - \mathrm{e}^{i ( \omega_n - \Omega ) t} \hat{b}_n^\dagger
		\right)
	\nonumber
	\\
	& &
	+ \sum_{n, l}
		\mathcal{M}_{nl}
		\left(
			\hat{b}_n \hat{b}_l \mathrm{e}^{-i ( \omega_n + \omega_l - \Omega ) t}
			- \hat{b}_n^\dagger \hat{b}_l^\dagger \mathrm{e}^{i ( \omega_n + \omega_l - \Omega ) t}
		\right)
	\nonumber
	\\
	& & 
	+ \sum_{n>l}
		\left(
			\mathcal{A}_{nl} + \mathcal{B}_{nl}
		\right)
		\left(
			\mathrm{e}^{-i ( \omega_n - \omega_l - \Omega ) t} \hat{b}_l^\dagger \hat{b}_n
			- \mathrm{e}^{i ( \omega_n - \omega_l - \Omega ) t} \hat{b}_n^\dagger \hat{b}_l
		\right)
\end{IEEEeqnarray}

with

\begin{IEEEeqnarray}{rCl}
	\IEEEyesnumber \IEEEyessubnumber*
	\mathcal{M}_{0n} & \equiv &
	- \frac{i}{2} \sqrt{n_0} A \int_{-L/2}^{L/2} \ud x
	\left(
		\delta \bar{\mathcal{V}} - \delta \bar{\mu}
	\right)
	\left( u_n + v_n \right)
	\\
	\mathcal{M}_{nl} & \equiv & 
	- \frac{i}{2} A \int_{-L/2}^{L/2} \ud x
	\left(
		\delta \bar{\mathcal{V}} - \delta \bar{\mu}
	\right)
	u_n v_l
	\\
	\mathcal{A}_{nl} & \equiv &
	- \frac{i}{2} A \int_{-L/2}^{L/2} \ud x
	\left(
		\delta \bar{\mathcal{V}} - \delta \bar{\mu}
	\right)
	u_n u_l
	\\
	\mathcal{B}_{nl} & \equiv &
	- \frac{i}{2} A \int_{-L/2}^{L/2} \ud x
	\left(
		\delta \bar{\mathcal{V}} - \delta \bar{\mu}
	\right)
	v_n v_l \; .
\end{IEEEeqnarray}

Using \eqref{eq:squeezing_mechanism_perturbing_potential} and to first order in $k_n \xi$, the matrix elements become (\cite{ratzel_dynamical_2018}, appendix \ref{sec:appendix_matrix_elements_time_evolution_box_trap})

\begin{IEEEeqnarray}{rCl} \label{eq:squeezing_mechanism_matrix_elements}
	\mathcal{M}_{0n}
	&\approx &
	i \sqrt{
		\frac{L^3 N_0 \xi}{\left( \sqrt{2} n \pi \right)^3}		
	}
	\left(
		\left(
			1- (-1)^n
		\right) \frac{a_\Omega}{L}
		-
		\left(
			1+ (-1)^n
		\right) \frac{\mathfrak{G}_\Omega}{2}
	\right)
	\nonumber
	\\*[5pt]
	\mathcal{M}_{nl}
	& \approx &
	\begin{cases}
		-i  \frac{\left( l^2 + n^2 \right) L^3}{
			2 \sqrt{2 l n} \left( l^2 - n^2 \right)^2 \pi^3 \xi
		}
		\left(
			\left( 1 - (-1)^{l+n} \right) \frac{a_\Omega}{L}
			-
			\left( 1 + (-1)^{l+n} \right) \frac{\mathfrak{G}_\Omega}{2}
		\right)
		& \text{for } n \neq l
		\\
		i \frac{\mathfrak{G}_\Omega L^3}{16 \sqrt{2} n^3 \pi^3 \xi}
		& \text{for } n=l
	\end{cases}
	\nonumber
	\\*[5pt]
	\mathcal{A}_{nl}
	& \approx &
	\mathcal{B}_{nl} \approx - \mathcal{M}_{nl} \; .
\end{IEEEeqnarray}

For processes on resonance such that $n = n_\Omega$ in $\mathcal{M}_{0n}$ (i.e. $l+n = n_\Omega$) in the other matrix elements with $n_\Omega \equiv L \Omega / (\pi c_s)$, the time evolution is then

\begin{IEEEeqnarray}{rCRl} \label{eq:squeezing_mechanism_time_evolution}
	\hat{U}_{I, res} &=&
	\mathrm{exp} & \Bigg[
		(-1)^{n_\Omega} \alpha_{n_\Omega} (t) \left( \hat{b}_{n_\Omega}^\dagger  - \hat{b}_{n_\Omega} \right)
		\nonumber
		\\
		& & &
		-
		\frac{1 + (-1)^{n_\Omega}}{2} \frac{r_{n_\Omega/2}}{2} \left(
			\hat{b}_{n_\Omega/2}^{\dagger 2} - \hat{b}_{n_\Omega/2}^2
		\right)
		\nonumber
		\\
		& & &
		- \sum_{n < n_\Omega/2}
			(-1)^{n_\Omega} r_{n, n_\Omega-n} (t)
			\left(
				 \hat{b}_n^\dagger \hat{b}_{n_\Omega-n}^\dagger - \hat{b}_n \hat{b}_{n_\Omega-n}
			\right)
		\nonumber
		\\
		& &	&	
		+ \sum_{n > n_\Omega}
			(-1)^{n_\Omega} \Theta_{n, n-n_\Omega} (t)
			\left(
				 \hat{b}_n^\dagger \hat{b}_{n_\Omega-n} - \hat{b}_{n_\Omega-n}^\dagger \hat{b}_n
			\right)
		\Bigg]
\end{IEEEeqnarray}

with the coefficients

\begin{IEEEeqnarray}{rClCrCl} \label{eq:squeezing_mechanism_evolution_parameters}
	\alpha_{n} (t) & \equiv & \frac{1}{\hbar} |\mathcal{M}_{0n}| t
	& \quad &
	r_{n} (t) & \equiv & \frac{2}{\hbar} | \mathcal{M}_{nn} | t
	\nonumber
	\\
	r_{n, l} (t) & \equiv & \frac{2}{\hbar} | \mathcal{M}_{nl} | t
	& &
	\Theta_{n, l} (t) & \equiv & \frac{2}{\hbar} | \mathcal{A}_{nl} | t \; .
\end{IEEEeqnarray}

As can be seen from \eqref{eq:squeezing_mechanism_time_evolution}, these coefficients parametrize (in the order they are written in above)

\begin{itemize}
	\item Displacement (see \eqref{eq:displacement_operator})
	\item Single-mode squeezing (see \eqref{eq:single-mode_squeezing_operator})
	\item \textit{Two-mode squeezing}
	\item Mode mixing
\end{itemize}

Thus, it is shown that the perturbation with a potential of the type \eqref{eq:squeezing_mechanism_perturbing_potential} on a BEC in a box-type cavity produces the desired two-mode squeezing. To get a better impression of the behaviour of its parameter $r_{n, l} (t)$ which we would like to maximize, note first from \eqref{eq:squeezing_mechanism_matrix_elements} that for an odd and for an even sum of mode numbers, only the $a_\Omega$ and the $\mathfrak{G}_\Omega$ term contributes to $r_{n, l}$, respectively. In other words: for $l+n = $odd, the linear periodic perturbation produces squeezing and for $l+n=$even, the harmonic periodic perturbation does. At first glance, this seems to indicate that for odd $l+n$, $r_{n, l}$ scales with the third power of the trap length, whereas it is only proportional to $L^2$ for an even sum of mode numbers. As we shall see in the next chapter, the proportionality of the upper bounds for $a_\Omega$ and $\mathfrak{G}_\Omega$ partly resolves this peculiarity.

By choosing $l+n=$odd, as is necessary for the proposed GW detector described in the previous chapter, one can also avoid single-mode squeezing by setting $\mathfrak{G}_\Omega=0$.

For a fixed difference between the mode numbers $l-n$, all the parameters in \eqref{eq:squeezing_mechanism_evolution_parameters} are monotonously decreasing for higher modes. $r_{n, l}$ is therefore maximal in $l=1$, $n=2$ or, for even $n_\Omega$, $l=1$, $n=3$.


\chapter{Implementation \label{sec:implementation}}

When trying to estimate which experimental setups would be best suited to maximize the two-mode squeezing induced by the mechanism depicted in the previous chapter, the work on this thesis took a turn that was, to the author, surprising, although it was to be expected in hindsight.

Naively, one can derive upper bounds for the magnitudes of the perturbing potential $a_\Omega$ and $\mathfrak{G}_\Omega$ in \eqref{eq:squeezing_mechanism_perturbing_potential} by demanding that $\delta \mathcal{V}$ only evokes a negligible deformation of the condensate and use these upper bounds to derive a maximum value for the two-mode squeeze factor $r_{n, l}$ given a particular set of modes $(n, l)$ and experimental parameters.

The application of the relations found in this manner to the experiment described in \cite{pollack_universality_2009}, however, led to a squeeze factor $r_{1, 2} \sim 2000$, corresponding to the production of $\sim 10^{868}$ phonons. This is of course unreasonable and would result in the immediate destruction of the condensate. However, some considerations made on the path to this result are still relevant for the work that followed. It will therefore be outlined in the next section.

The overly large squeeze factor from this naive approach made it clear that the limitations to the magnitude of the perturbation must lie not only in the deformation of the ground state, but also in the number of phonons produced: once it assumes the same order as the number $N_0$ of particles in the ground state, the Bose gas ceases to be a condensate. In this sense, $N_0$ can be seen as a budget of quasiparticles that are allowed to be excited by the squeezing mechanism. One can then apply the unitary transformation \eqref{eq:squeezing_mechanism_time_evolution} to an initial state, which is taken to be thermal, and derive the phonon number in the resulting state to find the maximum values for the coefficients \eqref{eq:squeezing_mechanism_evolution_parameters} and with them $a_\Omega$ and $\mathfrak{G}_\Omega$. The quality of squeezing in the final state is estimated by comparison to a squeezed vacuum state \eqref{eq:two-mode_squeezed_vacuum}. A description of this procedure can be found in section. \ref{sec:bounds_maximum_phonon_number}.

In sections \ref{sec:existing_experiments} and \ref{sec:proposed_experiment}, the results are then applied to existing experiments (or approximations thereof) with the outcome that they were obviously not designed with our squeezing scheme in mind and hence an experiment that would be better suited is proposed.

\section{Naive Approach \label{sec:naive_approach}}

Suppose the magnitudes $a_\Omega$ and $\mathfrak{G}_\Omega$ of the linear and the harmonic perturbation are limited solely by the requirement that the perturbed condensate wave function $\psi_p$ deviates from the unperturbed one to an extent that is negligible.

As a first approximation (and to get simple, analytical expressions for the bounds we want to derive), we can formulate this requirement by demanding that the mean deviation of the condensate density should be much smaller than the condensate particle number. For a box trap of volume $V=LA$, this means

\begin{equation} \label{eq:naive_approach_bound_condition}
	\int_V \ud \mathbf{r} \left| |\psi_0 |^2 - | \psi_{p} ( \mathbf{r} ) |^2 \right| \ll N_0 \; .
\end{equation}

We thus need to find the condensate wave functions for a box-type cavity with an additional linear and harmonic potential. To do so, we make use of the GPE in the Thomas-Fermi approximation \eqref{eq:thomas-fermi_solution} with the chemical potential $\mu = g n_0$ of the uniform ground state and the condition that the condensate particle number and thus the integral of $| \psi_{p} ( \mathbf{r} ) |^2$ over the trap volume remains $N_0$.

Consider first the linear perturbation, where the condensate density then fulfills

\begin{equation}
	| \psi_{p, lin} (x)|^2 = \frac{1}{g} \left( \mu - a_\Omega x \right)
\end{equation}

while the unperturbed ground state wave function is still uniform. Plugging this into the condition \eqref{eq:naive_approach_bound_condition} yields

\begin{IEEEeqnarray}{rCcL}
	\IEEEyesnumber \IEEEyessubnumber*
	\frac{A}{g} \int_{-L/2}^{L/2} \left| \frac{N_0}{V} - \frac{N_0}{V} + a_\Omega x \right|
	&=&
	\frac{A a_\Omega}{g} \frac{L^2}{4}
	& \ll N_0
	\\
	\Rightarrow a_\Omega & \ll & \frac{N_0}{V} \frac{4g}{L} &= \frac{4 \mu}{L} \; .
\end{IEEEeqnarray}

For the harmonic perturbation parametrized by $\mathfrak{G}_\Omega$, the same conditions lead to

\begin{equation}
	| \psi_{p, harm} (x)|^2 = \frac{1}{g} \left( \mu + \mathfrak{G}_\Omega (-x^2 + \frac{L^2}{12}) \right)
\end{equation}

and

\begin{IEEEeqnarray}{rCcL}
	\IEEEyesnumber \IEEEyessubnumber*
	\frac{A}{g} \int_{-L/2}^{L/2} \left| \mathfrak{G}_\Omega \left( -x^2 + \frac{L^2}{12} \right) \right|
	&=&
	\frac{2A \mathfrak{G}_\Omega L^3}{9 \sqrt{3} g}
	& \ll N_0
	\\
	\Rightarrow \mathfrak{G}_\Omega & \ll & \frac{N_0}{V} \frac{9 \sqrt{3} g}{2 L^2} &= \frac{9 \sqrt{3} \mu}{2 L^2} \; .
\end{IEEEeqnarray}

Due to the relative abundance of Bose-Einstein condensate experiments making use of harmonic traps in comparison to box type cavities, the author has decided to assume the existence of experiments of the latter nature with their remaining parameters (such as temperature, interaction strength, and particle number) being the same as in existing harmonic trap setups. This assumption, however, comes with some intricacies: if we imagine a BEC with repulsive interactions in an $x^2$ potential (or any non-hard-wall potential, for that matter), it is not hard to see that as its scattering length is increased, the gas will expand in its trap. The spatial dimensions of a gas in a quasi-infinite square potential well, on the other hand, are fixed. This means that the size of a box cavity approximating a harmonic trap depends not only on the trap frequencies, but also on the interaction strength, as well as the particle number and mass.

As an estimate for the size of the condensate, the Thomas-Fermi radius \eqref{eq:thomas-fermi_radius} is used, assuming a cylindrical hard-wall potential. For the overall life time of both the condensate and the phonons to use in \eqref{eq:squeezing_mechanism_evolution_parameters}, we can take the minimum of the inverse of the sum of the Landau and the Beliaev damping rates \eqref{eq:landau_and_beliaev_damping}\footnote{We now denote phonon modes not by their momentum $\mathbf{p}$, but by their mode number.} and the half-life time due to three-body losses \eqref{eq:three-body_half-life}

\begin{equation} \label{eq:phonon_life_time_overall}
	t_{life} = \mathrm{min} \left( \frac{1}{\gamma^{La}_l + \gamma^{Be}_l}, \, t_{1/2} \right) \; .
\end{equation}

Because we are squeezing in two modes and both $\gamma^{La}_l$ and $\gamma^{Be}_l$ are monotonously increasing with the mode number, the index $l$ denotes the higher one of the two modes considered.

\subsection{Example: the Hulet Group $^7$Li Experiment}

The $^7$Li experiment described in \cite{pollack_universality_2009}, performed by the Hulet group ad Rice University with the intention of finding signatures of Efimov physics provided a suitable candidate for a first test of the considerations described above due to the completeness of the set of experimental parameters given in the publication as well as the vast range of tunability of the interaction strength.

Parameters and approximate deduced characteristics for a BEC in a box trap with a test interaction strength $a_s=100 \, a_0$\footnote{$a_0 \approx 5.29 \cdot 10^{-11} \,$m is the Bohr radius.} are given in table \ref{table:harmonic_trap-box_trap_1}.

\begin{table}[hbtp]
\centering
\resizebox{\textwidth}{!}{
	\begin{tabular}{c|c|c|c|c|c|c}
		\specialcell{Harmonic \\ trap}
		&
		\specialcell{axial \\ trap \\ frequency \\ $[2 \pi \mathrm{s}^{-1}]$}
		&
		\specialcell{radial \\ trap \\ frequency \\ $[2 \pi \mathrm{s}^{-1}]$}
		&
		\specialcell{total \\ particle \\ number \\ $[ \, ]$}
		&
		\specialcell{condensate \\ particle \\ number \\ $[ \, ]$}
		&
		\specialcell{total \\ peak \\ density\footnote{Calculated by the author in the Thomas-Fermi approximation} \\ $[\mathrm{cm}^{-3}]$}
		&
		\specialcell{temperature\footnote{The temperature is given in terms of the critical temperature $T_c$, which in turn is taken at $a_s=200 \, a_0$.} \\ $[T_c]$}
		\\
		\hline
		&
		\rule{0pt}{3ex}
		$236$
		&
		$16$
		&
		$4 \cdot 10^5$
		&
		$3.6 \cdot 10^5$
		&
		$\approx 1.8 \cdot 10^{13}$
		&
		$0.5$
		\\
		\hline
		\hline
		\specialcell{Box \\ trap}
		&
		\specialcell{trap \\ length \\ $[\mu \mathrm{m}]$}
		&
		\specialcell{trap \\ radius \\ $[\mu \mathrm{m}]$}
		&
		\specialcell{total \\ particle \\ number \\ $[ \, ]$}
		&
		\specialcell{condensate \\ particle \\ number \\ $[ \, ]$}
		&
		\specialcell{total \\ average \\ density \\ $[\mathrm{cm}^{-3}]$}
		&
		\specialcell{temperature \\ $[n \mathrm{K}]$}
		\\
		\cline{2-7}
		&
		\rule{0pt}{3ex}
		$\approx 280$
		&
		$\approx 9$
		&
		$4 \cdot 10^5$
		&
		$3.6 \cdot 10^5$
		&
		$\approx 5.2 \cdot 10^{12}$
		&
		$\approx 22$
		\\
		\cline{2-7}
		&
		\specialcell{speed of \\ sound \\ $[\mathrm{mm}/\mathrm{s}]$}
		&
		\specialcell{first \\ phonon \\ mode \\ frequency \\ $[\mathrm{s}^{-1}]$}
		&
		\specialcell{phonon \\ life \\ time \\ $[\mathrm{s}]$}
		&
		\specialcell{maximum \\ harmonic \\ perturbation \\ $\mathfrak{G}_\Omega$ \\ $[\mathrm{kg \ m}/\mathrm{s}^2]$}
		&
		\specialcell{maximum \\ linear \\ perturbation \\ $a_\Omega$ \\ $[\mathrm{kg}/\mathrm{s}^2]$}
		&
		\specialcell{squeeze \\ factor \\ $r_{12}$ \\ $[ \, ]$}
		\\
		\cline{2-7}
		&
		\rule{0pt}{3ex}
		$\approx 5$
		&
		$\approx 58$
		&
		$\approx 10$
		&
		$\approx 3 \cdot 10^{-25}$
		&
		$\approx 4 \cdot 10^{-29}$
		&
		$\approx 2000$
		\end{tabular}
}
\caption{
	\footnotesize Approximation of the BEC in \cite{pollack_universality_2009} to a box trap of comparable parameters. The peak density in the harmonic trap was calculated by the author in the Thomas-Fermi approximation. The temperature was given in terms of the critical temperature $T_c$ at a scattering length $a_s = 200 a_0$ and the absolute temperature was deduced by the author following \cite{giorgini_condensate_1996}. For the calculations involving phonons, the first two modes were considered. The phonon life time was calculated considering Landau and Beliaev damping \eqref{eq:landau_and_beliaev_damping} as well as the limit to the overall condensate life time by three-body losses. Note the overly large squeeze factor which led to the considerations that follow in section \ref{sec:bounds_maximum_phonon_number}. \label{table:harmonic_trap-box_trap_1}
	}
\end{table}

\section{Phonon Budget and State Transformation \label{sec:bounds_maximum_phonon_number}}

In order to still be able to meaningfully speak of a Bose-Einstein condensate, we set the maximum number of phonons in the final state to a small --albeit, admittedly, somewhat arbitrary-- fraction of the condensate particle number $N_{phon} \leq N_0 / 50$. The task is then to find out how many phonons the time evolution \eqref{eq:squeezing_mechanism_time_evolution} creates given an initial thermal state. In order to do so, the covariance matrix formalism presents itself as a valuable tool once more.\footnote{
	In contrast to chapter \ref{sec:detector_proposal}, the \textit{complex} picture of the CM formalism is applied here (see appendix \ref{sec:CM_formalism_new}).
} This means that we need the CM of an infinite-mode thermal state \eqref{eq:CM_formalism_thermal_state_CM}, its particle number and a transformation in phase space reflecting the unitary time evolution \eqref{eq:squeezing_mechanism_time_evolution}. The thermal occupation number of a mode $l$ with energy $\hbar \omega_l$ is $(\mathrm{e}^{\beta \hbar \omega_l}-1)^{-1}$ (see appendix \ref{sec:CM_formalism_new}). In a box trap, we have $\omega_l = l \omega_1$ and therefore 

\begin{equation} \label{eq:thermal_state_infinite_mode_particle_number}
	N_{phon, i} = \sum_{l=1}^\infty \frac{1}{\mathrm{e}^{\beta \hbar l \omega_1}-1} = \frac{1}{2} - \frac{1}{\beta \hbar \omega_1} \left( \ln (\mathrm{e}^{\beta \hbar n \omega_1} - 1) + \psi_{\mathrm{Exp}(\beta \hbar \omega_1)} (1) \right) \; ,
\end{equation}

where $\psi_q (z)$ is the $q$-digamma function \cite{weisstein_polygamma_2002} and a transformation in phase space reflecting the unitary time evolution \eqref{eq:squeezing_mechanism_time_evolution}.

In general, the transformation of the first and second moments corresponding to a Gaussian unitary transformation in the Hilbert space is given by \cite{safranek_optimal_2016}

\begin{IEEEeqnarray}{rClCc}
	\IEEEyesnumber \IEEEyessubnumber*
	\hat{U} &=& \mathrm{e}^{
		\frac{i}{2} \hat{\mathbf{A}}^\dagger W \hat{\mathbf{A}} + \hat{\mathbf{A}}^\dagger K \boldsymbol{\gamma}
	}
	&
	&
	\\
	\Rightarrow \;
	S_U &=& \mathrm{e}^{i K W} 
	&
	\mathrm{and \quad}
	&
	\mathbf{b} = \left( \int_0^1 \ud t \mathrm{e}^{i K W t} \right) \boldsymbol{\gamma} \; ,
\end{IEEEeqnarray}

where $\hat{\mathbf{A}}$ is the vector of creation and annihilation operators, $K$ is the symplectic form and $W$ and $\boldsymbol{\gamma}$ encode the change in first and second moments, respectively. The CM $\sigma$ and the vector $\mathbf{d}$ of first moments then transform as

\begin{equation}
		\sigma \rightarrow S_U \sigma S_U^\dagger \; , \quad \mathbf{d} \rightarrow S_U \mathbf{d} + \mathbf{b} \; .
\end{equation}

The particle number of the final state can then be calculated with \cite{howl_active_2019}

\begin{equation} \label{eq:CM_particle_number_expectation_value}
	\langle \hat{N} \rangle = \frac{1}{4} \left(
		\mathrm{Tr} \left( \sigma \right) + \mathbf{d}^\dagger \mathbf{d} - 2N
	\right) \; .
\end{equation}

The derivation of the symplectic transformation corresponding to \eqref{eq:squeezing_mechanism_time_evolution} for all sets of modes at once would result in an infinite-dimensional matrix exponential and is not the way to go. Nonetheless, for a given $n_\Omega$, the part of our transformation that creates phonons can be split up into three types living on subspaces that are mutually noninteracting:

\begin{description}
	\item[Coherent displacement] in mode $n_\Omega$, which gets mixed into mode $2n_\Omega$
	\item[Single-mode squeezing] (for even $n_\Omega$) in mode $n_\Omega/2$, mixing into $3n_\Omega/2$
	\item[Two-mode squeezing] in modes $k$, $l$ with $k+l=n_\Omega$, mixing with modes $k+n_\Omega$ and $l+n_\Omega$
\end{description}

The maximum number $N_{phon, f}$ of phonons in the final state can thus be split up into the initial phonon number $N_{phon, i}$ and the phonons created be the three mechanisms above and we have

\begin{equation} \label{eq:implementation_phonon_budget}
	N_{phon, f} = N_{phon, i} + \Delta N_{phon, d} + \Delta N_{phon, 1-sq} + \Delta N_{phon, 2-sq} \leq \frac{N_0}{50}
\end{equation}

For a given set of experimental parameters,\footnote{That is, $L$ $N_0$ and $\xi$.} and a resonance mode $n_\Omega$, all coefficients in \eqref{eq:squeezing_mechanism_evolution_parameters} are uniquely related and can be expressed in terms of one two-mode squeeze factor. For later use in the subsequent sections, let us denote the modulus of the matrix elements in \eqref{eq:squeezing_mechanism_matrix_elements} without the factors corresponding to the perturbing potential as

\begin{IEEEeqnarray}{rCl}
	\IEEEyesnumber \IEEEyessubnumber*
	m_{0n} &=& \sqrt{
		\frac{L^3 N_0 \xi}{\left( \sqrt{2} n \pi \right)^3}		
	}
	\\
	m_{nl} \approx a_{nl} \approx b_{nl} &=&
	\begin{cases}
		\frac{\left( l^2 + n^2 \right) L^3}{
			2 \sqrt{2 l n} \left( l^2 - n^2 \right)^2 \pi^3 \xi
		}
		& \text{for } n \neq l
		\\
		\frac{\mathfrak{G}_\Omega L^3}{16 \sqrt{2} n^3 \pi^3 \xi}
		& \text{for } n=l \; .
	\end{cases}
\end{IEEEeqnarray}

With this and the transformations described below, the maximum squeeze factor $r_{l, n_\Omega-l}$ satisfying \eqref{eq:implementation_phonon_budget} has been computed for a set of box traps approximating existing experiments with a number of specific $n_\Omega$. At first glance, it may seem as squeezing at lower resonances such as $n_\Omega=3$ or $n_\Omega=4$ might be advantageous. For higher $n_\Omega$, there is more than one pair of modes whose sum is at resonance and therefore there will be multiple subspaces where two-mode squeezing occurs. In all of these subspaces, quasi-particles are created, accelerating the exhaustion of our phonon budget. Closer inspection shows this thought to be wrong: while there are multiple instances of two-mode squeezing for $n_\Omega \geq 5$, the amount of particles created strongly depends on the initial (thermal) occupation $(\mathrm{e}^{\beta \hbar \omega_l}-1)^{-1}$ of the mode $l$ in question, which decreases significantly with higher $l$. In addition, lower thermal occupation numbers also mean that the fidelity of the final state in comparison to the desired two-mode squeezed vacuum is drastically improved, as is shown section \ref{sec:quality}.

For arbitrary $n_\Omega$ the mode pair $(l, n_\Omega-l)$ subject to the highest squeeze factor is the one with the largest $l$ satisfying $l<n_\Omega/2$ and thus the smallest distance $(n_\Omega-l)-l$. This can be seen by comparing their respective matrix elements and finding that

\begin{equation}
	m_{l, n_\Omega-l}>m_{l', n_\Omega-l'} \quad \forall \; l' \, \in \, \mathbb{N} \, <l \; .
\end{equation}

We will refer to this pair of modes $(l_{max}, n_\Omega-l_{max})$ as the \textit{balanced pair} and choose the highest two-mode squeeze factor $r_{l_{max}, n_\Omega-l_{max}}$ as the one parameter\footnote{The calculations to express all the other parameters in terms of $r_{l_{max}, n_\Omega-l_{max}}$ can be found in the Mathematica file that was used to determine the symplectic transformation in appendix \ref{chap:app_mathematica_transformation}.} determining the strength of the transformation to express all other $r_{l', n_\Omega-l'}$ as

\begin{equation}
	r_{l', n_\Omega-l'} = \frac{m_{l', n_\Omega-l'}}{m_{l_{max}, n_\Omega-l_{max}}} r_{l_{max}, n_\Omega-l_{max}} \; .
\end{equation}


\subsection{Coherent Displacement and Mode Mixing}

The state transformation in the modes $(n_\Omega, 2n_\Omega)$ consists of displacement and mode mixing

\begin{equation}
	U = \mathrm{e}^{
		(-1)^{n_\Omega} \left(
			\alpha_{n_\Omega} (\hat{b}_{n_\Omega}^\dagger - \hat{b}_{n_\Omega} )
			+
			 \Theta_{2n_\Omega, n_\Omega} (\hat{b}_{2n_\Omega}^\dagger \hat{b}_{n_\Omega} - \hat{b}_{n_\Omega}^\dagger \hat{b}_{2n_\Omega} )
		\right)
	} \; ,
\end{equation}

yielding

\begin{IEEEeqnarray}{rClCrCl}
	\IEEEyesnumber \IEEEyessubnumber*
	W &=& \overbrace{
		- (-1)^{n_\Omega}  \Theta_{2n_\Omega, n_\Omega}
	}^{
		\equiv \varphi
	}
	\left( \sigma_z \otimes \sigma_y \right)
	&
	\quad
	&
	S_U &=& \mathbb{1} \otimes
	\begin{pmatrix}
		\cos \varphi & \sin \varphi
		\\
		-\sin \varphi & \cos \varphi
	\end{pmatrix}
	\\*[6pt]
	\boldsymbol{\gamma} &=& (-1)^{n_\Omega} \alpha_{n_\Omega}
	\begin{pmatrix}
		1
		\\
		1
	\end{pmatrix}
	\otimes
	\begin{pmatrix}
		1
		\\
		0
	\end{pmatrix}
	&
	&
	\mathbf{b}	 &=& \frac{\alpha_{n_\Omega}}{ \Theta_{2n_\Omega, n_\Omega}}
	\begin{pmatrix}
		1
		\\
		1
	\end{pmatrix}
	\otimes
	\begin{pmatrix}
		- \sin \varphi
		\\
		1 - \cos \varphi
	\end{pmatrix} \, ,
\end{IEEEeqnarray}

where $\sigma_z$ and $\sigma_y$ are Pauli matrices. The number of phonons created by the transformation above can then be found by applying it to the corresponding modes of a thermal state and recalling \eqref{eq:CM_particle_number_expectation_value}. For the overall phonon number in a transformed thermal state, one finds

\begin{equation}
\langle \hat{N} \rangle_{th, U} = \langle \hat{N} \rangle_{vac, U} + \langle \hat{N} \rangle_{th} \; .
\end{equation}

$\langle \hat{N} \rangle_{vac, U}$ and $\langle \hat{N} \rangle_{th}$ are the numbers of quasiparticles that would have been produced by the same transformation on a vacuum state and the initial thermal occupation numbers, respectively. The phonons created by displacement and mode mixing are then

\begin{equation}
	\Delta N_{phon, disp} = \langle \hat{N} \rangle_{vac, U} = 2
	\left( \frac{\alpha_{n_\Omega}}{ \Theta_{2n_\Omega ,n_\Omega}} \right)^2
	\sin^2 \left( \frac{ \Theta_{2n_\Omega, n_\Omega}}{2} \right) \; .
\end{equation}

As mentioned before, the ratio between the ratio $\alpha_{n_\Omega}/ \Theta_{2n_\Omega ,n_\Omega}$ is fixed for a given set of experimental parameters and $n_\Omega$:

Given such a set, equations \eqref{eq:squeezing_mechanism_matrix_elements} and  \eqref{eq:squeezing_mechanism_evolution_parameters} can be used to set a fixed ratio

\begin{equation}
	x_{disp} \equiv \frac{ \Theta_{2n_\Omega ,n_\Omega}}{\alpha_{n_\Omega}}= 2 \left| \frac{a_{2n_\Omega, n_\Omega}}{m_{0, n_\Omega}} \right|
	=
	\frac{5}{18} \frac{1}{\sqrt{N_0}}
	\left(
		\frac{\sqrt{2} L}{\pi n_\Omega \xi}
	\right)^{3/2}
\end{equation}

between the strengths of the displacement and the mode mixing and simplify 

\begin{equation}
	\Delta N_{phon, disp} = 2
	x_{disp}^2 \,
	\sin^2 \left( \frac{x_{disp} \alpha_{n_\Omega}}{2} \right) \; .
\end{equation}

\subsection{Single-mode Squeezing and Mode Mixing}

We start again by picking out the single-mode squeezing part of \eqref{eq:squeezing_mechanism_time_evolution} and the term of the exponent responsible for mixing with the mode that gets squeezed

\begin{equation} \label{eq:implementation_single-squeezing-mixing_unitary}
	U = \mathrm{e}^{
		- \frac{r_{n_\Omega/2}}{2} \left(
			\hat{b}_{n_\Omega/2}^{\dagger 2} - \hat{b}_{n_\Omega/2}^2
		\right)
		+
		\Theta_{3n_\Omega/2, n_\Omega/2}
		\left(
			\hat{b}_{3n_\Omega/2}^\dagger \hat{b}_{n_\Omega/2} - \hat{b}_{n_\Omega/2}^\dagger \hat{b}_{3n_\Omega/2}
		\right)
	} \; .
\end{equation}

While the transformation of the second moments

\begin{equation}
	W= - \frac{r_{n_\Omega/2}}{i}
	\begin{pmatrix}
		0 & 0 & 1 & 0
		\\
		0 & 0 & 0 & 0
		\\
		-1 & 0 & 0 & 0
		\\
		0 & 0 & 0 & 0
	\end{pmatrix}
	- \Theta_{3n_\Omega/2, n_\Omega/2} \: \sigma_z \otimes \sigma_y
\end{equation}

is nonvanishing, both squeezing and mode mixing leave the first moments of a state with $\mathbf{d}=0$ invariant since $\boldsymbol{\gamma}$ is zero. The symplectic transformation $S_U$ is a rather unwieldy expression and will not be displayed here.\footnote{The interested reader is again referred to appendix \ref{chap:app_mathematica_transformation}.} However, as in the previous section, we can take the ratio between the $r_{n_\Omega/2}$ and $\Theta_{3n_\Omega/2, n_\Omega/2}$, which is constant in this instance,

\begin{equation}
	x_{1sq} \equiv \frac{\Theta_{3n_\Omega/2, n_\Omega/2}}{r_{n_\Omega/2}}
	=
	\left|
		\frac{a_{3n_\Omega/2, n_\Omega/2}}{m_{n_\Omega/2, n_\Omega/2}}
	\right| = \frac{5}{4 \sqrt{3}}
\end{equation}

and use it to simplify the final quasiparticle number in the modes $n_\Omega/2$, $3n_\Omega/2$ after the transformation \eqref{eq:implementation_single-squeezing-mixing_unitary} as far as possible to

\begin{IEEEeqnarray}{rCl}
	\IEEEyesnumber \IEEEyessubnumber*
	\langle \hat{N} \rangle_{th, U} &=& \frac{1}{4} \left( \mathrm{Tr} (S_U \sigma_{th} S_U^\dagger) - 4 \right)
	\\
	&=&
	-1 + 2
	\frac{\coth \frac{\beta \hbar \omega}{2}}{1 + 2 \cosh \beta \hbar \omega} \times
	\\
	& &
	\left[
		\sqrt{\frac{3}{13}} \sin \left(
			\sqrt{\frac{13}{3}} \frac{r}{2}
		\right) \sinh r
		+
		\cosh \left(
			\beta \hbar \omega
		\right)
		\left(
			\langle \hat{N} \rangle_{vac, U} + 1
		\right)
	\right]\; ,
	\nonumber
\end{IEEEeqnarray}

where we have dropped the index $n_\Omega/2$ for $r$ and $\omega$ and the number of phonons created by \eqref{eq:implementation_single-squeezing-mixing_unitary} from a vacuum state is

\begin{equation}
	\langle \hat{N} \rangle_{vac, U} = \frac{1}{13}
	\left[
		25 - 12 \cos
		\left(
			\frac{1}{2} \sqrt{\frac{13}{3}} r
		\right)
	\right]
	\cosh r - 1
\end{equation}

\subsection{Two-mode Squeezing and Mode Mixing}

As it was the case for single-mode squeezing, if the first moments of the initial state vanish, they remain zero after the transformation. For an individual squeeze factor $r_{l, n_\Omega-n}$ the entire subspace that has to be considered due to mixing consists of the four modes $(l, n_\Omega-l, 2n_\Omega-l, n_\Omega+l)$ and in this subspace, we have the unitary

\begin{IEEEeqnarray}{rCl}
	U&= \mathrm{exp} \bigg[ &
		- (-1)^{n_\Omega} r_{l, n_\Omega-l}
		\left(
			 \hat{b}_l^\dagger \hat{b}_{n_\Omega-l}^\dagger - \hat{b}_l \hat{b}_{n_\Omega-l}
		\right)
		\nonumber
		\\
		& &
		+
		(-1)^{n_\Omega} \Theta_{n_\Omega +l, l}
		\left(
			 \hat{b}_{n_\Omega +l}^\dagger \hat{b}_{l} - \hat{b}_{l}^\dagger \hat{b}_{n_\Omega +l}
		\right)
		\nonumber
		\\
		& &
		+
		(-1)^{n_\Omega} \Theta_{2n_\Omega -l, n_\Omega-l}
		\left(
			 \hat{b}_{2n_\Omega -l}^\dagger \hat{b}_{n_\Omega-l} - \hat{b}_{n_\Omega-l}^\dagger \hat{b}_{2n_\Omega -l}
		\right)
	\bigg]
\end{IEEEeqnarray}

and with it the $W$ matrix\footnote{Using the shortened notation $r_{l, n_\Omega-l}=r$, $\Theta_{n_\Omega +l, l}=\Theta_1$ and $\Theta_{2n_\Omega -l, n_\Omega-l}=\Theta_2$.}

\begin{equation}
	W= \frac{(-1)^{n_\Omega}}{2i}
	\begin{pmatrix}
		0 & 0 &  \Theta_1 & 0 & 0 & r & 0 & 0
		\\
		0 & 0 & 0 &  \Theta_2 & r & 0 & 0 & 0
		\\
		- \Theta_1 & 0 & 0 & 0 & 0 & 0 & 0 & 0
		\\
		0 & - \Theta_2 & 0 & 0 & 0 & 0 & 0 & 0
		\\
		0 & -r & 0 & 0 & 0 & 0 & - \Theta_1 & 0
		\\
		-r & 0 & 0 & 0 & 0 & 0 & 0 & - \Theta_2
		\\
		0 & 0 & 0 & 0 &  \Theta_1 & 0 & 0 & 0
		\\
		0 & 0 & 0 & 0 & 0 &  \Theta_2 & 0 & 0
	\end{pmatrix} \; .
\end{equation}

Though the author was not able to find a reasonably short analytical expression for the symplectic transformation $S_U=\mathrm{e}^{iKW}$ or the resulting phonon number, both results can be evaluated for a given set of squeeze factors and thermal occupation numbers\footnote{Albeit with exceedingly long computation times for resonance conditions $n_\Omega \gtrsim 41$, see again appendix \ref{chap:app_mathematica_transformation}.} and simplified using the ratios between the squeeze factor and the mode mixing parameters

\begin{equation}
	x_{2sq, 1} \equiv \frac{\Theta_{n_\Omega +l, l}}{r_{l, n_\Omega-l}}
	=
	\left|
		\frac{a_{n_\Omega +l, l}}{m_{l, n_\Omega-l}}
	\right|
	=
	\sqrt{\frac{n-l}{n+l}} \frac{
		(n-2l)^2 (n^2 + 2ln + 2l^2)
	}{
		(n+2l)^2 (n^2 - 2ln + 2l^2)
	}
\end{equation}

and

\begin{equation}
	x_{2sq, 1} \equiv \frac{\Theta_{2n_\Omega -l, n_\Omega-l}}{r_{l, n_\Omega-l}}
	=
	\left|
		\frac{a_{2n_\Omega -l, n_\Omega-l}}{m_{l, n_\Omega-l}}
	\right|
	=
	\sqrt{\frac{l}{2n-l}} \frac{
		(n-2l)^2 (5n^2 - 6ln + 2l^2)
	}{
		(3n-2l)^2 (n^2 - 2ln + 2l^2)
	} \, .
\end{equation}

For the full transformation, $r_{n_\Omega/2}$ and $\alpha_{n_\Omega}$ were expressed in terms of the two-mode squeeze factor as

\begin{IEEEeqnarray}{rCl}
	\IEEEyesnumber \IEEEyessubnumber*
	r_{n_\Omega/2} &=&
	\frac{m_{n_\Omega/2, n_\Omega/2}}{m_{l, n_\Omega-l}} r_{l, n_\Omega-l}
	=
	\frac{\sqrt{l (n_\Omega-l)} (n_\Omega-2l)^2}{n_\Omega (n_\Omega^2 - 2 l n_\Omega + 2l^2)} r_{l, n_\Omega-l}
	\\
	\alpha_{n_\Omega} &=& \frac{m_{0, n_\Omega}}{m_{l, n_\Omega-l}} r_{l, n_\Omega-l}
	=
	\sqrt{\frac{N_0 \xi^3 \pi^3}{\sqrt{2} L^3}} \frac{\sqrt{l n_\Omega (n_\Omega-l)} (n_\Omega-2l)^2}{(n_\Omega^2 - 2 l n_\Omega - 2l^2)} r_{l, n_\Omega-l}
\end{IEEEeqnarray}

\section{Quality of the Final State \label{sec:quality}}

The initial state considered in the GW detector proposal \cite{sabin_phonon_2014} from chapter \ref{sec:detector_proposal} is a two-mode squeezed vacuum state. The result of the mechanism we are considering, however, is a thermal state that was subjected to squeezing and mode mixing. In order to quantify its value for quantum metrology applications that assume a squeezed vacuum, it is therefore necessary to not only provide the maximum squeeze factor that is allowed for by the phonon budget and the highest possible perturbation, but one also has to consider the overlap between the final state and the ideal one of the squeezed vacuum. As a measure for this overlap, we can make use of the Uhlmann fidelity once again.

It then quickly becomes apparent that a quantity decisive for the quality of the final state is the thermal occupation number before the transformation, which we will express by $\beta \hbar \omega_1/2$.\footnote{
	$\omega_1$ is the first phonon frequency and the factor $1/2$ is for convenience because of the CM elements of the thermal state, see \eqref{eq:CM_formalism_thermal_state_CM}. We keep considering a box trap where $\omega_n = n \omega_1$.
} For a first impression, one can calculate the fidelity between a squeezed vacuum and a squeezed thermal state in the first two phonon modes with equal squeeze factors to find

\begin{equation}
	\mathcal{F} (\sigma_{sq2}, \sigma_{therm, sq}) = 4 \mathrm{e}^{-3 \beta \hbar \omega_1/2} \ \sinh \left( \frac{\beta \hbar \omega_1}{2} \right) \sinh \left( \beta \hbar \omega_1 \right) \; .
\end{equation}

\begin{figure}[phbt]
	\centering
	\begin{subfigure}[b]{0.3\textwidth}
		\centering
		\includegraphics[width=\textwidth]{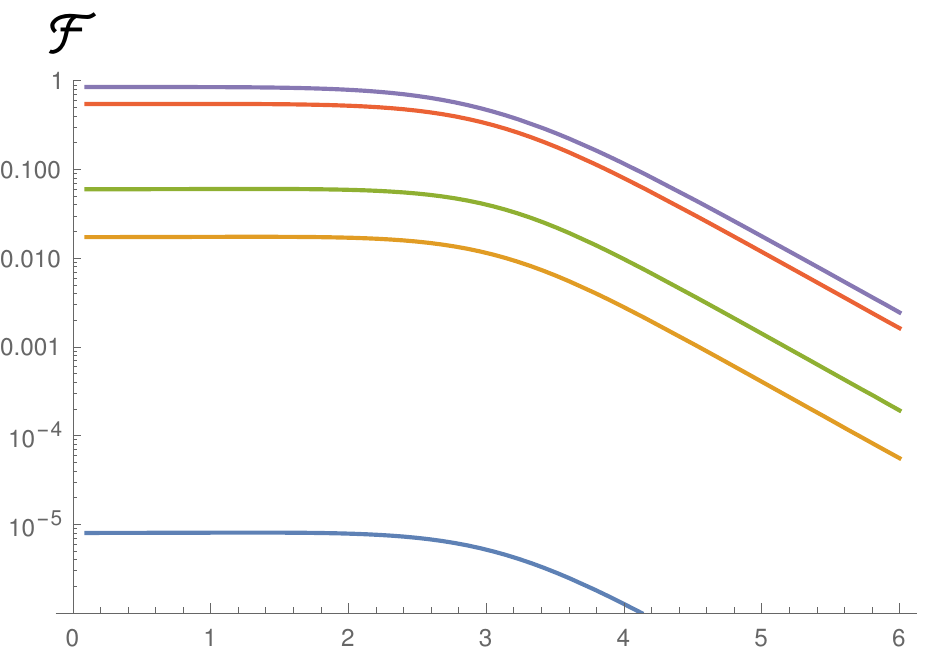}
		\caption{}
	\end{subfigure}
	\hfill
	\begin{subfigure}[b]{0.3\textwidth}
		\centering
		\includegraphics[width=\textwidth]{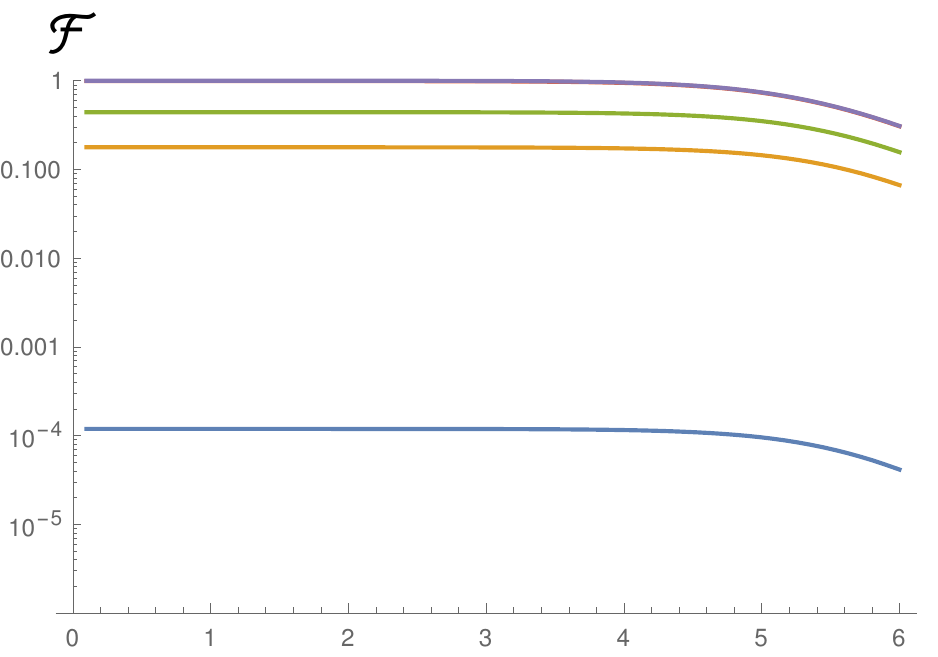}
		\caption{}
	\end{subfigure}
	\hfill
	\begin{subfigure}[b]{0.38\textwidth}
		\centering
		\includegraphics[width=\textwidth]{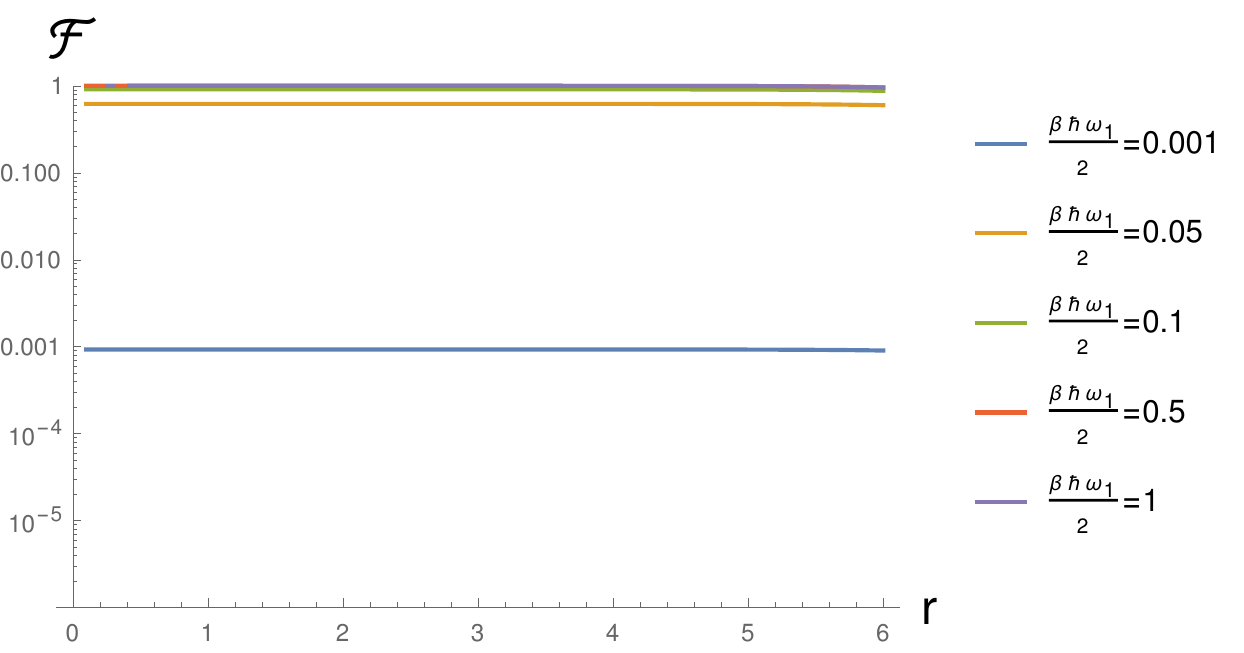}
		\caption{}
	\end{subfigure}
	\\
	\begin{subfigure}[b]{0.3\textwidth}
		\centering
		\includegraphics[width=\textwidth]{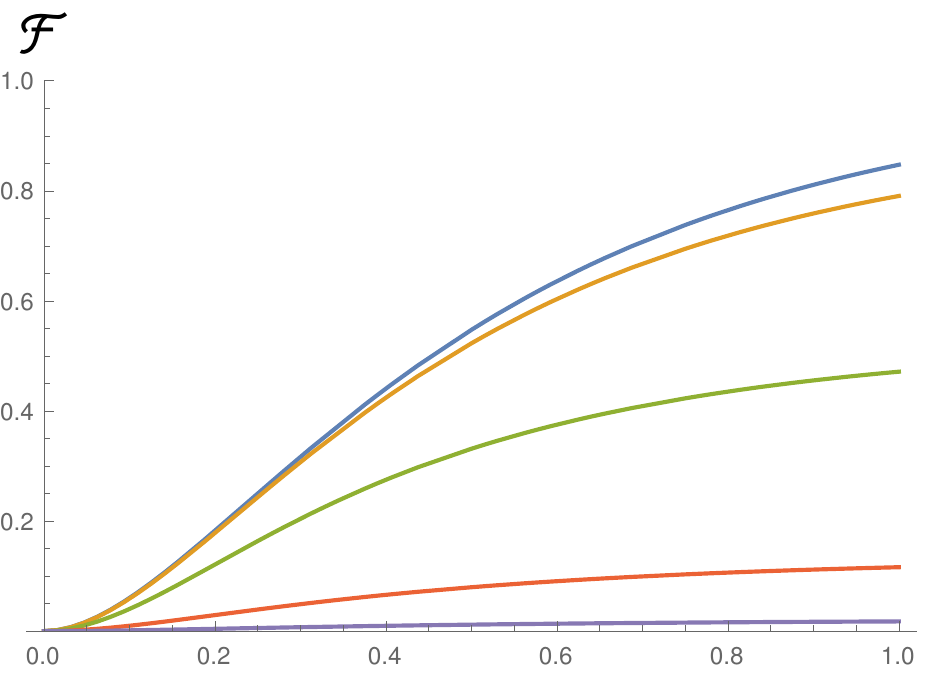}
		\caption{}
	\end{subfigure}
	\hfill
	\begin{subfigure}[b]{0.3\textwidth}
		\centering
		\includegraphics[width=\textwidth]{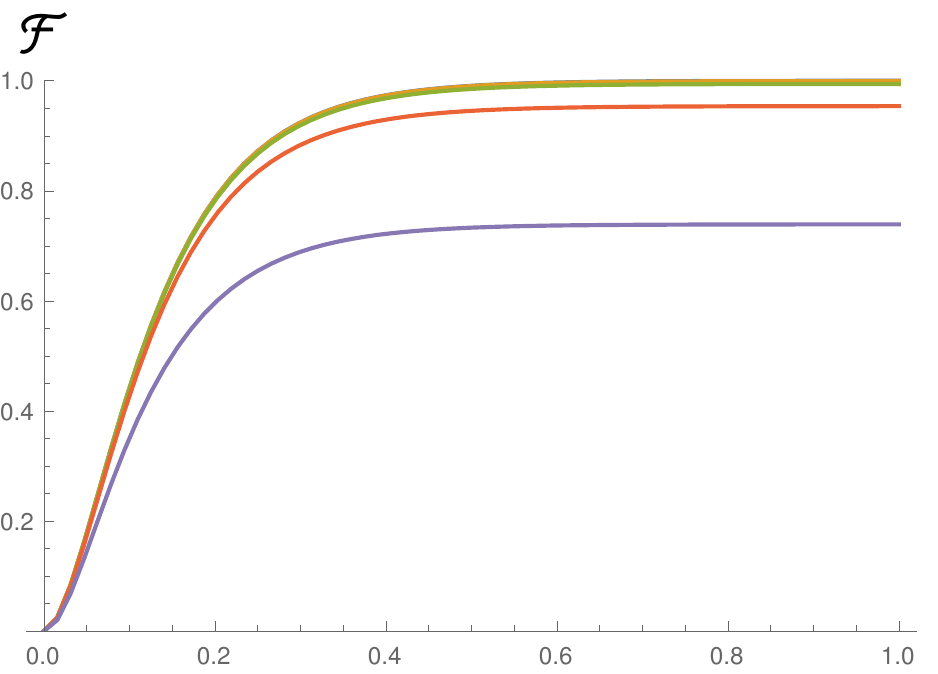}
		\caption{}
		\label{fig:quality_plots_nOm_11_2}
	\end{subfigure}
	\hfill
	\begin{subfigure}[b]{0.38\textwidth}
		\centering
		\includegraphics[width=\textwidth]{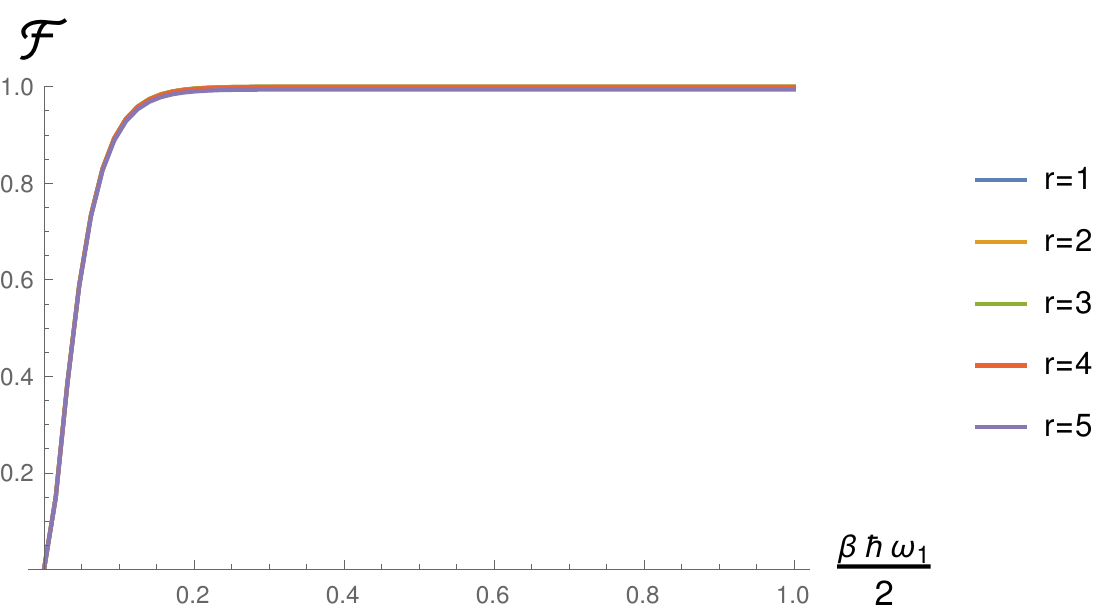}
		\caption{}
	\end{subfigure}
	\\
	\begin{subfigure}[b]{0.3\textwidth}
		\centering
		\includegraphics[width=\textwidth]{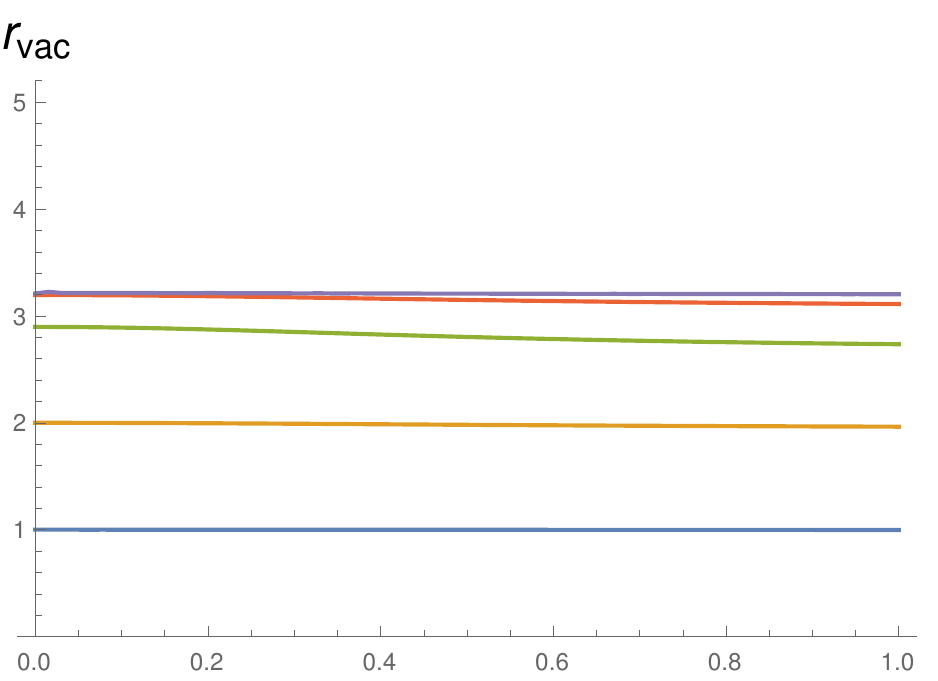}
		\caption{}
	\end{subfigure}
	\hfill
	\begin{subfigure}[b]{0.3\textwidth}
		\centering
		\includegraphics[width=\textwidth]{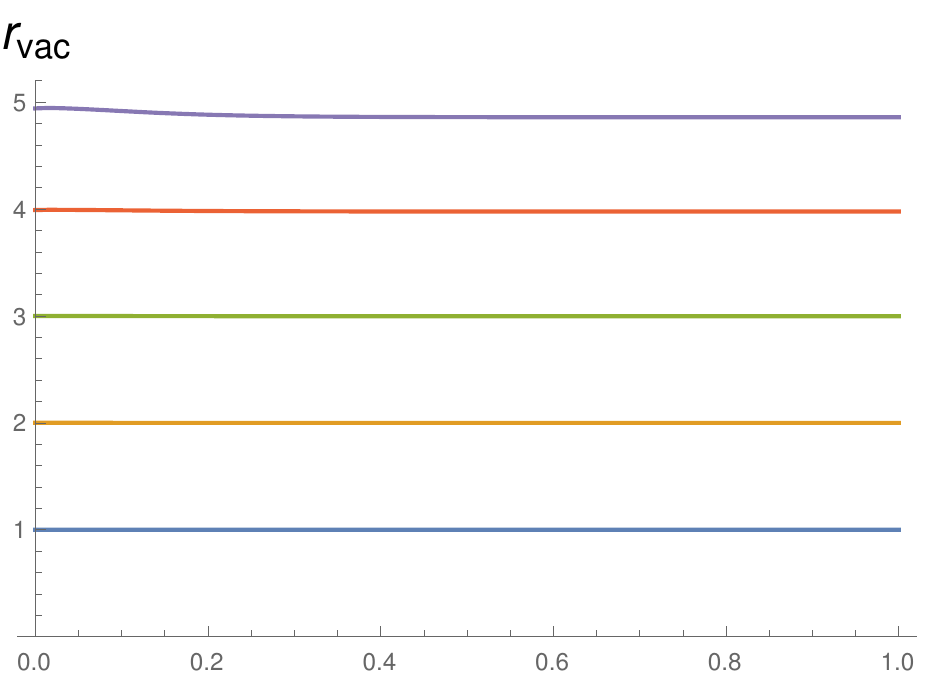}
		\caption{}
	\end{subfigure}
	\hfill
	\begin{subfigure}[b]{0.38\textwidth}
		\centering
		\includegraphics[width=\textwidth]{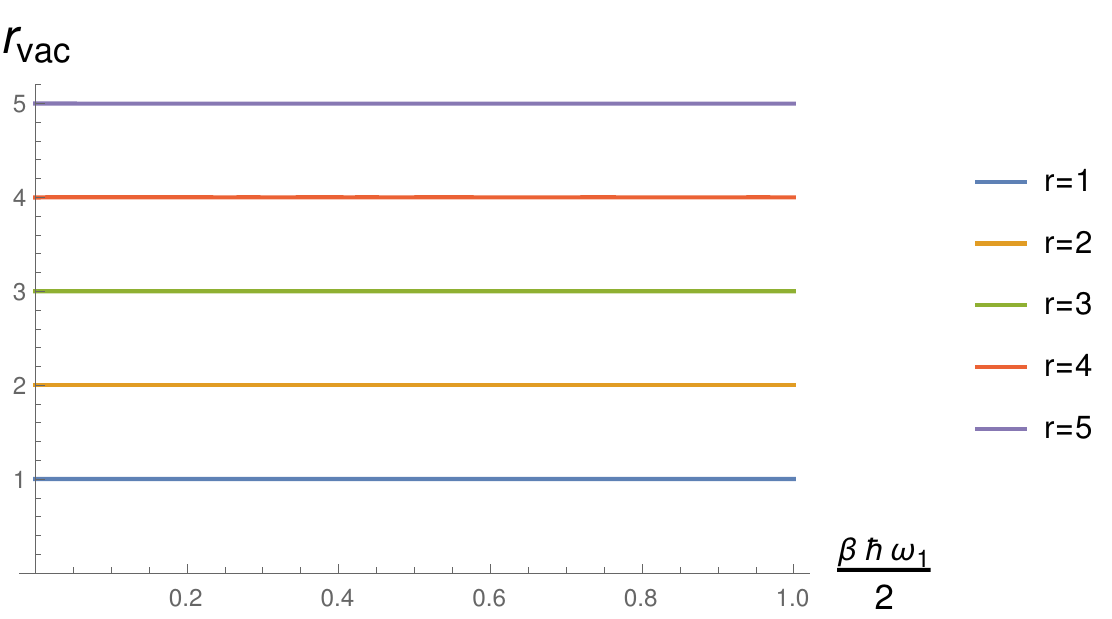}
		\caption{}
	\end{subfigure}
	\\
	\caption{
		{\footnotesize	On the quality of the final state in the balanced mode pair after the full symplectic transformation of the squeezing scheme. The three columns correspond to the three resonance conditions $n_\Omega=3$, $n_\Omega=11$ and $n_\Omega=31$ and the rows are
		\\
		-- the Uhlmann fidelity between the final state and the nearest squeezed vacuum state as a function of the two-mode squeeze factor $r_{l, n_\Omega-l}$ of the transformation for various frequency-temperature ratios $\beta \hbar \omega_1/2$
		\\
		-- the Uhlmann fidelity as a function of the frequency-temperature ratio over a range of squeeze factors $r_{l, n_\Omega-l}$
		\\
		-- the squeeze factor of the nearest vacuum state as a function of the frequency-temperature ratio over a range of squeeze factors. $r_{l, n_\Omega-l}$
		\\
		Note how at low resonances, $r_{vac}$ reaches an asymptotic limit that is seemingly independent of both the squeeze factor of the transformation and $\beta \hbar \omega_1/2$. From the second row and particularly (e), we can see that in the temperature-frequency range $\beta \hbar \omega_1/2 \approx 0.22$ of the experiment proposed in section \ref{sec:proposed_experiment}, squeezing with factors $r \gtrsim 4$ becomes efficient only for higher-lying resonances $n_\Omega \gtrsim 11$.
		\label{fig:quality_plots}
		}
	}
\end{figure}

Owing to the rather involved nature of the state transformation, an expression for the fidelity between the actual final state and the squeezed vacuum is not quite as easy to find. Nonetheless, we can maximize \linebreak $\mathcal{F} \left( \sigma_{out} (r_{l, n_\Omega-l}, \beta \hbar \omega_1/2), \, \sigma_{sq2} (r_{vac}) \right)$ numerically over $r_{vac}$ in order to make some qualitative observations. For this purpose, see figure \ref{fig:quality_plots}. One such observation is that while the parameter $r_{l, n_\Omega-l}$ of the state transformation might increase, the corresponding squeeze factor of the nearest squeezed vacuum seems to asymptotically reach a maximum even for large $r_{l, n_\Omega-l}$ at low resonance conditions. Furthermore, at low-lying resonances, the reachable fidelity not only strongly depends on $\beta \hbar \omega_1/2$, but also on the squeeze factor. This is most probably a consequence of the imbalanced initial population of the modes in the thermal state. Two-mode squeezing, particularly strong squeezing with $r_{l, n_\Omega-l} \gtrsim 2$ thus becomes efficient only for higher mode pairs.

\section{Some Guidelines for Optimization}

The complicated nature of the state transformation and with it the long computation times of final states and their properties in connection with the vast parameter space makes it difficult to optimize over experimental parameters and modes to squeeze in. Nonetheless, we can make some qualitative observations on which degrees of freedom could be utilized to maximize the allowed squeeze factor and the quality of the final state.

At the same time, we must keep in mind the limitations to the phonon and condensate life times, posed by three-body losses\footnote{In order to keep expressions simple, we assume for now to be far from possible Efimov resonances.} \eqref{eq:three-body_half-life}, Landau damping \eqref{eq:landau_damping_limits} and Beliaev damping \eqref{eq:zero_temperature_beliaev_damping_rate}. Requiring reasonable phononic life times could then be used to bound the parameter space in which the final states can be optimized.

Finally, we have to remind ourselves of the assumptions that we have made along the way and that have to be fulfilled in order for the derivations so far to be applicable: in section \ref{sec:uniform_weakly_interacting}, we assumed to be dealing with a \textit{dilute and weakly interacting} Bose gas, where $|a_s| n_0^{1/3} \ll 1$. Further on, we have considered the \textit{ultracold regime} $k_B T \ll \mu$ and lastly, the quasi-particles were treated as phonons, implying $k_n \xi \ll 1$.

To gain some more intuition on how an experiment could be improved for squeezing, let us consider the cases where one of the possible causes of phonon decoherence becomes dominant and derive the resulting maximum squeeze factors \eqref{eq:squeezing_mechanism_evolution_parameters} from the naive approach \ref{sec:naive_approach} in each case as well as the conditions listed above. For a better overview, these quantities are expressed in terms of typical characteristics that are given when describing an experiment in table \ref{table:optimization}.

Low damping rates and low thermal occupation numbers call for high particle densities, which are, on the other hand, bounded from above by the diluteness condition and three-body losses.

High mode numbers $l$ also have the advantage of low quasi-particle numbers in the initial state and thus the quality of the state after the transformation. They are, however, limited by their shorter life time as well as the condition that they can be approximated as phonons.

Assuming that the scattering length can be tuned using a Feshbach resonance and in the absence of an Efimov resonance, increasing $a_s$ will lead to drastically enhanced three-body losses, which scale as $a_s^4$. A compromise must therefore be found when maximizing $a_s$ to reduce the thermal population .

An increase in the trap length $L$ will lead to higher squeeze factors permitted by the naive approach, but at the same time it reduces the phonon frequencies and thereby lowers the energy levels of the quasi-particle state. This means, again, that occupation numbers in the initial state are increased, lowering the amount of phonons we are allowed to create and the fidelity of the final state.

The temperature, finally, is simple: improved cooling is beneficial in every single aspect considered in table \ref{table:optimization}.

\begin{table}[hbtp]
\centering
\resizebox{\textwidth}{!}{
	\begin{tabular}{c|c|c|c|c}
		\rule[-2.5ex]{0pt}{7ex}
		&
		$(\gamma^{La}_{k_B T \ll \mu})^{-1}$ \eqref{eq:landau_damping_limits}
		&
		$(\gamma^{La}_{k_B T \gg \mu})^{-1}$ \eqref{eq:landau_damping_limits}
		&
		$(\gamma^{Be, 0})^{-1}$ \eqref{eq:zero_temperature_beliaev_damping_rate}
		&
		$t_{1/2}$ \eqref{eq:three-body_half-life}
		\\
		\hline
		\hline
		\rule{0pt}{3ex}
		$t \approx$
		\rule[-2ex]{0pt}{1ex}
		&
		$\frac{640}{3 \pi} \frac{\hbar^7 \beta^4 L n_0^3 a_s^2}{l m^3}$
		&
		$\frac{8}{\pi} \frac{\beta L}{a_s l}$
		&
		$\frac{640}{3 \pi^6} \frac{m L^5 n_0}{\hbar l^5}$
		&
		$\frac{1}{50} \frac{m}{\hbar n_0^2 a_s^4}$
		\\
		\hline
		\rule{0pt}{3ex}
		$r \propto$
		\rule[-2ex]{0pt}{1ex}
		&
		$\frac{\hbar^8 \beta^4 L^2 n_0^{9/2} a_s^{7/2}}{l^2 m^4}$
		&
		$\frac{\hbar^2 L^2 \beta n_0^{3/2} a_s^{1/2}}{l^2 m}$
		&
		$\frac{L^6 n_0^{5/2} a_s^{3/2}}{l^6}$
		&
		$\frac{L}{l n_0^{1/2} a_s^{5/2}}$
		\\
		\hline
		\hline
		\rule{0pt}{3ex}
		$\beta \omega_l$
		\rule[-2ex]{0pt}{1ex}
		&
		\multicolumn{2}{c||}{
			$2 \sqrt{\pi} \frac{\hbar l \sqrt{n_0 a_s}}{L m}$
		}
		&
		\specialcell{weak \\ interactions}
		&
		$|a_s| n_0^{1/3} \ll 1$
		\\
		\hline
		\hline
		\specialcell{phonon \\ regime}
		&
		\multicolumn{2}{c||}{
			$\sqrt{\frac{\pi l^2}{L^2 n_0 a_s}} \ll 1$
		}
		&
		\specialcell{ultracold \\ regime}
		&
		$\frac{1}{4 \pi} \frac{m}{\hbar^2 \beta \, n_0 \, a_s} \ll 1$
	\end{tabular}
}
\caption{
	\footnotesize Approximate values for the limitations to the phonon and condensate life times and resulting proportionalities of the maximum two-mode squeeze factor from SECTION \ref{sec:naive_approach} in the case where the respective damping or particle losses become dominant.
	\\
	The two bottom rows show a measure $\beta \omega_l$ for the thermal occupation of the initial state, which should be minimal, and restrictions from the diluteness condition and the assumptions of the phononic and the ultracold regime.
	\\
	To see where one could optimize and which boundaries will be encountered, all quantities above are given in terms of the parameters characterizing an individual experiment. \label{table:optimization}
	}
\end{table}

\section{Existing Experiments \label{sec:existing_experiments}}

In face of the sheer number of BEC experiments around the globe, the task of picking a representative selection to examine with respect to the possible application of the squeezing scheme presented was not quite easy. The choice was made with the intention of giving an overview of various species, where each particular one of the experiments listed below has a distinguishing property, be it the measurement of Efimov parameters \cite{pollack_universality_2009, berninger_universality_2011}, extremely low absolute temperatures \cite{leanhardt_cooling_2003} or high particle numbers \cite{van_der_stam_large_2007} or because they were, in fact, performed in box traps \cite{gaunt_bose-einstein_2013}. For experiments using harmonic potentials, a condensate of comparable size in a cylindrical hard-wall potential was assumed by deriving the length and diameter from the Thomas-Fermi radius \eqref{eq:thomas-fermi_radius}.

After the investigation of the selected experiments for their potential for phonon squeezing, it became clear that they were not designed with our method in mind, albeit for varying reasons. The following paragraphs provide a brief description of each condensate, including the (approximate) numbers most decisive for their squeezability:

\begin{itemize}
	\item The temperature $T$
	\item The initial particle number  $N_0$
	\item The initial number of thermal phonons $N_{phon, i}$
	\item A resonance $n_\Omega$ whose balanced pair (see section \ref{sec:bounds_maximum_phonon_number}) has reasonable life times $t_{life}$ while still being in the phonon regime
	\item The highest possible corresponding two-mode squeeze factor $r_{l_{max}, n_\Omega -l_{max}}$, which is the lower one of the squeeze factors permitted for by the naive ansatz and the phonon budget.
	\item The Uhlmann fidelity $\mathcal{F}$ of the closest squeezed vacuum state and its squeeze factor $r_{vac}$
\end{itemize}

$N_{phon, i}$ was calculated using \eqref{eq:thermal_state_infinite_mode_particle_number}, $t_{life}$ by means of \eqref{eq:phonon_life_time_overall}, where no low or high temperature limits were taken and the integrals for the Landau and Beliaev damping rates were computed numerically. Computation of $\mathcal{F}$ was done by transforming the initial thermal state, dropping all subspaces except the one with the highest two-mode squeeze factor and numerically maximizing $\mathcal{F}$ between the final and a squeezed vacuum state over its squeeze factor $r_{vac}$.

\paragraph{Westbrook Group $^4$He$^*$ Experiment (table \ref{table_westbrook_4He_experiment})}

The 2012 experiment by Jaskula et al. \cite{jaskula_acoustic_2012} with metastable $^4$He$^*$ has some similarities with the mechanism presented in this thesis: the stiffness of the harmonic trapping potential was modulated to produce pairs of squeezed phonons (although they are, in the spirit of chapter \ref{sec:squeezed_states}, referred to as twin beam states). In the axial direction, the potential is very shallow (axial and radial trapping frequencies are $7$ Hz and $1500$ Hz), which, in combination with the relatively high temperature and low particle number, takes it far from the ultracold regime and makes it an improper candidate. As a consequence, the number of thermal phonons before attempting to squeeze would already exceed our bound of $N_0/50$ for  a similarly dimensioned box cavity.


\paragraph{Hulet Group $^7$Li Experiment (table \ref{table:hulet_Li_experiment})}

The setup used in \cite{pollack_universality_2009} was chosen as an example primarily because of the completeness of the data given in the publication and its measurement of Efimov parameters by (for positive scattering lengths) observing local minima of three-body losses. For the computation of the values in  table \ref{table:hulet_Li_experiment}, $a_s$ was assumed to be tuned to the first minimum of the three-body recombination rate at $\approx 117 a_0$, with $a_0$ being the Boltzmann radius. The elongation of the potential in question conforms well with the assumptions made in section \ref{sec:squeezing mechanism}. Because the temperature in \cite{pollack_universality_2009} was given only in terms of a fraction of the critical temperature, the formula derived by Giorgini et al. \cite{giorgini_condensate_1996} for the critical temperature in interacting Bose gases was applied to derive the absolute temperature.

\paragraph{Ketterle Group $^{23}$Na Experiment (table \ref{table:ketterle_Na_experiment})}

On the search for extremely cold Bose gases, the 2003 experiment by Wolfgang Ketterle's group \cite{leanhardt_cooling_2003} was brought to the author's attention, but it does come with some caveats: while the absolute temperature reached was a record setting $450$pK, the trap was also very shallow and pancake-shaped. This means that the state before the transformation was highly populated and transverse phonon modes would have to be accounted for, which was not done in this thesis. In addition, even at this low temperature, $k_B T/\mu \approx 30$.


\paragraph{van der Straten Group $^{23}$Na Experiment (table \ref{table:van_der_straten_Na_experiment})}

In the early phase of discussing this thesis, $^{23}$Na presented itself as a promising candidate for squeezing due to the large particle numbers that have been achieved, see for example \cite{van_der_stam_large_2007}. The data that the results in table \ref{table:van_der_straten_Na_experiment} are based on were kindly provided in personal correspondence by Peter van der Straten. The extensive size of the condensate and the low, non-tunable scattering length keep collective excitations well in the phonon regime, but their life time is strongly limited by high temperature Landau damping and thermal population numbers do not allow for much squeezing.


\paragraph{Hadzibabic Group $^{39}$K Experiment (table \ref{table_hadzibabic_39K_experiment})}

The $^{39}$K and $^{87}$Rb setups (of which only the former is still in use) located at the University of Cambridge are two of the rare examples of BECs in a uniform potential and thus of particular interest for this thesis. Given the parameters provided in \cite{gaunt_bose-einstein_2013}, they are, however, not well suited for our purposes: the comparably low particle density $n_0 \approx 10^{12} \, \mathrm{cm}^{-3}$ increases the healing length and thus pushes quasi-particles out of the phononic regime at a test scattering length $a_s = 100 a_0$. Increasing the interaction strength, on the other hand, would further reduce phonon life times.

\paragraph{Hadzibabic Group $^{87}$Rb Experiment (table \ref{table_hadzibabic_87Rb_experiment})}

Since the early days of BECs, $^{87}$Rb has been the workhorse among the atomic species. This box trap setup, though, is now discontinued. In the light of our intentions, it is far from the ultracold limit at $k_B T/\mu \approx 13$ and suffers from short phonon life times, which are again, a consequence of Landau damping and prevent us from going beyond $n_\Omega=3$. 


\paragraph{Grimm Group $^{133}$Cs Experiment (table \ref{table:grimm_133Cs_experiment})}

The Innsbruck $^{133}$Cs BEC \cite{kraemer_evidence_2006, berninger_universality_2011}, like the Houston $^7$Li one \cite{pollack_universality_2009}, was used successfully to measure Efimov parameters. In \cite{kraemer_evidence_2006}, only an upper bound for $\eta_+<2$ could be found for the decay parameter at positive scattering lengths. In the calculations made here, we will follow the argument made in the paper cited and take the theoretical assumption from \cite{braaten_three-body_2001} that $\eta_+ = \eta_-$ as a rough estimate. Dips in $C_3$ \eqref{eq:three-body_a^4_scaling} are then not pronounced enough to result in local minima of the three-body loss rate. Nonetheless, the scattering length is assumed to be tuned to an Efimov resonance at $a_s = 210 a_0$.

%

\begin{table}[p!hbt]
\centering
\resizebox{0.5\textwidth}{!}{
	\begin{tabular}{c|c|c|c}
		$T \, [ \mathrm{nK} ]$
		\rule[-2.5ex]{0pt}{7ex}
		&
		$N_0$
		&
		$N_{phon, i}$
		&
		$r_{l_{max}, n_\Omega -l_{max}}$
		\\
		\hline
		\rule{0pt}{3ex}
		$200$
		\rule[-2ex]{0pt}{1ex}
		&
		$10^5$
		&
		$7800$
		&
		$0$
		\\
		\hline
		\hline
		\rule{0pt}{3ex}
		$n_\Omega$
		\rule[-2ex]{0pt}{1ex}
		&
		$t_{life} \, [\mathrm{s}]$
		&
		$\mathcal{F}$
		&
		$r_{vac}$
		\\
		\hline
		\rule{0pt}{3ex}
		$3$
		\rule[-2ex]{0pt}{1ex}
		&
		$ 1$
		&
		$-$
		&
		$-$
	\end{tabular}
}
\caption{\footnotesize Westbrook Group $^4$He$^*$ Experiment: even very high quasi-particle modes can still be considered phonons here, but high temperature Landau damping with its $l^{-1}$ dependence already limits the life time of the lowest mode to $\approx 2$s. The large value of $N_{phon, i}$ does not allow for the creation of additional phonons within the bounds we have set.
	\label{table_westbrook_4He_experiment}}
\end{table}

\begin{table}[p!hbt]
\centering
\resizebox{0.5\textwidth}{!}{
	\begin{tabular}{c|c|c|c}
		$T \, [ \mathrm{nK} ]$
		\rule[-2.5ex]{0pt}{7ex}
		&
		$N_0$
		&
		$N_{phon, i}$
		&
		$r_{l_{max}, n_\Omega -l_{max}}$
		\\
		\hline
		\rule{0pt}{3ex}
		$ 22$
		\rule[-2ex]{0pt}{1ex}
		&
		$3.6 \cdot 10^5$
		&
		$ 220$
		&
		$ 2.9$
		\\
		\hline
		\hline
		\rule{0pt}{3ex}
		$n_\Omega$
		\rule[-2ex]{0pt}{1ex}
		&
		$t_{life} \, [\mathrm{s}]$
		&
		$\mathcal{F}$
		&
		$r_{vac}$
		\\
		\hline
		\rule{0pt}{3ex}
		3
		\rule[-2ex]{0pt}{1ex}
		&
		$ 3$
		&
		$0.002$
		&
		$ 2.9$
	\end{tabular}
}
\caption{\footnotesize Hulet Group $^7$Li Experiment: 
	the gas is not in the ultracold regime with $k_B T/\mu \approx 0.96$. The low fidelity is due to the initial thermal population. Squeezing in higher modes is not an option due to the shorter phonon life times, which are limited by Landau damping. Both of these issues could potentially be resolved by tightening the trap in the axial direction.
	\label{table:hulet_Li_experiment}}
\end{table}

\begin{table}[p!hbt]
\centering
\resizebox{0.5\textwidth}{!}{
	\begin{tabular}{c|c|c|c}
		$T \, [ \mathrm{nK} ]$
		\rule[-2.5ex]{0pt}{7ex}
		&
		$N_0$
		&
		$N_{phon, i}$
		&
		$r_{l_{max}, n_\Omega -l_{max}}$
		\\
		\hline
		\rule{0pt}{3ex}
		$0.45$
		\rule[-2ex]{0pt}{1ex}
		&
		2500
		&
		84
		&
		$-$
		\\
		\hline
		\hline
		\rule{0pt}{3ex}
		$n_\Omega$
		\rule[-2ex]{0pt}{1ex}
		&
		$t_{life} \, [\mathrm{s}]$
		&
		$\mathcal{F}$
		&
		$r_{vac}$
		\\
		\hline
		\rule{0pt}{3ex}
		$-$
		\rule[-2ex]{0pt}{1ex}
		&
		$-$
		&
		$-$
		&
		$-$
	\end{tabular}
}
\caption{\footnotesize Ketterle Group $^{23}$Na Experiment: 
	the quasi-particles have reasonably long life times, but their large wave numbers and healing length of the condensate place even the lowest modes outside the phonon regime. The small number of atoms in the condensate does not allow for the additional creation of quasi-particles.
	\label{table:ketterle_Na_experiment}}
\end{table}

\begin{table}[p!hbt]
\centering
\resizebox{0.5\textwidth}{!}{
	\begin{tabular}{c|c|c|c}
		$T \, [ \mathrm{nK} ]$
		\rule[-2.5ex]{0pt}{7ex}
		&
		$N_0$
		&
		$N_{phon, i}$
		&
		$r_{l_{max}, n_\Omega -l_{max}}$
		\\
		\hline
		\rule{0pt}{3ex}
		$250$
		\rule[-2ex]{0pt}{1ex}
		&
		$10^7$
		&
		$10^5$
		&
		$1.2$
		\\
		\hline
		\hline
		\rule{0pt}{3ex}
		$n_\Omega$
		\rule[-2ex]{0pt}{1ex}
		&
		$t_{life} \, [\mathrm{s}]$
		&
		$\mathcal{F}$
		&
		$r_{vac}$
		\\
		\hline
		\rule{0pt}{3ex}
		3
		\rule[-2ex]{0pt}{1ex}
		&
		$1.4$
		&
		$10^{-8}$
		&
		$0.1$
	\end{tabular}
}
\caption{\footnotesize van der Straten Group $^{23}$Na Experiment: 
	due to the high temperature, there is not much room for squeezing and the final state is far from the ideal of a squeezed vacuum. Targeting higher modes is not an option due to Landau damping.
	\label{table:van_der_straten_Na_experiment}}
\end{table}

\begin{table}[p!hbt]
\centering
\resizebox{0.5\textwidth}{!}{
	\begin{tabular}{c|c|c|c}
		$T \, [ \mathrm{nK} ]$
		\rule[-2.5ex]{0pt}{7ex}
		&
		$N_0$
		&
		$N_{phon, i}$
		&
		$r_{l_{max}, n_\Omega -l_{max}}$
		\\
		\hline
		\rule{0pt}{3ex}
		$10$
		\rule[-2ex]{0pt}{1ex}
		&
		$4 \cdot 10^5$
		&
		$110$
		&
		$2.5$
		\\
		\hline
		\hline
		\rule{0pt}{3ex}
		$n_\Omega$
		\rule[-2ex]{0pt}{1ex}
		&
		$t_{life} \, [\mathrm{s}]$
		&
		$\mathcal{F}$
		&
		$r_{vac}$
		\\
		\hline
		\rule{0pt}{3ex}
		$3$
		\rule[-2ex]{0pt}{1ex}
		&
		$0.7$
		&
		$2 \cdot 10^{-3}$
		&
		$0.8$
	\end{tabular}
}
\caption{\footnotesize Hadzibabic Group $^{39}$K Experiment: 
	the values above were computed assuming collective excitations can still be treated as phonons. Low life times keep us from squeezing in higher modes. Improved cooling and higher particle densities could help cure this problem. The room for improvement is limited by three-body losses.
	\label{table_hadzibabic_39K_experiment}}
\end{table}

\begin{table}[p!hbt]
\centering
\resizebox{0.5\textwidth}{!}{
	\begin{tabular}{c|c|c|c}
		$T \, [ \mathrm{nK} ]$
		\rule[-2.5ex]{0pt}{7ex}
		&
		$N_0$
		&
		$N_{phon, i}$
		&
		$r_{l_{max}, n_\Omega -l_{max}}$
		\\
		\hline
		\rule{0pt}{3ex}
		$45$
		\rule[-2ex]{0pt}{1ex}
		&
		$4 \cdot 10^5$
		&
		$1200$
		&
		$1.9$
		\\
		\hline
		\hline
		\rule{0pt}{3ex}
		$n_\Omega$
		\rule[-2ex]{0pt}{1ex}
		&
		$t_{life} \, [\mathrm{s}]$
		&
		$\mathcal{F}$
		&
		$r_{vac}$
		\\
		\hline
		\rule{0pt}{3ex}
		$3$
		\rule[-2ex]{0pt}{1ex}
		&
		$0.3$
		&
		$4 \cdot 10^{-5}$
		&
		$1.9$
	\end{tabular}
}
\caption{\footnotesize Hadzibabic Group $^{87}$Rb Experiment:
	potential for optimization for the purposes of our squeezing mechanism lies in increasing the particle density and possibly the trap length, mostly though in lowering the temperature.
	\label{table_hadzibabic_87Rb_experiment}}
\end{table}

\begin{table}[p!hbt]
\centering
\resizebox{0.5\textwidth}{!}{
	\begin{tabular}{c|c|c|c}
		$T \, [ \mathrm{nK} ]$
		\rule[-2.5ex]{0pt}{7ex}
		&
		$N_0$
		&
		$N_{phon, i}$
		&
		$r_{l_{max}, n_\Omega -l_{max}}$
		\\
		\hline
		\rule{0pt}{3ex}
		$10$
		\rule[-2ex]{0pt}{1ex}
		&
		$10^5$
		&
		$400$
		&
		$1.9$
		\\
		\hline
		\hline
		\rule{0pt}{3ex}
		$n_\Omega$
		\rule[-2ex]{0pt}{1ex}
		&
		$t_{life} \, [\mathrm{s}]$
		&
		$\mathcal{F}$
		&
		$r_{vac}$
		\\
		\hline
		\rule{0pt}{3ex}
		$11$
		\rule[-2ex]{0pt}{1ex}
		&
		$0.3$
		&
		$0.04$
		&
		$1.9$
	\end{tabular}
}
\caption{\footnotesize Grimm Group $^{133}$Cs Experiment: in contrast to the experiments above, $r_{l_{max}, n_\Omega -l_{max}}$ is bounded by the naive ansatz rather than the phonon budget in this instance. Phonons soon fall victim to Landau damping. Choosing $n_\Omega=3$ instead would bring life times up to $\approx 1$s, but at the same time decrease the fidelity by more than an order of magnitude.
	\label{table:grimm_133Cs_experiment}}
\end{table}

\newpage

\section{Proposal: a Uniform $^7$Li BEC \label{sec:proposed_experiment}}

Clearly, experiments similar to the ones listed above leave little chance to produce states that come close to our goals of reasonable life times, high squeeze factors and good overlap with a squeezed vacuum state. This poses the question what a setup that can achieve these goals could look like and brings us back to the considerations on optimization and constraints from table \ref{table:optimization}.

As far as the atomic species is concerned, there are two features of $^7$Li that make it a promising candidate: its low mass, which allows for higher phonon frequencies, and the fact that the Efimov dip in $L_3$ at $a_s \approx 117 a_0$\footnote{
	Do not let the fact that $a_+$ was measured to be $\approx 119 a_0$ confuse you. $a_+$ is the scattering length at which $C_3$ is minimized, but due to the additional factor $a_s^4$ in the rate coefficient $L_3$, the actual minimum of the three-body loss rate is at slightly lower scattering lengths.}
with a decay parameter $\eta_+ \approx 0.008$ measured in \cite{pollack_universality_2009} can be exploited to achieve high densities without suffering from excessive three-body decay.

Because the balanced mode pair is more balanced the closer the modes are and also in order to avoid producing unwanted phonons by single-mode squeezing, we will restrict ourselves to targeting only odd $n_\Omega$.

The crucial quantity, as before, will be the temperature of the condensate. We will assume the low, but reasonable value of $T=10$nK. If we then imagine the same number $N_0 \approx 3.6 \cdot 10^5$ of atoms that was quoted in \cite{pollack_universality_2009} in an elongated cylindrical box trap with length $L= 150 \, \mu$m and radius $r = 5 \, \mu$m, all the conditions in table \ref{table:optimization} are met for balanced phonon pairs up to $n_\Omega=41$.

To squeeze to the highest factor $r_{15, 16}$ allowed for by the particle budget with the resonance condition $n_\Omega=31$, we can remind ourselves of equations \eqref{eq:squeezing_mechanism_matrix_elements} and \eqref{eq:squeezing_mechanism_evolution_parameters} to find the appropriate linear perturbation $a_\Omega$ after choosing a squeezing time $t$

\begin{equation}
	a_\Omega = \frac{\hbar \sqrt{2ln} \left(l^2-n^2 \right)^2 \pi^2 \xi \, r_{ln}}{2 \left(l^2 + n^2 \right) L^2 t} \; .
\end{equation}

For the setup in question, the healing length is $\xi \approx 450$nm. With the choice to take $t=1$s to squeeze, we get for the magnitude of the linear perturbation $a_\Omega \approx 6.9 \cdot 10^{-30} \mathrm{J} \mathrm{m}^{-1}$, which is two orders of magnitude below the maximum derived in section \ref{sec:naive_approach} and hence produces only negligible perturbations in the ground state.

In the light of the gravitational wave detector proposal described in chapter \ref{sec:detector_proposal}, we can now get an estimate for the sensitivity using eq. \eqref{eq:pumped-up_SU11_sensitivity} with a single detector $N_d=1$, integration time $\tau$ of one year and a squeeze angle $\theta = \sqrt{0.092}$ \cite{howl_active_2019}. As an ad-hoc ansatz to incorporate the overlap of the produced state with the ideal input state of a squeezed vacuum, we will take the Uhlmann fidelity of the final state after squeezing and a squeezed vacuum as an additional factor on the quantum Fisher information of the GW detector scheme. For $n_\Omega=31$ and thus the balanced pair of modes $(15, 16)$, we find

\begin{equation}
	\Delta h_{+, 0} \sim 10^{-16} \; .
\end{equation}

Mind that this is before serious attempts of optimizing the squeezing scheme and before \textit{any} attempt to improve the experimental parameters with respect to the GW detector. Thus, even though the strain sensitivity above does not allow for measurement of the sources considered in \cite{howl_quantum_2020}, there is a vast parameter space that has to be explored.

\begin{table}[hbpt]
\centering
\resizebox{0.8\textwidth}{!}{
	\begin{tabular}{c||c|c|c|c|c}
		$T = 10 \mathrm{nK}$
		\rule[-2.5ex]{0pt}{7ex}
		&
		$N_{phon, i}$
		&
		$\omega_{n_\Omega} [\mathrm{Hz}]$
		&
		$r_{l_{max}, n_\Omega -l_{max}}$
		&
		$t_{life} [\mathrm{s}]$
		&
		$\mathcal{F}$
		\\
		\hline
		\hline
		\rule{0pt}{3ex}
		$n_\Omega = 3$
		\rule[-2ex]{0pt}{1ex}
		&
		$9.5$
		&
		$870$
		&
		$3.8$
		&
		$74$
		&
		$0.016$
		\\
		\hline
		\rule{0pt}{3ex}
		$n_\Omega = 11$
		\rule[-2ex]{0pt}{1ex}
		&
		$9.5$
		&
		$3200$
		&
		$4.5$
		&
		$74$
		&
		$0.45$
		\\
		\hline
		\rule{0pt}{3ex}
		$n_\Omega = 31$
		\rule[-2ex]{0pt}{1ex}
		&
		$9.5$
		&
		$8900$
		&
		$4.8$
		&
		$66$
		&
		$0.93$
		\\
		\hline
		\hline
		$T = 5 \mathrm{nK}$
		\rule[-2.5ex]{0pt}{7ex}
		&
		$N_{phon, i}$
		&
		$\omega_{n_\Omega} [\mathrm{Hz}]$
		&
		$r_{l_{max}, n_\Omega -l_{max}}$
		&
		$t_{life} [\mathrm{s}]$
		&
		$\mathcal{F}$
		\\
		\hline
		\hline
		\rule{0pt}{3ex}
		$n_\Omega = 3$
		\rule[-2ex]{0pt}{1ex}
		&
		$3.3$
		&
		$870$
		&
		$4.2$
		&
		$74$
		&
		$0.024$
		\\
		\hline
		\rule{0pt}{3ex}
		$n_\Omega = 11$
		\rule[-2ex]{0pt}{1ex}
		&
		$3.3$
		&
		$3200$
		&
		$4.7$
		&
		$74$
		&
		$0.7$
		\\
		\hline
		\rule{0pt}{3ex}
		$n_\Omega = 31$
		\rule[-2ex]{0pt}{1ex}
		&
		$3.3$
		&
		$8900$
		&
		$4.8$
		&
		$74$
		&
		$0.99$
	\end{tabular}
}
\caption{\footnotesize Results on two-mode squeezing in the proposed $^7$Li BEC for the initial thermal phonon population number $N_{phon, i}$, the resonance frequency $\omega_{n_\Omega} [\mathrm{Hz}]$, the two-mode squeeze factor for the balanced mode pair $r_{l_{max}, n_\Omega -l_{max}}$, the life time $t_{life}$ of said mode pair and the Uhlmann fidelity $\mathcal{F}$ between the transformed state and a two-mode squeezed vacuum state.
	\\
	For $T=10 \mathrm{nK}$, the state after the transformation begins to become useful with regard to quantum metrology applications at $n_\Omega \gtrsim 11$, which puts the resonance frequencies in the kHz range. While in lower modes, three-body losses are the dominant limit to the phonon life times, at higher modes Landau damping starts to become relevant.
	\\
	If cooling can be improved to reach $T=5 \mathrm{nK}$, Landau damping loses its relevance up to $n_\Omega = 31$ and the $t_{life}$ becomes independent of the mode number. More importantly, though, thermal phonons are reduced further and thus the fidelity of the output state becomes useful for even lower lying resonances.
	\\
	With both temperatures, the most striking difference between the three resonances depicted lies in the overlap of the final state with a squeezed vacuum state.
	\label{table:proposed_experiment}}
\end{table}

\chapter{Conclusion and Outlook \label{chap:conclusion}}

As does a majority of the work in present-day physics, this thesis stands on the shoulders of giants. This is pointed out here because the thesis builds upon the findings of an extensive range of fields, ranging from ultracold atoms and general relativity over quantum optics and quantum metrology to Efimov physics, some of them prominent and well-established, others recent and more obscure. It was therefore the intention of the author to provide the reader with the concepts that proved most essential for finding the presented results and refer to comprehensive sources wherever this would have been beyond the scope of the thesis.

The process leading up to it involved multiple unexpected turns, most notably the relevance of Efimov physics. Although at first, it presented itself merely as a technique to find the necessary relation between three-body losses and the interaction strength, its compelling features seeming out of reach, it turned out to be a promising tool retain reasonable life times of a condensate while simultaneously minimizing the thermal population of phonon modes and thus allowing for stronger squeezing and higher fidelities of the states produced.

Motivated by the initial proposal \cite{sabin_phonon_2014}, the ensuing debate \cite{schutzhold_interaction_2018, howl_comment_2018} the ongoing work \cite{howl_active_2019, howl_quantum_2020-1, howl_quantum_2020} on a BEC-based gravitational wave detector and with it the potential expansion of GW astronomy into previously inaccessible realms with apparatus that are more compact than the currently active interferometric GW detectors, the aim was to contribute to this debate and development by pointing out the details of the production of the necessary squeezed states. This was accomplished by adapting an idea originally intended to \textit{detect} a potential by the transformation it evokes \cite{ratzel_dynamical_2018} to \textit{controlling} said potential to \textit{produce} squeezed phonons. Though hope might have originally been to simply maximize the squeeze factor contained in said transformation, it soon became apparent that the situation required a more intricate look. This led to what is, in the eyes of the author, the major achievement of this thesis: the consideration of the full transformation of the phononic states populating a uniform BEC under a periodic harmonic or linear perturbation.

With these theoretical findings, a mechanism for two-mode phonon squeezing was presented, which, albeit not well suited for the existing experiments that were taken as examples, can be highly effective in setups that seem within reach of present-day technology. It can thus be incorporated into the efforts going towards gravitational wave detectors, but conceivable applications also span over the entire field of frequency metrology.

A central question that remains to be addressed is the one of optimization. While qualitative arguments on where experiments could be improved and where the boundaries of the parameter space lie were given, the involved nature of the state transformation and with it exceedingly long evaluation times did not allow the author to tackle this task. Potential answers are twofold: one direction might be to find simplifications that are applicable in certain limiting cases, the other is for a more experienced programmer to streamline the computation process, allowing for quicker evaluation of the transformation.

For the incorporation of the results presented into GW detectors, the necessity of squeezed vacuum states as initial phonon states remains to be cleared. Work on this issue has been done and can be found in \cite{sabin_thermal_2016}. Furthermore, optimization of BEC parameters for the proposed detector scheme is currently being pursued and can now be done while at the same time taking phonon squeezing into account.

\listoffigures


\bibliographystyle{siam} 
\bibliography{./thesis_arxiv.bib}

\appendix

\part*{Appendix \label{part:appendix}}
\addcontentsline{toc}{part}{Appendix}	

\chapter{Bogoliubov Transformations \label{sec:bogoliubov_trafos}}

The general definition of a Bogoliubov transformation is more broad than the one introduced in \ref{sec:uniform_weakly_interacting}. While only the most important points for this work are given here, the interested reader is referred to Birrell and Davies' book \cite{birrell_quantum_1984} for a far more in-depth picture and, put into the context of relativistic quantum metrology, \cite{ahmadi_quantum_2014}.

Imagine a massless, real scalar quantum field (a boson field) $\hat{\Phi}$ satisfying the Klein-Gordon equation

\begin{equation} \label{eq:app_bogo_klein-gordon_massless}
	\Box \hat{\Phi} = 0 
\end{equation}

with periodic boundary conditions. One can then find a complete, discrete set of modes solving \eqref{eq:app_bogo_klein-gordon_massless} such that $\hat{\Phi}$ can be expanded as

\begin{equation} \label{eq:app_bogo_field_expansion}
	\hat{\Phi} = \sum_i \hat{a}_i \varphi_i + \hat{a}^\dagger_i \varphi^*_i \; ,
\end{equation}

with the creation and annihilation operator satisfying the canonical commutation relations

\begin{IEEEeqnarray}{rCl}
	\IEEEyesnumber \label{eq:canonical_commutation_relations} \IEEEyessubnumber*
	[ \hat{a}_i, \, \hat{a}^\dagger_j ] &=& \delta_{ij}
	\\
	\, [ \hat{a}_i, \, \hat{a}_j ] &=& 0 = [ \hat{a}^\dagger_i, \, \hat{a}^\dagger_j ]
\end{IEEEeqnarray}

The modes $\varphi_i$ are orthonormal

\begin{equation}
	\left( \varphi_i , \varphi_j \right) = \delta_{ij}
\end{equation}

with respect to the Klein-Gordon inner product\footnote{The integral here goes over a spatial hypersurface for a fixed time t} \cite{birrell_quantum_1984}

\begin{IEEEeqnarray}{rCl}
	\left( \varphi_i , \varphi_j \right)
	& \equiv &
	- i \int \ud \mathbf{x} \,  \varphi_i \partial_t \varphi_j^* - \left( \partial_t \varphi_i \right) \varphi_j^*
	\nonumber
	\\
	& \equiv &
	- i \int \ud \mathbf{x} \, \varphi_i \overleftrightarrow{\partial_t} \varphi_j^* \; .
\end{IEEEeqnarray}

The expansion \eqref{eq:app_bogo_field_expansion} is, in general, not unique. $\hat{\Phi}$ can be defined in terms of another set of ladder operators satisfying the same commutation relations and orthonormal modes

\begin{IEEEeqnarray}{rCl}
	\hat{\Phi} &=& \sum_j \hat{b}_j \bar{\varphi}_j + \hat{b}^\dagger_j \bar{\varphi}^*_i
	\\	
	& = & \sum_i \hat{a}_i \varphi_i + \hat{a}^\dagger_i \varphi^*_i \; . \nonumber
\end{IEEEeqnarray}

The modes and ladder operators are then connected by the Bogoliubov transformation \cite{birrell_quantum_1984}

\begin{IEEEeqnarray}{rCClLl}
	\IEEEyesnumber \label{eq:app_bogo_trafo_definition} \IEEEyessubnumber*
	\bar{\varphi}_j &=& \sum_i &
	\alpha_{ji} \varphi_i
	& + &
	\beta_{ji} \varphi^*_i
	\\
	\varphi_i &=& \sum_j &
	\alpha^*_{ji} \bar{\varphi}_j
	& - &
	\beta_{ji} \bar{\varphi}^*_i
	\\*[5pt]
	\IEEEeqnarraymulticol{6}{c}{
		\mathrm{with} \;
		\alpha_{ji} \equiv \left( \bar{\varphi}_j, \varphi_i \right)
		, \;
		\beta_{ji} \equiv - \left( \bar{\varphi}_j, \varphi_i^* \right)
		}
	\\*[5pt]
	\label{eq:app_bogo_example_particle_creation}
	\Rightarrow \; \hat{a}_i &=& \sum_j &
	\alpha_{ji} \hat{b}_j
	&+&
	\beta_{ji}^* \hat{b}_j^\dagger
	\\
	\hat{b}_j &=& \sum_i &
	\alpha_{ji}^* \hat{a}_i
	&-&
	\beta_{ji}^* \hat{a}_i^\dagger \; .
\end{IEEEeqnarray}

$\alpha_{ji}$ and $\beta_{ji}$ are called the \textit{Bogoliubov coefficients}. A Bogoliobov transformation can ---but does not have to--- correspond to a coordinate change between different observers, an illustrative example for which is the Unruh effect \cite{birrell_quantum_1984}, where an inertial observer's vacuum state appears as a thermal state to a uniformly accelerated observer. In general nonvanishing coefficients $\beta_{ji} \neq 0$ are responsible for \textit{particle creation}, as is particularly well illustrated in \eqref{eq:app_bogo_example_particle_creation}: while the vacuum state $\hat{b}_i | \bar{0} \rangle = 0$ has a zero eigenvalue for the annihilation operator $\hat{b}_j$, for another observer's annihilation operator related to $\hat{b}_j$ by \eqref{eq:app_bogo_example_particle_creation}, we have $\hat{a}_i | \bar{0} \rangle \neq 0$ if $\beta_{ji} \neq 0$.

\chapter{The Covariance Matrix Formalism \label{sec:CM_formalism_new}}

When dealing with continuous variable systems and especially with Gaussian states, the covariance matrix (CM) formalism \cite{adesso_entanglement_2007, safranek_quantum_2015, safranek_optimal_2016} is a remarkably helpful tool. For once, this is because it lives on the \textit{phase space} which admits to a direct sum structure as opposed to the tensor product structure of the Hilbert space. As a result, when considering only subsystems, there is no need for infinite-dimensional traces.

Gaussian states are uniquely determined by their first and second moments. In the CM formalism, they are therefore described by a vector and a covariance matrix encoding the former and the latter, respectively. Gaussian transformations\footnote{
	Gaussian transformations preserve the ``Gaussianity'' of a state. They are given by unitaries whose exponents are at most quadratic in the creation and annihilation operators.
}
are then given by the set of \textit{symplectic transformations}, which can be derived from their corresponding Hilbert space representation.

The representation of the CM formalism is, however, not unique. First and second moments can be defined in terms of expectation values of the creation and annihilation operators (this is referred to as the \textit{complex picture}) or the quadratures\footnote{Even within the real picture, there are ambiguities in the definition of the quadratures and their ordering in the CM.} (leading to the \textit{real picture}). The following sections will give the identities most important for the purposes of this thesis and conclude with the reconciliation of the pictures presented in \cite{adesso_entanglement_2007, safranek_quantum_2015, safranek_optimal_2016}.

\section{The Symplectic Form}

Consider the bosonic creation and annihilation operators $\hat{a}_n$, $\hat{a}_n^\dagger$ for the mode $n$. The operators for all modes in an $N$-dimensional Hilbert space can then be collected in the vector $\hat{\mathbf{A}} \equiv \left( \hat{a}_1, \ldots \hat{a}_N, \hat{a}_1^\dagger , \ldots , \hat{a}_N^\dagger \right)^T$ and their commutation relations are encoded in the \textit{symplectic form} $K$ \cite{safranek_quantum_2015}

\begin{equation}
	K_{ij} \equiv \left[ \hat{\mathbf{A}}_i, \hat{\mathbf{A}}_j^\dagger \right] \; \Rightarrow \; K =
	\begin{pmatrix}
		1 & 0
		\\
		0 & -1
	\end{pmatrix}
	\otimes \mathbb{1}_N \; .
\end{equation}

$K$ is unitary and hermitian and thus $KK=\mathbb{1}$.

\section{Gaussian States: First and Second Moments}

In the complex picture of the CM formalism, the first and second moments of a state are expressed in terms of expectation values of the creation and annihilation operators as

\begin{IEEEeqnarray}{rCl}
	\IEEEyesnumber \label{eq:CM_CM_definition_new} \IEEEyessubnumber*
	\mathbf{d} & \equiv & \langle \hat{\mathbf{A}} \rangle
	\\
	\sigma_{ij} & \equiv &
	\langle
		\hat{\mathbf{A}}_i \hat{\mathbf{A}}_j^\dagger
		+
		\hat{\mathbf{A}}_j^\dagger \hat{\mathbf{A}}_i
	\rangle
	-
	2 \langle \hat{\mathbf{A}}_i \rangle \langle \hat{\mathbf{A}}_j^\dagger \rangle
\end{IEEEeqnarray}

The first moments, encoded in the vector $\mathbf{d}$ can be set to zero using local unitary operations and are thus mostly ignored in quantum metrological considerations. The relevant properties for these applications, however, lie in the \textit{covariance matrix} $\sigma$.

As for the Hilbert space, where the density matrix $\rho$ of a physical state must fulfill

\begin{equation}
	\mathrm{Tr} \, \rho = 1 \, , \; \rho \geq 0 \; ,
\end{equation}

there is the \textit{bona fide} criterion

\begin{equation} \label{eq:CM_bona_fide}
	\sigma + K \geq 0
\end{equation}

for covariance matrices, which also contains information about the commutation relations of the creation and annihilation operators and thereby includes the uncertainty relation.

Once the question of how to represent states in the phase space is settled, the natural next question is how they transform. The unitary transformations $\rho \rightarrow U \rho U^\dagger$ on the Hilbert space that preserve the Gaussian nature of a state are mirrored in the phase space by

\begin{equation} \label{eq:CM_symplectic_trafo_definition_new}
	\sigma \, \rightarrow \, S \sigma S^\dagger \; ,
\end{equation}

where the linear transformation $S$ leaves the symplectic form $K$ invariant

\begin{equation} \label{eq:CM_symplectic_form_invariance_new}
	S K S^\dagger = K
\end{equation}

and thus keeps the canonical commutation relations intact. Note that \eqref{eq:CM_symplectic_form_invariance_new} also implies $|S|=1$. The symplectic transformations in the equations above form are a complex representation of the real symplectic group $Sp (2N, \mathbb{R})$ \cite{safranek_quantum_2015}.

As previously mentioned, Gaussian transformations are unitaries whose exponents are, at most, quadratic in the creation and annihilation operators.

\begin{equation}
	\hat{U} = \mathrm{e}^{
		\frac{i}{2} \hat{\mathbf{A}}^\dagger W \hat{\mathbf{A}} + \hat{\mathbf{A}}^\dagger K \gamma
	} \quad \mathrm{with} \quad
	W = W^\dagger =
	\begin{pmatrix}
		X & Y
		\\
		Y^* & X^*
	\end{pmatrix}
	\; , \;
	\gamma =
	\begin{pmatrix}
		\tilde{\gamma}
		\\
		\tilde{\gamma}^*
	\end{pmatrix}
\end{equation}

Under the transformation represented by $\hat{U}$, $\sigma$ and $\mathbf{d}$ become \cite{safranek_optimal_2016}

\begin{equation} \label{eq:CM_transformation_behaviour}
	\sigma \rightarrow S_U \sigma S_U^\dagger \; , \quad \mathbf{d} \rightarrow S_U \mathbf{d} + \mathbf{b}
\end{equation}

with

\begin{equation} \label{eq:CM_symplectic-unitary-correspondence}
	S_U = \mathrm{e}^{i K W} \quad \mathrm{and} \quad \mathbf{b} = \left( \int_0^1 \ud t \mathrm{e}^{i K W t} \right) \gamma \; .
\end{equation}

The first part of \eqref{eq:CM_symplectic-unitary-correspondence} is used chapter \ref{sec:implementation} to derive the symplectic transformation corresponding to the time evolution \eqref{eq:squeezing_mechanism_time_evolution}.

One symplectic transformation of particular importance is the diagonalization of $\sigma$ by virtue of Williamson's theorem \cite{williamson_algebraic_1936}:

\begin{equation}
	\sigma = S_D D S_D^\dagger \quad \mathrm{such \, that} \quad
	D = \mathbb{1}_2 \otimes \mathrm{diag} \left( \lambda_1, \ldots \, \lambda_N \right) \, ,
\end{equation}

where $\{ \lambda_i \}$ are the \textit{symplectic eigenvalues}. $D$ corresponds to a diagonal product state in the Hilbert space.

The bona fide condition \eqref{eq:CM_bona_fide} also implies an expression of the uncertainty principle in terms of the symplectic spectrum of a CM \cite{adesso_entanglement_2007}.

\begin{equation}
	\lambda_i \geq 1 \; \forall \, i = 1,  \ldots, n
\end{equation}

Pure states saturate the uncertainty principle and hence all the symplectic eigenvalues of a pure state are equal to one.

\section{Quantum Metrology in the CM Formalism \label{sec:RQM_new}}

When trying to estimate a parameter $\epsilon$ that gets encoded into a state by a transformation, the central quantity is the quantum Fisher information (QFI) $H_\epsilon$. It puts an upper bound on the precision of the estimation of $\epsilon$ in the form of the quantum Cramér-Rao bound \cite{braunstein_statistical_1994}

\begin{equation}
	\langle \left( \Delta \hat{\epsilon} \right)^2 \rangle \geq \frac{1}{M H_\epsilon} \; .
\end{equation}

In the equation above, $\hat{\epsilon}$ is an estimator for the parameter in question, $M$ is the number of independent measurements and the bound is meant after optimization over all possible measurements.

The QFI tells us how far two states that are separated by an infinitesimal change $\ud \epsilon$ lie from each other and it is thus not surprising that it is related to a fidelity between $\sigma_\epsilon$ and $\sigma_{\epsilon + \ud \epsilon}$, more precisely the Uhlmann fidelity $\mathcal{F} \left( \sigma_\epsilon, \sigma_{\epsilon + \ud \epsilon} \right)$ \cite{ahmadi_quantum_2014}

\begin{equation} \label{eq:RQM_QFI_from_uhlmann_fidelity_new}
	H_\epsilon = \frac{
		8 \left( 1 - \sqrt{
			\mathcal{F} \left( \sigma_\epsilon, \sigma_{\epsilon + \ud \epsilon} \right)
		} \right)
	}{
		\ud \epsilon^2
	} \; .
\end{equation}

In the Hilbert space, the Uhlmann fidelity between two states $\rho_1$ and $\rho_2$ is defined as

\begin{equation}
	\mathcal{F} (\rho_1, \rho_2) \equiv \left(
		\mathrm{Tr} \sqrt{
			\sqrt{\rho_1} \rho_2 \sqrt{\rho_1}
		}
	\right)^2
\end{equation}

For two-mode Gaussian states, an explicit expression for the equivalent $\mathcal{F} ( \sigma_1, \sigma_2)$ has been given in \cite{marian_uhlmann_2012}. In the complex picture of the CM formalism, we will use the expression given in \cite{safranek_quantum_2015}:

\begin{IEEEeqnarray}{rCl} \label{eq:RQM_uhlmann_fidelity_new}
	\IEEEyesnumber \IEEEyessubnumber*
	\mathcal{F} \left( \sigma_1, \sigma_2 \right)
	&=&
	\frac{
		4 \mathrm{e}^{
			\delta \mathbf{d}^\dagger \left( \sigma_1 + \sigma_2 \right)^{-1} \delta \mathbf{d}
		}
	}{
		\sqrt{\Gamma} + \sqrt{\Lambda} - \sqrt{\left( \sqrt{\Gamma} + \sqrt{\Lambda} \right)^2 - \Delta}
	}
	\\
	\IEEEeqnarraymulticol{3}{l}{
		\mathrm{with \, the \, determinants}
	} \nonumber
	\\
	\Delta & \equiv & \left| \sigma_1 + \sigma_2 \right|
	\\
	\Gamma & \equiv & \left|
		\mathbb{1} + K \sigma_1 K \sigma_2
	\right|
	\\
	\Lambda & \equiv & \left| \sigma_1 + K \right| \, \left| \sigma_2 + K \right|
	\\
	\IEEEeqnarraymulticol{3}{l}{
		\mathrm{and}
	} \nonumber
	\\
	\delta \mathbf{d} & \equiv & \mathbf{d}_1 - \mathbf{d}_2 \; .
\end{IEEEeqnarray}

\section{Examples of Covariance Matrices and Symplectic Transformations}

\paragraph{The vacuum state,} whose first moments vanish, is a pure, completely uncorrelated state. Its CM must therefore be diagonal with all symplectic eigenvalues $\lambda_i=1$, which is simply an identity matrix. In $N$ modes, we have

\begin{equation}
	\sigma_{vac} = \mathbb{1}_{2N} \; .
\end{equation}

This also means that any state that is derived by a Gaussian transformation of the vacuum (such as the squeezed vacuum state, see chapter \ref{sec:squeezed_states}) is found by taking $\sigma=SS^\dagger$.

\paragraph{The thermal state} is also completely uncorrelated and thus diagonal. It is, however, a mixed state. To find the symplectic eigenvalues, consider the substate of a single mode $i$

\begin{equation}
	\sigma_{th, i} = \begin{pmatrix}
		\lambda_i & 0
		\\
		0 & \lambda_i
	\end{pmatrix} \; ,
\end{equation}

take the expectation value of the particle number operator $\hat{N} \equiv \sum_{i=1}^N \hat{a}_i^\dagger \hat{a}_i$ \cite{howl_active_2019}

\begin{equation} \label{eq:CM_particle_number_expectation_value_appendix}
	\langle \hat{N} \rangle = \frac{1}{4} \left(
		\mathrm{Tr} \left( \sigma \right) + \mathbf{d}^\dagger \mathbf{d} - 2N
	\right)
\end{equation}

And remember the average occupation number for a single-mode thermal state in mode $i$, $\langle \hat{N} \rangle_{th} = 1/(\mathrm{e}^{\beta \hbar \omega_i}-1)$ with the inverse temperature $\beta$ and mode frequency $\omega_i$. Equating this and \eqref{eq:CM_particle_number_expectation_value_appendix}, we find

\begin{IEEEeqnarray}{rCl}
	\IEEEyesnumber \IEEEyessubnumber*
	\langle \hat{N} \rangle_{th} &=& \frac{1}{4} \left( 2 \lambda_i - 2 \right)
	\\
	\lambda_i &=& 2 \langle \hat{N} \rangle_{th} + 1
	\\
	\Rightarrow \lambda_i &=& \frac{2}{\mathrm{e}^{\beta \hbar \omega_i}-1} + 1
	=
	\coth \frac{\beta \hbar \omega_i}{2}
\end{IEEEeqnarray}

and we have derived the CM of an $N$-mode thermal state

\begin{equation} \label{eq:CM_formalism_thermal_state_CM}
	\sigma_{th} = \mathbb{1}_2 \otimes \mathrm{diag}
	\left(
		\coth \frac{\beta \hbar \omega_1}{2}, \ldots, \coth \frac{\beta \hbar \omega_N}{2}
	\right) \; .
\end{equation}

\paragraph{Single-mode squeezing:} starting with the real single-mode squeezing operator (see eq.\eqref{eq:single-mode_squeezing_operator}) in the Hilbert space $\hat{U} = \mathrm{Exp} \left( -\frac{r}{2} ( \hat{a}^{\dagger 2} - \hat{a}^2 ) \right)$, we want to apply \eqref{eq:CM_symplectic-unitary-correspondence} and find

\begin{equation}
	W_{sq1}= i \, r \begin{pmatrix}
		0 & 1
		\\
		-1 & 0
	\end{pmatrix}
\end{equation}

and the symplectic transformation

\begin{equation}
	S_{sq1} = \mathrm{Exp} \left[
		-r \begin{pmatrix}
			0 & 1
			\\
			-1 & 0
		\end{pmatrix}
	\right]
	=
	\begin{pmatrix}
		\cosh r & -\sinh r
		\\
		-\sinh r & \cosh r
	\end{pmatrix} \; .
\end{equation}

As mentioned above, the covariance matrix of the squeezed vacuum state is obtained by simply taking $S_{sq1} S_{sq1}^\dagger$:

\begin{equation}
	\sigma_{sq1} = 
	\begin{pmatrix}
		\cosh 2r & -\sinh 2r
		\\
		-\sinh 2r & \cosh 2r
	\end{pmatrix} \; .
\end{equation}

\paragraph{Two-mode squeezing} is produced by the two-mode squeezing operator \eqref{eq:squeezed_states_two-mode_squeezing_operator} which we take to be real for simplicity's sake to get

\begin{equation}
	W_{sq2}= i \, r \begin{pmatrix}
		0 & 0 & 0 & 1
		\\
		0 & 0 & 1 & 0
		\\
		0 & -1 & 0 & 0
		\\
		-1 & 0 & 0 & 0
	\end{pmatrix}
\end{equation}

and

\begin{equation}
	S_{sq2}= \begin{pmatrix}
		\cosh r & 0 & 0 & -\sinh r
		\\
		0 & \cosh r & -\sinh r & 0
		\\
		0 & -\sinh r & \cosh r & 0
		\\
		-\sinh r & 0 & 0 & \cosh r
	\end{pmatrix} \; .
\end{equation}

The squeezed vacuum state, finally, has the CM

\begin{equation}
	\sigma_{sq2}= \begin{pmatrix}
		\cosh 2r & 0 & 0 & -\sinh 2r
		\\
		0 & \cosh 2r & -\sinh 2r & 0
		\\
		0 & -\sinh 2r & \cosh 2r & 0
		\\
		-\sinh 2r & 0 & 0 & \cosh 2r
	\end{pmatrix} \; .
\end{equation}

\paragraph{Bogoliubov transformations} can be easily derived by considering \eqref{eq:app_bogo_example_particle_creation} and remembering the transformation behaviour of $\mathbf{d}$ \eqref{eq:CM_transformation_behaviour}. One then finds

\begin{equation} \label{eq:CM_symplectic_trafo_from_bogo_new}
	S_{Bog} = \begin{pmatrix}
		\alpha & \beta
		\\
		\beta^* & \alpha^*
	\end{pmatrix}
	\; ,
\end{equation}

where $\alpha$ and $\beta$ are the matrices made up of the Bogoliubov coefficients $\alpha_{ij}$, $\beta_{ij}$.

\section{The Real and the Complex Picture \label{sec:CM_real_complex}}

When reading about the CM formalism and its applications in different sources, one encounters different pictures of it, which are nonetheless equivalent. Primarily, one can divide them into the real and the complex picture as described in \cite{safranek_quantum_2015}. However, there is still ambivalence in both the definition of the quadratures in the real picture as well as the ordering of the subspaces, as can be seen by comparing the pictures listed above with the one found in Adesso's dissertation \cite{adesso_entanglement_2007}.

The table on the next page shows how these pictures are reconciled and connected. The indices $r$ and $ad$ denote the real pictures in $\cite{safranek_quantum_2015}$ and \cite{adesso_entanglement_2007}, respectively. Note that, due to the differing norm between their quadratures, the transformation $T_{ad \rightarrow r}$ from Adesso's formalism to the $r$ picture is \textit{not} unitary.

\begin{landscape}
\begin{table}[ht]
\centering
\resizebox{1.5\textwidth}{!}{
	\begin{tabular}{l||c|c|c|c|c}
		& Complex formalism
		&
		$U_{r \rightarrow c} = \frac{1}{\sqrt{2}} \begin{pmatrix}
			\mathbb{1} & i \mathbb{1}
			\\
			\- \mathbb{1} & -i \mathbb{1}
		\end{pmatrix}$
		& Real formalism
		&
		$T_{ad \rightarrow r, 2} = \frac{1}{\sqrt{2}}
		\begin{pmatrix}
			1 & 0 & 0 & 0
			\\
			0 & 0 & 1 & 0
			\\
			0 & 1 & 0 & 0
			\\
			0 & 0 & 0 & 1
		\end{pmatrix}
		$
		& Adesso's formalism \cite{adesso_entanglement_2007}
		\\
		\hline \hline
		\hspace{5pt}
		\specialcell{Basic \\ operators}
		&
		\specialcell{$\hat{a}_i$ \\ $\hat{a}_i^\dagger$}
		&
		&
		\specialcell{
			$\hat{x}_{r, i} = \frac{1}{\sqrt{2}} (\hat{a}_i + \hat{a}_i^\dagger )$
			\\
			$\hat{p}_{r, i} = -\frac{i}{\sqrt{2}} (\hat{a}_i - \hat{a}_i^\dagger )$
		}
		&
		&
		\specialcell{
			$\hat{x}_{ad, i} = \hat{a}_i + \hat{a}_i^\dagger$
			\\
			$\hat{p}_{ad, i} = -i ( \hat{a}_i + \hat{a}_i^\dagger )$
		}
		\\
		\hline
		\hspace{5pt}
		$1^{st}$ order vector
		&
		$\hat{\mathbf{A}} = \left( \hat{a}_1, \ldots \hat{a}_N, \hat{a}_1^\dagger , \ldots , \hat{a}_N^\dagger \right)^T$
		&
		$\hat{\mathbf{A}} = U_{r \rightarrow c} \hat{\mathbf{Q}}$
		&
		$\hat{\mathbf{Q}} = \left( \hat{x}_{r, 1}, \ldots, \hat{x}_{r, N}, \hat{p}_{r, 1}, \ldots, \hat{p}_{r, N} \right)^T$
		&
		$\hat{\mathbf{Q}} = T_{ad \rightarrow r} \hat{\mathbf{R}}$
		&
		$\hat{\mathbf{R}} = \left(\hat{x}_{ad, 1}, \hat{p}_{ad, 1}, \ldots \hat{x}_{ad, N}, \hat{p}_{ad, N} \right)^T$
		\\
		\hline
		\hspace{5pt}
		$1^{st}$ moments
		&
		$\mathbf{d} = \langle \hat{\mathbf{A}} \rangle$
		&
		&
		$\mathbf{d}_r = \langle \hat{\mathbf{Q}} \rangle$
		&
		&
		$\mathbf{R} = \langle \hat{\mathbf{R}} \rangle$
		\\
		\hline
		\hspace{5pt}
		\specialcell{
			Symplectic
			\\
			form
		}
		&
		\specialcell{
			$K_{ij} \equiv \left[ \hat{\mathbf{A}}_i, \hat{\mathbf{A}}_j^\dagger \right]$
			\\
			$K= \begin{pmatrix}
				\mathbb{1}_N & 0
				\\
				0 & \mathbb{1}_N
			\end{pmatrix}$
		}
		&
		
		&
		\specialcell{
			$i \Omega_{r, ij} \equiv \left[ \hat{\mathbf{Q}}_i, \hat{\mathbf{Q}}_j \right]$
			\\
			$\Omega_{r, ij}= \begin{pmatrix}
				0 & \mathbb{1}_N
				\\
				-\mathbb{1}_N & 0
			\end{pmatrix}$
		}
		&
		
		&
		\specialcell{
			$2 i \Omega_{ad, ij} \equiv \left[ \hat{\mathbf{R}}_i, \hat{\mathbf{R}}_j \right]$
			\\
			$\Omega_{ad} = \bigoplus_{i=1}^N \omega$
			\\
			$\omega = \begin{pmatrix}
				0 & 1
				\\
				-1 & 0
			\end{pmatrix}$
			}
		\\
		\hline
		CM
		&
		\specialcell{
			$\sigma_{ij}= \left\langle
				\hat{\mathbf{A}}_i \hat{\mathbf{A}}_j^\dagger + \hat{\mathbf{A}}_j^\dagger \hat{\mathbf{A}}_i
			\right\rangle$
			\\
			$- 2 \left\langle \hat{\mathbf{A}}_i \right\rangle \left\langle \hat{\mathbf{A}}_j^\dagger \right\rangle$
		}
		&
		$\sigma = U_{r \rightarrow c} \sigma_r U_{r \rightarrow c}^\dagger$
		&
		\specialcell{
			$\sigma_{r, ij}= \left\langle
				\hat{\mathbf{Q}}_i \hat{\mathbf{Q}}_j + \hat{\mathbf{Q}}_j \hat{\mathbf{Q}}_i
			\right\rangle$
			\\
			$- 2 \left\langle \hat{\mathbf{Q}}_i \right\rangle \left\langle \hat{\mathbf{Q}}_j \right\rangle$
		}
		&
		$\sigma_r = T_{ad \rightarrow r} \sigma_{ad} T_{ad \rightarrow r}^{-1}$
		&
		\specialcell{
			$\sigma_{ad, ij}= \frac{1}{2} \left\langle
				\hat{\mathbf{R}}_i \hat{\mathbf{R}}_j + \hat{\mathbf{R}}_j \hat{\mathbf{R}}_i
			\right\rangle$
			\\
			$-  \left\langle \hat{\mathbf{R}}_i \right\rangle \left\langle \hat{\mathbf{R}}_j \right\rangle$
		}
		\\
		\hline
		\specialcell{
			Structure
			\\
			of 1$^{st}$
			\\
			moments, $\sigma$
		}
		&
		\specialcell{
			$\mathbf{d} = \begin{pmatrix}
				\tilde{\mathbf{d}}
				\\
				\tilde{\mathbf{d}}^*
			\end{pmatrix},
			\;
			\sigma = \begin{pmatrix}
				X & Y
				\\
				Y^* & X^*
			\end{pmatrix}$
			\\
			$\sigma^\dagger = \sigma$
			\\
			$
			\Leftrightarrow \,
			X^\dagger = X, \, Y^T=Y$
		}
		&
		&
		\specialcell{
			$
			\mathbf{d}_r= \begin{pmatrix}
				\mathbf{x}
				\\
				\mathbf{p}
			\end{pmatrix}, \;
			\sigma_r = \begin{pmatrix}
				X_r & Y_r
				\\
				Y_r^T & Z_r
			\end{pmatrix}
			$
			\\
			$
			\sigma_r^T = \sigma_r
			$
			\\
			$
			\Leftrightarrow \; X_r^T = X_r, \, Z_r^T=Z_r
			$
		}
		&
		&
		\specialcell{
			$
			\mathbf{d}_{ad} = \bigoplus_{i=1}^N
			\begin{pmatrix}
				\hat{x}_{ad, i}
				\\
				\hat{p}_{ad, i}
			\end{pmatrix}, \;
			\sigma_{ad} = \begin{pmatrix}
				X_{ad} & Y_{ad}
				\\
				Y_{ad}^T & Z_{ad}
			\end{pmatrix}
			$
			\\
			$
			\sigma_{ad}^T = \sigma_{ad}
			$
			\\
			$
			\Leftrightarrow \; X_{ad}^T = X_{ad}, \, Z_{ad}^T=Z_{ad}
			$
		}
		\\
		\hline
		Diagonalization
		&
		\specialcell{
			$\sigma = SDS^\dagger, \;
			S= \begin{pmatrix}
				\alpha & \beta
				\\
				\beta^* & \alpha^*
			\end{pmatrix}$
			\\
			$
			D= \mathbb{1} \otimes L
			$
			\\
			$
			L= \mathrm{diag} \left( \lambda_1 , \ldots \lambda_N \right)
			$
			\\
			$
			S \tilde{\in} \mathrm{Sp} (2N, \mathbb{R})
			$
			\\
			$SKS^\dagger = K$
		}
		&
		$S=U_{r \rightarrow c} S_r U_{r \rightarrow c}^\dagger$
		&
		\specialcell{
			$\sigma_r = S_r D_r S_r^T$
			\\
			$D_r = D$
			\\
			$
			S_r =
			\begin{pmatrix}
				\alpha_r & \beta_r
				\\
				\gamma_r & \delta_r
			\end{pmatrix}
			$
			\\
			$=
			\begin{pmatrix}
				\mathrm{Re} ( \alpha + \beta ) & -\mathrm{Im} (\alpha - \beta )
				\\
				\mathrm{Im} ( \alpha + \beta ) & \mathrm{Re} (\alpha - \beta)
			\end{pmatrix}
			$
			\\
			$S_r \in \mathrm{Sp} (2N, \mathbb{R})$
			\\
			$S_r \Omega_r S_r^T = \Omega_r$
		}
		&
		$S_r = T_{ad \rightarrow r, 2} S_{ad} T_{ad \rightarrow r, 2}^{-1}$
		&
		\specialcell{
			$
			\sigma_{ad} = S_{ad}^T \nu S_{ad}
			$
			\\
			$
			\nu = \bigoplus_{i=1}^N
			\begin{pmatrix}
				\nu_i & 0
				\\
				0 & \nu_i
			\end{pmatrix}
			$
			\\
			$
			S_{ad} = \begin{pmatrix}
				\alpha_{ad, 11} & \alpha_{ad, 12}
				\\
				\alpha_{ad, 21} & \alpha_{ad, 22}
			\end{pmatrix}
			$
			\\
			$\alpha_{ad, ij}=
			\begin{pmatrix}
				\mathrm{Re} (\alpha_{ij} + \beta_{ij}) & - \mathrm{Im} (\alpha_{ij} - \beta_{ij})
				\\
				-\mathrm{Im} (\alpha_{ij} + \beta_{ij}) & \mathrm{Re} (\alpha_{ij} - \beta_{ij})
			\end{pmatrix}					
			$
			\\
			$S_{ad} \in \mathrm{Sp} (2N, \mathbb{R})$
			\\
			$S_{ad} \Omega_{ad} S_{ad}^T = \Omega_{ad}$
		}
	\end{tabular}
}
\caption{\footnotesize
	The complex representation of the covariance matrix formalism, the real representation and a variant of the latter as it is used in eg. \cite{adesso_entanglement_2007}. Columns between the descriptions of the representations describe how to transform between them. The $2$ in the index of the transformation $T_{ad \rightarrow r, 2}$ denotes the number of modes that the transformation is displayed for. The tilde in $S \tilde{\in} \mathrm{Sp} (2N, \mathbb{R})$ indicates that in the complex formalism, $S$ is an element of a \textit{complex representation} of the real symplectic group.
	\label{table:CM_formalism_real_complex}
}
\end{table}
\end{landscape}

\chapter{Example of the Variance of Combinations of Observables in Two-mode Squeezed States \label{sec:appendix_two-mode_squeezed_variance}}

As an example, take the combination of quadratures $\hat{x}_\gamma^A \pm \hat{x}_\gamma^B$ as they are defined in chapter \ref{sec:squeezed_states} and where the upper indices $A, B$ denote the respective subspace. Since $\langle mm | (\hat{x}_\gamma^A \pm \hat{x}_\gamma^B) |nn \rangle = 0$, the expectation value for a squeezed state $ _2\langle \zeta | ( \hat{x}_\gamma^A \pm \hat{x}_\gamma^B ) | \zeta \rangle_2$ must vanish as well. For the variance, we thus only need to calculate $ \langle ( \hat{x}_\gamma^A \pm \hat{x}_\gamma^B )^2 \rangle_{\zeta_2}$. To do so, expand the square of the quadratures and use the bosonic commutation relations for creation and annihilation operators to find\footnote{
$\hat{a}^{(\dagger)}$ and $\hat{b}^{(\dagger)}$	denote the creation and annihilation operators in modes $A$ and $B$.
}

\begin{IEEEeqnarray}{rCl}
	( \hat{x}_\gamma^A \pm \hat{x}_\gamma^B )^2 &=&
	\mathrm{e}^{-2i \gamma}
	\left[
		\frac{1}{2} \left( \hat{a}^2 + \hat{b}^2 \right)
		\pm \hat{a} \hat{b}
	\right]
	+
	\mathrm{e}^{2i \gamma}
	\left[
		\frac{1}{2} \left( \hat{a}^{\dagger 2} + \hat{b}^{\dagger 2} \right)
		\pm \hat{a}^\dagger \hat{b}^\dagger
	\right]
	\nonumber
	\\
	& & +
	\mathbb{1} + \hat{a}^\dagger \hat{a} + \hat{b}^\dagger \hat{b}
	\pm
	\left[
		\hat{a} \hat{b}^\dagger + \hat{a}^\dagger \hat{b}
	\right] \; .
\end{IEEEeqnarray}

The only nonvanishing elements $\langle mm | (\hat{x}_\gamma^A \pm \hat{x}_\gamma^B)^2 |nn \rangle$ are then

\begin{equation}
	\langle mm | (\hat{x}_\gamma^A \pm \hat{x}_\gamma^B)^2 |nn \rangle
	=
	\pm \left[
		n \mathrm{e}^{-2i \gamma} \delta_{m, n-1} + (n+1) \mathrm{e}^{2i \gamma} \delta_{m, n+1}
	\right]
	+
	(1+2n) \delta_{n, n}
\end{equation}

and the variance becomes

\begin{IEEEeqnarray}{rLl}
	\Delta^2 ( \hat{x}_\gamma^A \pm \hat{x}_\gamma^B )
	&= \frac{1}{\cosh^2 r} \Bigg[ &
	\pm \mathrm{e}^{i(\theta - 2 \gamma)} \sum_{n=1}^{\infty} (-1)^{2n-1} n \tanh^{2n-1}r
	\nonumber
	\\
	& &
	\pm \mathrm{e}^{-i(\theta - 2 \gamma)} \sum_{n=0}^{\infty} (-1)^{2n+1} (n+1) \tanh^{2n+1}r
	\nonumber
	\\
	& &
	+ \sum_{n=0}^{\infty} (-1)^{2n} (2n+1) \tanh^{2n}r
	\Bigg]
	\nonumber
	\\
	&= \frac{1}{\cosh^2 r} \Bigg[ &
	 \pm \left(
	 	\mathrm{e}^{i(\theta - 2 \gamma)} + \mathrm{e}^{-i(\theta - 2 \gamma)}
	 \right)
	 \sum_{n=0}^{\infty} (-1)^{2n+1} (n+1) \tanh^{2n+1}r
	 \nonumber
	 \\
	 & &
	 + \sum_{n=0}^{\infty} (-1)^{2n} (2n+1) \tanh^{2n}r	 
	 \Bigg]
	\nonumber
	\\*[3pt]
	& = &
	\cosh (2r) \mp \cos (\theta - 2 \gamma) \sinh (2r) \; .
\end{IEEEeqnarray}

For $\theta = 2 \gamma$, the variance is thus minimized to

\begin{equation}
\Delta^2 ( \hat{x}_{\theta/2}^A \pm \hat{x}_{\theta/2}^B ) = \mathrm{e}^{\mp 2r} \; .
\end{equation}

Analogously, for the joint measurement of the quadratures $( \hat{p}_\gamma^A \pm \hat{p}_\gamma^B )$, we get

\begin{equation}
	\Delta^2 ( \hat{p}_\gamma^A \pm \hat{p}_\gamma^B ) = \cosh(2r) \pm \cos( \theta - 2 \gamma ) \sinh (2r)
\end{equation}

and at $\theta = 2 \gamma$

\begin{equation}
\Delta^2 ( \hat{p}_{\theta/2}^A \pm \hat{p}_{\theta/2}^B ) = \mathrm{e}^{\pm 2r} \; .
\end{equation}

\chapter{Derivations for the Squeezing Mechanism \label{sec:appendix_box_trap}}

\section{BdG Equations in a Box Trap \label{sec:appendix_BdG_eqns_box_trap}}

For a box type cavity of length $L$, cross section $A$ and volume $V=LA$, the condensate inside the trap is uniform $\psi_0 = \sqrt{n_0}$ and the chemical potential is $\mu = U_0 n_0$ (see section \ref{sec:uniform_weakly_interacting}). The BdG equations \eqref{eq:squeezing_mechanism_bdg_arbitrary_potential} thus become

\begin{IEEEeqnarray}{rCl}
	\IEEEyesnumber \label{eq:squeezing_apendix_BdG_eqns_box_potential_derivation} \IEEEyessubnumber*
	\hbar \omega_n u_n ( \mathbf{r} ) &=& \left( - \frac{\hbar^2}{2m} \nabla^2 - U_0 n_0 + 2 U_0 n_0 \right) u_n ( \mathbf{r} ) + U_0 n_0 v_n ( \mathbf{r} )
	\\
	-\hbar \omega_n v_n ( \mathbf{r} ) &=& \left( - \frac{\hbar^2}{2m} \nabla^2 - U_0 n_0 + 2 U_0 n_0 \right) v_n ( \mathbf{r} ) + U_0 n_0 u_n ( \mathbf{r} )
	\\
	\omega_n u_n ( \mathbf{r} ) &=& \frac{\hbar}{2m} \left[ \left( -  \nabla^2 + \frac{2m U_0 n_0}{\hbar^2} \right) u_n ( \mathbf{r} ) + \frac{2m U_0 n_0}{\hbar^2} n_0 v_n ( \mathbf{r} ) \right]
	\\
	- \omega_n v_n ( \mathbf{r} ) &=& \frac{\hbar}{2m} \left[ \left( -  \nabla^2 + \frac{2m U_0 n_0}{\hbar^2} \right) v_n ( \mathbf{r} ) + \frac{2m U_0 n_0}{\hbar^2} n_0 u_n ( \mathbf{r} ) \right]
\end{IEEEeqnarray}

with von Neumann boundary conditions at $x= \pm L/2$. $2m U_0 n_0/ \hbar^2$ can be recognized as the healing length $\xi$ \eqref{eq:healing_length_definition}. Furthermore, the perturbing potential $\delta \mathcal{V}$ depends only on $x$ (and on the time, of course). Therefore, only modes in the $x$-direction are considered \cite{ratzel_dynamical_2018} and the stationary BdG equations simplify to

\begin{IEEEeqnarray}{rCl}
	\IEEEyesnumber \label{eq:squeezing_apendix_BdG_eqns_box_potential_final} \IEEEyessubnumber*
	\omega_n u_n ( x) &=& \frac{\hbar}{2m} \left[ \left( -  \nabla^2 + \xi \right) u_n ( x) + \xi n_0 v_n ( x) \right]
	\label{eq:squeezing_apendix_BdG_eqns_box_potential_un}
	\\
	- \omega_n v_n ( x) &=& \frac{\hbar}{2m} \left[ \left( -  \nabla^2 + \xi \right) v_n ( x) + \xi n_0 u_n ( x) \right]
\end{IEEEeqnarray}

with the mode solutions \cite{ratzel_dynamical_2018}

\begin{IEEEeqnarray}{rClrCl}
	\IEEEyesnumber \label{eq:squeezing_apendix_BdG_mode_solutions} \IEEEyessubnumber*
	u_n &=& \alpha_n \varphi_n
	&
	v_n &=& \beta_n \varphi_n
	\\
	\alpha_n &=& \sqrt{\frac{1}{V}} \sqrt{\frac{1}{\sqrt{2}k_n \xi} + 1} \qquad
	&
	\beta_n &=& - \sqrt{\frac{1}{V}} \sqrt{\frac{1}{\sqrt{2}k_n \xi} - 1}
	\\
	\varphi_n &=& \cos \left( k_n (x+\frac{L}{2}) \right)
	&
	k_n &=& \frac{n \pi}{L}
\end{IEEEeqnarray}

from \ref{eq:squeezing_apendix_BdG_eqns_box_potential_un}, the Bogoliubov dispersion relation can be recovered in the limit $k_n \xi \ll 1$

\begin{IEEEeqnarray}{rCl}
\IEEEyesnumber \label{eq:squeezing_apendix_c_s_derivation} \IEEEyessubnumber*
	\omega_n &=& \frac{\hbar}{2 m \alpha_n \xi^2} \left[
		\left(
			\overbrace{
				(k_n \xi)^2
			}^{
				\approx 0
			} +1
		\right) \alpha_n + \beta_n
	\right]
	\\
	&=& \frac{\hbar}{2m \xi^2}
	\Big(
		1 +
		\underbrace{
			\frac{\beta_n}{\alpha_n}
		}_{
			-1 + \sqrt{2}k_n \xi + \mathcal{O}(k_n \xi)^2
		}
	\Big)
	\\
	\Rightarrow \omega_n &=& c_s k_n
\end{IEEEeqnarray}

by comparison to \eqref{eq:zeroth_order_speed_of_sound}.

\section{Derivation of the Interaction Picture Hamiltonian and the Time Evolution Operator for the Squeezing Mechanism \label{sec:appendix_matrix_elements_time_evolution_box_trap}}

Take the interaction Hamiltonian before changing into the interaction picture

\begin{equation}
	H^{(1)} = A \sin \Omega t \int \ud x \, \left( \delta \bar{\mathcal{V}} - \delta \bar{\mu} \right) \left( \sqrt{n_0} \left( \delta \hat{\psi} + \delta \hat{\psi}^\dagger \right) + \delta\hat{\psi}^\dagger \delta \hat{\psi} \right)
\end{equation}

and expand $\delta \hat{\psi}, \delta \hat{\psi}^\dagger$ in the mode solutions

\begin{IEEEeqnarray}{rRl}
	H^{(1)}_{int} = A \sin \Omega t
	& \int \ud x \left( \delta \bar{\mathcal{V}} - \delta \bar{\mu} \right) \Bigg[ &
	\sqrt{n_0} \sum_n \left( u_n + v_n \right) \left( \hat{b}_n + \hat{b_n}^\dagger \right)
	\nonumber
	\\
	& + &
	\sum_{n, m} u_n v_m \left( \hat{b}_n \hat{b}_m + \hat{b}_n^\dagger \hat{b}_m^\dagger \right)
	\nonumber
	\\
	& + &
	\sum_{n, m} u_n u_m \hat{b}_n \hat{b}_m^\dagger + v_n v_m \hat{b}_n^\dagger \hat{b}_m \Bigg] \; .
\end{IEEEeqnarray}

Dropping the ground state energy and thereby a global phase in the time evolution, we can use the diagonalized Bogoliubov Hamiltonian \eqref{eq:squeezing_mechanism_diagonalized_bogo_hamiltonian} to get for the quasi-particle creation and annihilation operators in the interaction picture $\hat{b}_{n, I}= \mathrm{e}^{-i \omega_n t} \hat{b}_n$ and $\hat{b}_{n, I}^\dagger = \mathrm{e}^{i \omega_n t} \hat{b}_n^\dagger$ and therefore

\begin{IEEEeqnarray}{rRl}
	\label{eq:squeezing_appendix_interaction_hamiltonian}
	H^{(1)}_{int, I} 
	& = A \sin \Omega t \Bigg[ &
	\sqrt{n_0} \sum_n \int \ud x \, \left(
		\delta \bar{\mathcal{V}} - \delta \mu
	\right) \left( u_n + v_n \right)
	\left(
		\mathrm{e}^{-i \omega t} \hat{b}_n + \mathrm{e}^{i \omega t} \hat{b}_n^\dagger
	\right)
	\nonumber
	\\
	& + &
	\sum_{n, m} \int \ud x \,
	\left(
		\delta \bar{\mathcal{V}} - \delta \mu
	\right)
	\bigg\{ u_n v_m
	\left(
		\mathrm{e}^{-i ( \omega_n + \omega_n ) t} \hat{b}_n \hat{b}_m + \mathrm{e}^{i ( \omega_n + \omega_m ) t} \hat{b}_n^\dagger \hat{b}_m^\dagger
	\right)
	\nonumber
	\\
	& & +
	u_n u_m \mathrm{e}^{-i ( \omega_n - \omega_m ) t} \hat{b}_n \hat{b}_m^\dagger
	+ v_n v_m \mathrm{e}^{i ( \omega_n - \omega_m ) t} \hat{b}_n^\dagger \hat{b}_m
	\bigg\}
	\Bigg] \; .
\end{IEEEeqnarray}

The index $I$ denotes the interaction picture here. In order to take the rotating wave approximation (RWA), pick out the term containing the $u_n u_m$ factor, use $\sin x = -\frac{i}{2} (e^{ix} + \mathrm{e}^{-ix})$ and split the sum up into contributions with $n>m$, $n<m$ and $n=m$. For the latter, one can immediately see that it is neglected in the RWA because the phase from the mode frequencies alone cancels out. For the former two parts of the sum, we can group the pre-factors to $\tilde{\mathcal{A}}_{nm} \equiv A \int \ud x \, \left(
		\delta \bar{\mathcal{V}} - \delta \mu
	\right) u_n v_m = \tilde{\mathcal{A}}_{mn}$. With that, we can write

\begin{IEEEeqnarray}{rCl}
	\IEEEyesnumber \label{eq:squeezing_appendix_RWA} \IEEEyessubnumber*
	\sum_{n>m} \tilde{\mathcal{A}}_{nm} \mathrm{e}^{-i ( \omega_n - \omega_m )t} \hat{b}_n \hat{b}_m^\dagger \sin \Omega t
	&+&
	\sum_{n<m} \tilde{\mathcal{A}}_{nm} \mathrm{e}^{-i ( \omega_n - \omega_m )t} \hat{b}_n \hat{b}_m^\dagger \sin \Omega t
	\\
	=
	\sum_{n>m} \tilde{\mathcal{A}}_{nm} \mathrm{e}^{-i ( \omega_n - \omega_m )t} \hat{b}_n \hat{b}_m^\dagger \sin \Omega t
	&+&
	\sum_{m<n} \tilde{\mathcal{A}}_{mn} \mathrm{e}^{-i ( \omega_m - \omega_n )t} \hat{b}_m \hat{b}_n^\dagger \sin \Omega t
	\\
	=
	\sum_{n>m} \tilde{\mathcal{A}}_{nm} \mathrm{e}^{-i ( \omega_n - \omega_m )t} \hat{b}_n \hat{b}_m^\dagger \sin \Omega t
	&+&
	\sum_{n>m} \tilde{\mathcal{A}}_{nm} \mathrm{e}^{i ( \omega_n - \omega_m )t} \hat{b}_n^\dagger \hat{b}_m \sin \Omega t
	\\
	\IEEEeqnarraymulticol{3}{l}{
		=
		\sum_{n>m} \underbrace{
			- \frac{i}{2} \tilde{\mathcal{A}}_{nm}
		}_{
			\equiv \mathcal{A}_{nm}
		}
		\left(
			\mathrm{e}^{-i ( \omega_n - \omega_m - \Omega ) t} \hat{b}_m^\dagger \hat{b}_n
			- \mathrm{e}^{i ( \omega_n - \omega_m - \Omega ) t} \hat{b}_n^\dagger \hat{b}_m
		\right)
	} \; .
	\label{eq:squeezing_appendix_RWA_step}
\end{IEEEeqnarray}

An analogous procedure is necessary for the term proportional to $v_n v_m$. For the first two lines in \eqref{eq:squeezing_appendix_interaction_hamiltonian}, the mode frequencies are either separate or added in the exponents and the sum does not need to be split up to take the RWA. One then gets

\begin{IEEEeqnarray}{rCl}
	H^{(1)}_{int, I}
	&=&
	\sum_n
		\mathcal{M}_{0n} (u_n + v_n)
		\left(
			\mathrm{e}^{-i ( \omega_n - \Omega ) t} \hat{b}_n - \mathrm{e}^{i ( \omega_n - \Omega ) t} \hat{b}_n^\dagger
		\right)
	\nonumber
	\\
	& &
	+ \sum_{n, m}
		\mathcal{M}_{nm}
		\left(
			\hat{b}_n \hat{b}_m \mathrm{e}^{-i ( \omega_n + \omega_m - \Omega ) t}
			- \hat{b}_n^\dagger \hat{b}_m^\dagger \mathrm{e}^{i ( \omega_n + \omega_m - \Omega ) t}
		\right)
	\nonumber
	\\
	& & 
	+ \sum_{n>m}
		\left(
			\mathcal{A}_{nm} + \mathcal{B}_{nm}
		\right)
		\left(
			\mathrm{e}^{-i ( \omega_n - \omega_m - \Omega ) t} \hat{b}_m^\dagger \hat{b}_n
			- \mathrm{e}^{i ( \omega_n - \omega_m - \Omega ) t} \hat{b}_n^\dagger \hat{b}_m
		\right)
\end{IEEEeqnarray}

with the matrix elements defined as

\begin{IEEEeqnarray}{rCl}
	\IEEEyesnumber \label{eq:squeezing_appendix_matrix_elements} \IEEEyessubnumber*
	\mathcal{M}_{0n} & \equiv &
	- \frac{i}{2} \sqrt{n_0} A \int_{-L/2}^{L/2} \ud x
	\left(
		\delta \bar{\mathcal{V}} - \delta \bar{\mu}
	\right)
	\left( u_n + v_n \right)
	\\
	\mathcal{M}_{nm} & \equiv & 
	- \frac{i}{2} A \int_{-L/2}^{L/2} \ud x
	\left(
		\delta \bar{\mathcal{V}} - \delta \bar{\mu}
	\right)
	u_n v_m
	\\
	\mathcal{A}_{nm} & \equiv &
	- \frac{i}{2} A \int_{-L/2}^{L/2} \ud x
	\left(
		\delta \bar{\mathcal{V}} - \delta \bar{\mu}
	\right)
	u_n u_m
	\\
	\mathcal{B}_{nm} & \equiv &
	- \frac{i}{2} A \int_{-L/2}^{L/2} \ud x
	\left(
		\delta \bar{\mathcal{V}} - \delta \bar{\mu}
	\right)
	v_n v_m \; .
\end{IEEEeqnarray}

In the phonon regime $k_n \ll \xi^{-1}$, we have $u_n \approx - v_n \approx (\sqrt{2} k_n \xi)^{-1/2} \cos (k_n (x+L/2))$ and therefore $\mathcal{M}_{nm} \approx - \mathcal{B}_{nm} \approx - \mathcal{B}_{nm}$

For a uniform ground state, the time-independent part of perturbation to the chemical potential \eqref{eq:squeezing_mechanism_chemical_potential_perturbation} reduces to

\begin{IEEEeqnarray}{rCl}
	\delta \bar{\mu} &=&
	\frac{
		A \int_{-L/2}^{L/2} \ud x \,
			n_0 \delta \bar{\mathcal{V}} (x)
	}{
		A \int \ud x \, n_0	
	}
	\nonumber
	\\
	&=&
	\frac{1}{L}
	\int_{-L/2}^{L/2} \ud x \,
		a_\Omega x + \mathfrak{G}_\Omega \frac{x^2}{2}
	\nonumber
	\\
	&=& \frac{\mathfrak{G}_\Omega L^2}{24} \; .
\end{IEEEeqnarray}

For the matrix element $\mathcal{M}_{0n}$, this means

\begin{IEEEeqnarray}{rCl}
	\mathcal{M}_{0n} &=&
	- \frac{i}{2} \sqrt{n_0} A \left( \alpha_n +  \beta_n \right) \cdot
	\nonumber
	\\ & & \cdot
	\int_{-L/2}^{L/2} \ud x \,
		\left( a_\Omega x + \mathfrak{G}_\Omega \left( \frac{x^2}{2} + \frac{L^2}{24} \right) \right) \cos \left( \frac{n \pi}{L} \left( x+ \frac{L}{2} \right) \right)
	\nonumber
	\\
	&\approx &
	i \sqrt{
		\frac{L^3 N_0 \xi}{\left( \sqrt{2} n \pi \right)^3}		
	}
	\left(
		\left(
			1- (-1)^n
		\right) \frac{a_\Omega}{L}
		-
		\left(
			1+ (-1)^n
		\right) \frac{\mathfrak{G}_\Omega}{2}
	\right) \; ,
\end{IEEEeqnarray}

where the coefficient $(\alpha_n + \beta_n )$ has been expanded up to $\mathcal{O}(k_n \xi)^{1/2}$.

Apart from differing coefficients, the matrix elements $\mathcal{M}_{nm}$, $\mathcal{A}_{nm}$ and $\mathcal{B}_{nm}$ all share the same integral which we shall denote by $\mathcal{C}_{nm}$.

\begin{IEEEeqnarray}{rCl} \label{eq:squeezing_appendix_integral}
	\mathcal{C}_{nm} &=&
	\int_{-L/2}^{L/2} \ud x \,
		\left(
			a_\Omega x + \mathfrak{G}_\Omega \left( \frac{x^2}{2} + \frac{L^2}{24} \right)
		\right)
		\cos \left( \frac{n \pi}{L} \left( x+ \frac{L}{2} \right) \right)
		\cos \left( \frac{m \pi}{L} \left( x+ \frac{L}{2} \right) \right)
		\nonumber
		\\
		&=&
		\begin{cases}
			- \frac{\left( m^2 + n^2 \right) L^3}{\left( m^2 - n^2 \right)^2 \pi^2}
			\left(
				\left( 1 - (-1)^{m+n} \right) \frac{a_\Omega}{L}
				-
				\left( 1 + (-1)^{m+n} \right) \frac{\mathfrak{G}_\Omega}{2}
			\right) & \text{for } n \neq m
			\\
			\mathfrak{G}_\Omega \frac{L^3}{8 n^2 \pi^2} & \text{for } n=m
		\end{cases}
\end{IEEEeqnarray}

Expanding again for small $k_n \xi$ and $k_m \xi$, the coefficients of the remaining matrix elements become

\begin{equation}
	\alpha_n \beta_m \approx - \alpha_n \alpha_m \approx - \beta_n \beta_m \approx - \frac{1}{\sqrt{2 m n} \pi \xi A} \; .
\end{equation}

Combine with \eqref{eq:squeezing_appendix_matrix_elements} and \eqref{eq:squeezing_appendix_integral} to get

\begin{IEEEeqnarray}{rCl}
	\mathcal{M}_{nm}
	& \approx &
	\begin{cases}
		-i  \frac{\left( m^2 + n^2 \right) L^3}{
			\sqrt{2 m n} \left( m^2 - n^2 \right)^2 \pi^3 \xi
		}
		\left(
			\left( 1 - (-1)^{m+n} \right) \frac{a_\Omega}{L}
			-
			\left( 1 + (-1)^{m+n} \right) \frac{\mathfrak{G}_\Omega}{2}
		\right)
		& \text{for } n \neq m
		\\
		i \frac{\mathfrak{G}_\Omega L^3}{16 \sqrt{2} n^3 \pi^2 \xi}
		& \text{for } n=m
	\end{cases}
	\nonumber
	\\
	\mathcal{A}_{nm}
	& \approx &
	\mathcal{B}_{nm} \approx - \mathcal{M}_{nm}
\end{IEEEeqnarray}

\chapter{Mathematica Code for the Symplectic Transformation and the Resulting Phonon Numbers and Uhlmann Fidelity \label{chap:app_mathematica_transformation}}



\section*{Supplementary material}

# Assumptions

*In[ • ]:=*
```
$Assumptions = l ∈ Reals && l > 0 && n ∈ Integers && n > l &&
    θ ∈ Reals && α ∈ Reals && β > 0 && ℏ > 0 && ω > 0 && r > 0 && L > 0 &&
    ξ > 0 && N0 > 0 && nOm ∈ Integers && nOm > 0 && x > 0 && x ∈ Reals
```

*Out[ • ]=*
l ∈ ℝ && l > 0 && n ∈ ℤ && n > l && θ ∈ ℝ && α ∈ ℝ && β > 0 && ω > 0 &&
    r > 0 && L > 0 && ξ > 0 && N0 > 0 && nOm ∈ ℤ && nOm > 0 && x > 0 && x ∈ ℝ

*In[ • ]:=* (* take r for the squeeze factor, x=βℏω/2 *)

# Setting the directory to save the evaluated results in

*In[ • ]:=*
```
SetDirectory[NotebookDirectory []]
```

# Factors between the squeezing, displacement and mode mixing parameters for a given $n_\Omega$

## "small" matrix elements (the modulus of the ones without the $a_\Omega$ and the $G_\Omega$ factors)

*In[ • ]:=*
```
m0n[L_, N0_, ξ_, n_] := Sqrt[L^3 * N0 * ξ / (Sqrt[2] * n * Pi)^3]
mnl[L_, ξ_, l_, n_] :=
   (l^2 + n^2) * L^3 / (2 * Sqrt[2 * l * n] * (l^2 - n^2)^2 * Pi^3 * ξ)
mnn[L_, ξ_, n_] := L^3 / (16 * Sqrt[2] * n^3 * Pi^3 * ξ)
```

129

# 2-mode squeezing and the mixing parameters

Factor between the 2 - mode squeeze factor and the parameter for mixing with the lower one of the two higher modes

*In[ • ]:=* `xr2θ1[n_, l_] := mnl[L, ξ, n + l, l] / mnl[L, ξ, l, n - l]`

Definition of the parameter for mixing with the lower one of the two higher modes in terms of the squeeze factor and the mode numbers

*In[ • ]:=* `θr21[n_, l_, r2mode_] := xr2θ1[n, l] * r2mode`

Factor between the 2 - mode squeeze factor and the parameter for mixing with the lower one of the two higher modes

*In[ • ]:=* `xr2θ2[n_, l_] := mnl[L, ξ, 2 n - l, n - l] / mnl[L, ξ, l, n - l]`

Definition of the parameter for mixing with the lower one of the two higher modes in terms of the squeeze factor and the mode numbers

*In[ • ]:=* `θr22[n_, l_, r2mode_] := xr2θ2[n, l] * r2mode`

# Displacement and the mixing parameter as a function of the 2-mode squeeze factor

Factor between the 2 - mode squeeze factor and the displacement parameter

*In[ • ]:=* `xrα[N0_, L_, ξ_, n_, l_] := 1 / 2 * m0n[L, N0, ξ, n] / mnl[L, ξ, l, n - l]`

Definition of the displacement parameter in terms of the squeeze factor, the mode numbers and experimental parameters

*In[ • ]:=* `αnOm[L_, N0_, ξ_, n_, l_, r2mode_] := xrα[N0, L, ξ, n, l] * r2mode`

> Factor between the displacement parameter and the corresponding mode mixing parameter

*In[•]:=* `xαθ[L_, N0_, ξ_, n_] := 2 * mnl[L, ξ, 2 * n, n] / m0n[L, N0, ξ, n]`

> Definition of the mode mixing parameter between the displaced mode and the corre-
> sponding higher mode in terms of the 2-mode squeeze factor, the mode numbers and
> experimental parameters

*In[•]:=* `θα[L_, N0_, ξ_, n_, l_, r2mode_] :=`
   `xαθ[L, N0, ξ, n] * αnOm[L, N0, ξ, n, l, r2mode]`

# 1-mode squeezing and the mixing parameter as a function of the 2-mode squeeze factor

> Factor between the 2-mode squeeze factor and the 1-mode squeeze factor

*In[•]:=* `xrr[n_, l_] := mnn[L, ξ, n / 2] / mnl[L, ξ, l, n - l]`

> Definition of the 1-mode squeeze factor in terms of the 2-mode squeeze factor and the
> mode numbers

*In[•]:=* `r1mode[n_, l_, r2mode_] := xrr[n, l] * r2mode`

> Definition of the parameter for mixing between the 1-mode squeezed mode and some
> higher mode.

*In[•]:=* `xr1θ := mnl[L, ξ, 3 / 2 * n, 1 / 2 * n] / mnn[L, ξ, 1 / 2 * n]`

# Time Evolution

## Definition of Squeezed States and Symplectic

# Squeezing Transformation for Later Comparison

Definition of a 2-mode squeezing symplectic transformation and squeezed vacuum state in the complex CM picture

```
In[ ∘ ]:=  S2sqC[r_] :=
    MatrixExp[-r*{{0, 0, 0, 1}, {0, 0, 1, 0}, {0, 1, 0, 0}, {1, 0, 0, 0}}]
σ2sqC[r_] := Transpose[S2sqC[r]].S2sqC[r]
```

---

# Thermal States and their particle numbers
# Particle number for a general n-mode Gaussian state (independent of the chosen CM picture)

Definition of a single-mode thermal state in mode n in the complex CM picture
Definition of the single-mode and the overall particle number for an infinite mode number thermal state with $\omega_n = n*\omega_1$

```
In[ ∘ ]:=  σth1C[β_, ω_, n_] := Coth[β*ℏ*ω*n/2]*IdentityMatrix[2]
ntherm1[β_, ω_, n_] := 1/(Exp[β*ℏ*ω*n]-1)
ntherm1x[x_, n_] := 1/(Exp[2*x*n]-1)
```

And the combined average particle number of modes n=1,... ∞

Mathematica refuses to do the sum with an exponential function, so we have to do this

```
In[ ∘ ]:=  Assuming[x ∈ Reals && x > 1,
    {Refine[Sum[1/(x^n-1), {n, 1, Infinity}]]}] // FullSimplify
```

$$Out[ ∘ ]= \left\{ \frac{-2\,Log[-1+x]+Log[x]-2\,QPolyGamma[0, 1, x]}{2\,Log[x]} \right\}$$

Replace x⟶Exp[y] and refine

*In[ • ]:=* `Assuming[y ∈ Reals && y > 0,`

  `{Refine[` $\dfrac{-2\,\text{Log}[-1+\text{Exp}[y]]+\text{Log}[\text{Exp}[y]]-2\,\text{QPolyGamma}[0,1,\text{Exp}[y]]}{2\,\text{Log}[\text{Exp}[y]]}$ `]}] //`

  `FullSimplify`

*Out[ • ]=* $\left\{\dfrac{y-2\,\text{Log}[-1+e^{y}]-2\,\text{QPolyGamma}[0,1,e^{y}]}{2\,y}\right\}$

Replace $y \longrightarrow \beta^{*}\hbar^{*}\omega$

*In[ • ]:=* `nthermtot[x_] :=` $\dfrac{(2\,x)-2\,\text{Log}[-1+e^{(2\,x)}]-2\,\text{QPolyGamma}[0,1,e^{(2\,x)}]}{2\,(2\,x)}$

### Definition of the thermal state in modes (l, $n_\Omega - l$, $n_\Omega$+l, 2$n_\Omega$-l)

**Keep in mind: $x = \beta\hbar\omega_1/2$**

*In[ • ]:=* `σth4mode[nOm_, l_, x_] :=`
  `KroneckerProduct[IdentityMatrix[2], DiagonalMatrix[`
    `{Coth[l * x], Coth[(nOm - l) * x], Coth[(nOm + l) * x], Coth[(2 nOm - l) * x]}]]`
  `σth2mode[nOm_, x_] := KroneckerProduct[IdentityMatrix[2],`
    `DiagonalMatrix[{Coth[nOm * x], Coth[(3 * nOm) * x]}]]`

### Definition of the particle number for an n-mode Gaussian state with CM $\sigma$ and 1st moments vector d

*In[ • ]:=* `NGauss[σ_, d_, n_] := 1/4 * (`
      `Tr[σ] + d.d - 2 * n`
  `)`

# Uhlmann Fidelity for two-mode Gaussian states with vanishing first moments in the complex CM formalism

**IMPORTANT REMARK: Something is ill-defined here: I am getting imaginary values for the Uhlmann fidelity further down. Either the function is wrong or the CM does not corre-**

## Define the Determinants that enter the Uhlmann Fidelity and the symplectic form for two modes

```
In[ • ]:=  Kc2 := KroneckerProduct [{{1, 0}, {0, -1}}, IdentityMatrix [2]]
         Δ[σ1_, σ2_] := Det[σ1 + σ2]
         Γ[σ1_, σ2_] := Det[IdentityMatrix [4] + Kc2.σ1.Kc2.σ2]
         Λ[σ1_, σ2_] := Det[σ1 + Kc2] * Det[σ2 + Kc2]
         UhlFid[σ1_, σ2_] := 4 / (
               (Sqrt[Γ[σ1, σ2]] + Sqrt[Λ[σ1, σ2]])
                 -
                 Sqrt[
                     (Sqrt[Γ[σ1, σ2]] + Sqrt[Λ[σ1, σ2]])^2 - Δ[σ1, σ2]
                 ]
             )
```

# Definition of the symplectic transformation in the modes affected by 2-mode Squeezing and mode mixing
# Definition of the resulting state for a given set of modes , $r_{l, n\Omega-l}$ and $\beta\hbar\omega_1/2$

## W-matrix for the Unitary Time Evolution

Matrix encoding the unitary transformation for $n_\Omega$=2, U=Exp[i/2*A$^+$.W.A + A$^+$.K.$\gamma$] in the complex covariance matrix formalism
Symplectic form in the complex covariance matrix formalism

```
In[ • ]:=  W2Sq[x_, y_, z_] := -I * {
              {0, 0, y, 0, 0, x, 0, 0},
              {0, 0, 0, z, x, 0, 0, 0},
              {-y, 0, 0, 0, 0, 0, 0, 0},
              {0, -z, 0, 0, 0, 0, 0, 0},
              {0, -x, 0, 0, 0, 0, -y, 0},
              {-x, 0, 0, 0, 0, 0, 0, -z},
              {0, 0, 0, 0, y, 0, 0, 0},
              {0, 0, 0, 0, 0, z, 0, 0}}
           Kc[n_] := KroneckerProduct[{{1, 0}, {0, -1}}, IdentityMatrix[n]]
```

## Transformed state for a given set of modes , $r_{l,\,n\Omega-l}$ and $\beta\hbar\omega_1/2$

Redefine the symplectic transformation as full matrix exponential right away

```
In[ • ]:=  Symp2SqueezMix[r_, nOm_, l_] :=
              MatrixExp[I * Kc[4].W2Sq[r, xr2θ1[nOm, l] * r, xr2θ2[nOm, l] * r]]
```

Evaluate it for particular modes and use that evaluation to define the transformations

## Table of Symplectic Transformations

The inner table is a table of the symplectic trafos for a given $n_\Omega$, starting with the maximum l satisfying $l < n_\Omega/2$ and going down from there.

The outer table is counting through possible $n_\Omega$, starting with $n_\Omega = 3$.

this means that for a symplectic transformation for 2-mode squeezing with a particular $n_\Omega$, we have to call

Symp2SqueezMixTable[r]    [[$n_\Omega$-2]] [[ Ceiling[ $n_\Omega$/2-1] ]]

where r is the squeeze factor for the maximum l in $r_{l,n_\Omega-l}$

```
(*Symp2SqueezMixTable [r_]=Table[Table[Symp2SqueezMix [
    mnl[L, ξ, k, nOm-k]/mnl[L, ξ, Ceiling[nOm/2-1], nOm-Ceiling[nOm/2-1]]*r,
    nOm, k
  ], {k, Ceiling[nOm/2-1], 1, -1}], {nOm, 3, 11}]
DumpSave["Symp2SqueezMixTable .mx", Symp2SqueezMixTable ]*)
```

Out[ ∘ ]=

{ ⋯ 1 ⋯ }

large output    show less    show more    show all    set size limit …

Out[ ∘ ]= {Symp2SqueezMixTable }

Inf[ ∘ ]= `Get["Symp2SqueezMixTable31 .mx", Symp2SqueezMixTable ]`

## Table of Tables of the Out States

The one we will actually use is $\sigma$out2SqueezMixtabletable[r, x][[nOm]] [[l, counting down from lmax=Ceiling[nOm/2-1] and stopping at 1]]

Do not run this, it takes long depending on up to where you evaluate it.
The Get below gets it from the storage and is an initialization cell.

```mathematica
σout2SqueezMix[r_, nOm_, l_, x_] := Symp2SqueezMixTable [r][[nOm - 2]][[1]].
  σth4mode[nOm, l, x].Symp2SqueezMixTable [r][[nOm - 2]][[1]]†
σout2SqueezMixtable [r_, nOm_, x_] :=
  Table[σout2SqueezMix[r, nOm, l, x], {l, Ceiling[nOm / 2 - 1], 1, -1}]
(*σout2SqueezMixtabletable [r_, x_]=
  Table[σout2SqueezMixtable [r, nOm, x], {nOm, 3, 11}]
    DumpSave["σout2SqueezMixtabletable .mx", σout2SqueezMixtabletable ]*)
```

Out[ ]=

$$\left\{\left\{\left\{ \boxed{\cdots\ 1\ \cdots} \right\}, \boxed{\cdots\ 6\ \cdots}, \left\{ \frac{888\,125\ \sqrt{5}\ (74\,127\,111 + 31\ \sqrt{5\,717\,216\,466\,761})\,e^{-\frac{461}{16\pi}\sqrt{\frac{2}{5\,(74\,127\,111\ \cdots)}}r}}{31\,(177\,233\,710\,469\,591\ +74\,127\,111\ \sqrt{5\,717\,216\,466\,761}\ )} + \frac{888\,125\ \cdots}{31\,(17} \right. \right.\right.$$

| large output | **show less** | **show more** | **show all** | **set size limit …** |

Out[ ]=  {σout2SqueezMixtabletable }

In[ ]=  `Get["NEWsigmaOut2SqueezMixTable31 .mx", σout2SqueezMixtabletable ]`

Final and produced particle number for a given $n_\Omega$

In[ ]:=
```mathematica
N2SqueezTherm [r_, x_, nOm_] :=
  Sum[NGauss[σout2SqueezMixtabletable [r, x][[nOm - 2]][[l]], {0, 0, 0, 0}, 4],
    {l, 1, Ceiling[nOm / 2 - 1]}]
ΔN2SqueezTherm [r_, x_, nOm_] := N2SqueezTherm [r, x, nOm] -
  Sum[ntherm1x[x, l] + ntherm1x[x, nOm - l] + ntherm1x[x, nOm + l] +
    ntherm1x[x, 2 * nOm - l], {l, 1, Ceiling[nOm / 2 - 1]}]
```

# Definition of the symplectic transformation in the modes ($n_\Omega$, $2\,n_\Omega$) affected by displacement and mode mixing

W-matrix for the Unitary Time Evolution and symplectic transformation

## vector γ for the change in first moments

Depends only on the mode mixing parameter, the displacement only affects the first moments.

```
In[•]:=  WDispMix[θ_] :=
          1 / I * θ * {{0, -1, 0, 0}, {1, 0, 0, 0}, {0, 0, 0, 1}, {0, 0, -1, 0}}
        SDispMix[θ_] := MatrixExp[I * Kc[2].WDispMix[θ]]
        γDispMix[α_] := α * {1, 0, 1, 0}
```

## For the change b in the 1st moments vector d

We assume vanishing 1st moments in the in state.
Take the integral over the symplectic transformation and multiply it onto the γ vector

```
In[•]:=  bDispMix[α_, θ_] := Integrate[SDispMix[θ * t], {t, 0, 1}].(γDispMix[α])
```

## Number of particles created

Take the norm first and use the result in the definition below so that the calculation does not take ages every time

```
In[•]:=  Norm[bDispMix[α, θ]]^2 / 4 // FullSimplify
```

$$Out[•]= -\frac{\alpha^2 (-1 + \text{Cos}[\theta])}{\theta^2}$$

```
In[•]:=  ΔNDispMix[L_, N0_, ξ_, nOm_, l_, r2mode_] :=
          -\frac{αnOm[L, N0, ξ, nOm, l, r2mode]^2 (-1 + Cos[θα[L, N0, ξ, nOm, l, r2mode]])}{θα[L, N0, ξ, nOm, l, r2mode]^2}
```

# Definition of the symplectic transformation in the modes $(n_\Omega/2, \ 3\,n_\Omega/2)$ affected by single-mode squeezing and mode mixing

The ratio between the single-mode squeeze factor and the corresponding mode mixing parameter is always $\frac{5}{4\sqrt{3}}$

The definition for the symplectic transformation accounts for no single-mode squeezing

happening for odd $n_\Omega$.

```
In[ ]:=  W1SqueezMix[r1mode_] :=
          -r1mode / I * {{0, 0, 1, 0}, {0, 0, 0, 0}, {-1, 0, 0, 0}, {0, 0, 0, 0}} +
            1 / I * ———— * r1mode *
                    4 √3
             {{0, -1, 0, 0}, {1, 0, 0, 0}, {0, 0, 0, 1}, {0, 0, -1, 0}}
          S1SqueezMix[nOm_, r1mode_] :=
          1 / 2 * ((1 + (-1)^nOm) * MatrixExp[I * Kc[2].W1SqueezMix[r1mode]]
              + (1 - (-1)^nOm) IdentityMatrix[4])
```

```
In[ ]:=  S1SqueezMix[3, r] // MatrixForm // FullSimplify
```

```
Out[ ]//MatrixForm=
        ⎛ 1  0  0  0 ⎞
        ⎜ 0  1  0  0 ⎟
        ⎜ 0  0  1  0 ⎟
        ⎝ 0  0  0  1 ⎠
```

## Transformed state for a given set of modes , $r_{n\Omega/2}$ and $\beta\hbar\omega_1/2$

**Keep in mind: $x = \beta\hbar\omega_1/2$ in the argument of the thermal state**

Remark: Mathematica was acting up and giving the out state an imaginary part 0. in every
entry for whatever reason. Testing it with $\mathrm{Im}[\sigma\mathrm{out1SqueezMix}[4, 6, 0.15]][[1, 1]]==0$
returned true, so I am defining it as just the real part because I know for a fact that it is real.

```
In[ ]:=  σout1SqueezMix[nOm_, r_, x_] :=
          Re[S1SqueezMix[nOm, r].σth2mode[nOm / 2, x].S1SqueezMix[nOm, r]ᵀ]
```

## Number of particles created

1st moments do not get changed by squeezing and mode mixing.

```
In[ ]:=  N1SqueezTherm[nOm_, l_, r2mode_, x_] :=
          NGauss[σout1SqueezMix[nOm, r1mode[nOm, l, r2mode], x], {0, 0}, 2]
        ΔN1SqueezTherm[nOm_, l_, r2mode_, x_] :=
          N1SqueezTherm[nOm, l, r2mode, x] - ntherm1x[x, nOm / 2] - ntherm1x[x, 3 nOm / 2]
```

# Number of particles created by the time

# evolution

Particles created are sum of the ones created by 2-mode squeezing, single-mode squeezing and displacement (each with mode mixing)

> Keep in mind: $x = \beta \hbar \omega_1 / 2$

```
In[ ]:= ΔNtot[L_, N0_, ξ_, r2mode_, nOm_, x_] :=
          N[ΔNDispMix[L, N0, ξ, nOm, Ceiling[nOm/2 - 1], r2mode]] +
           ΔN1SqueezTherm[nOm, Ceiling[nOm/2 - 1], r2mode, x] +
           ΔN2SqueezTherm[r2mode, x, nOm]
        Ntotout[L_, N0_, ξ_, r2mode_, nOm_, x_] :=
          ΔNtot[L, N0, ξ, r2mode, nOm, x] + nthermtot[x]
        r2modeMaxOut[L_, N0_, ξ_, nOm_, x_] := FindRoot[
            Ntotout[L, N0, ξ, r2modeMax, nOm, x] == N0/50, {r2modeMax, 3}][[1, 2]]
```

```
In[ ]:= Fidr2modeMax[r2modeMax_, nOm_, x_] := NMaximize[
          Re[UhlFid[Drop[Drop[σout2SqueezMixtabletable[r2modeMax, x][[nOm - 2]][[1]],
             {3, 4, 1}, {3, 4, 1}], {-2, -1, 1}, {-2, -1, 1}], σ2sqC[-r]]], r]
```

140




\end{document}